\documentclass{aa}  

\usepackage{graphicx}
\usepackage{txfonts}
\begin{document}

 \title{Optical and UV surface brightness of translucent dark nebulae.} 
\subtitle{Dust albedo, radiation field and fluorescence emission by H$_2$ 
\thanks{Based on observations collected at the Centro Astronómico Hispano Alemán (CAHA) at Calar Alto, 
operated jointly by the Max-Planck Institut für Astronomie and the Instituto de Astrofísica de Andalucía (CSIC)\newline
 Based on observations collected at the European Organisation for Astronomical Research in the Southern 
Hemisphere}}

   \author{K. Mattila\inst{1}
          \and
           M. Haas \inst{2}
          \and
          L.\,K. Haikala \inst{3}
          \and 
          Y-S. Jo \inst{4}
          \and
          K. Lehtinen\inst{1,8}
          \and
          Ch. Leinert\inst{5}
          \and
          P. V\"ais\"anen \inst{6,7}
                         }

%             }
\institute{Department of Physics, University of Helsinki, P.O. Box 64, FI-00014 Helsinki, Finland
\email{kalevi.mattila@helsinki.fi} 
\and
Astronomisches Institut, Ruhr-Universit\"at Bochum,  Universit\"atsstrasse 150, D-44801 Bochum, Germany 
\and
Instituto de Astronomía y Ciencias Planetarias de Atacama, Universidad de Atacama, Copayapu 485, Copiapo, Chile  
\and
Korea Astronomy and Space Science Institute (KASI), 776 Daedeokdae-ro, Yuseong-gu, Daejeon, 305-348, Korea
\and
Max-Planck-Institut f\"ur Astronomie, K\"onigstuhl 17, D-69117 Heidelberg, Germany
\and
South African Astronomical Observatory, P.O. Box 9 Observatory, Cape Town, South Africa
\and
Southern African Large Telescope, P.O. Box 9 Observatory, Cape Town, South Africa
\and
Finnish Geospatial Research Institute FGI, Geodeetinrinne 2, FI-02430 Masala, Finland
}
   \date{Received April xx, 2018; accepted April yy, 2018}

\newcommand{\cgs}{$10^{-9}$\,erg~cm$^{-2}$s$^{-1}$sr$^{-1}$\AA$^{-1}$}\,

% \abstract{}{}{}{}{} 
% 5 {} token are mandatory
 
  \abstract
   {Dark nebulae display a surface brightness because dust grains scatter light of the general 
interstellar radiation field (ISRF). High-galactic-latitudes dark nebulae are seen as bright nebulae 
when surrounded by transparent areas which have less scattered light from the general galactic dust layer.}
   {Photometry of the bright dark nebulae LDN~1780, LDN~1642 and LBN~406 shall be used to derive scattering 
properties of dust and to investigate the presence of UV fluorescence emission by molecular hydrogen 
and  the extended red emission (ERE).}
   {We used multi-wavelength optical photometry and imaging at ground-based telescopes and archival 
imaging and spectroscopic UV data from the spaceborn $GALEX$ and SPEAR/FIMS instruments. In the analysis we 
used Monte Carlo RT and both observational data and synthetic models for the ISRF in the solar neighbourhood.
The line-of-sight extinctions through the clouds have been determined using near infrared excesses of background
stars and the 200/250 $\mu$m far infrared emission by dust as measured using the $ISO$ and $Herschel$ space observatories.}
{The optical surface brightness of the three target clouds can be explained in terms of scattered light.
The dust albedo ranges from  $\sim$0.58 at 3500\,\AA\, to  $\sim$0.72 at 7500\,\AA\,.
{ The SED of LDN~1780 is explained in terms of optical depth and background 
scattered light effects instead of the original published suggestion in terms of ERE.}  
The FUV surface brightness of  LDN~1780 cannot be explained by scattered light only. 
In LDN~1780 H$_2$ fluorescent emission in the wavelength range 1400 -- 1700 \AA\, has been detected 
and analysed. }
{Our albedo values are in good agreement with the predictions of the dust model of 
Weingartner and Draine and with the THEMIS CMM model for evolved core-mantle grains.
The distribution of  H$_2$ fluorescent emission in LDN~1780 shows a pronounced dichotomy with a strong preference for 
its southern side where enhanced illumination is impinging from the Sco\,OB2 association and the O star $\zeta$ Oph. 
A good correlation is found between the  H$_2$ fluorescence and a previously mapped 21-cm excess emission.
The H$_2$ fluorescence emission in LDN~1780 has been modelled using a PDR code; the resulting values for
H$_2$ column density  and  the total gas density are consistent with the estimates derived from CO observations 
and optical extinction along the line of sight. }


   \maketitle
%
%-------------------------------------------------------------------
\section{Introduction}

In studies of the physics and chemical composition of the interstellar grains there has been recent renewal 
of interest in the observational results of the albedo $a$ and the asymmetry parameter of the scattering 
function $g=\langle{\rm cos}\theta\rangle$ of the grains over the optical, UV and near-IR wavelength regions, 
see for example \citet{jones,ysard, togi,lee08,lim15} and \citet{murthy16}. Surface brightness of dark nebulae 
offers a good observational means for their 
determination. A review of earlier results has been presented by \citet{gordon04}. 

While studying the scattered light of dark nebulae there are two other surface brightness components which have
to be taken into account. They are also important and interesting on their own right: 

Far UV fluorescence emission by molecular hydrogen was introduced by  \citet{williams} who 
suggested it as an additional component to explain the high dust albedo in the FUV as detected by 
\citet{lillie}. { \citet{witt89} reported the first detection outside the solar systen. 
Its presence in diffuse ISM was observationally 
confirmed by \citet{martin} and in a dark nebula by \citet{hurwitz}.} An all-sky survey performed with the
Far Ultraviolet Imaging Spectrograph (FIMS), also known as the Spectroscopy of Plasma 
Evolution from Astrophysical Radiation (SPEAR) 
\citep{edelstein06a,seon11} has revealed its wide distribution and importance for ISM studies \citep{jo17}. 
Detailed studies have already been made of a number of individual ISM targets, including dark nebulae, 
and have been presented for example by \citet{park09,jo11} and \citet{lim15}.     

Extended red emission (ERE) is a photoluminence phenomenon of interstellar grains
{ first observed in the Red Rectangle by \citet{cohen75}, \citet{greenstein77} and  
\citet{schmidt80},  and  in reflection nebulae by \citet{witt84}.} 
It has been since then studied in a number of dusty interstellar environments, for a review see 
\cite{witt04}. Its contribution is mainly between { wavelengths of $\lambda \sim$5400 to 9000\,\AA\,. }
Spectrophotometric observations of LDN~1780 by \citet{mattila79} were interpreted by \citet{chlewicki},
and  
\citet{gordon}, \citet{smith02} and others in terms of ERE. Later on the detection of ERE was announced 
by \citet{witt08} in several high-latitude dark nebulae.

The target clouds of this study,
LDN~1780 (including LDN~1778, \citealt{lynds}), LDN~1642 and LBN~406 \citep{lynds65}, hereafter 
called the Draco nebula, are also well known as bright dark nebulae. For their basic properties
see Table~\ref{table:0}. Most of their diffuse light is 
considered to originate as scattering of the interstellar radiation field (ISRF) photons by dust 
grains in the clouds. While all dark nebulae exhibit the scattered light component it is not
plainly visible in the cases where the nebula is projected against a bright low-galactic-latitude sky. 
The character of LDN~1780 surface brightness was discussed already by \citet{struve}. 

Photometric observations of LDN~1780 and their interpretation in terms of grain scattering properties 
were presented by  \citet{mattila70a, mattila70b}. H$\alpha$ surface brightness of LDN~1780 was suggested to be due 
to in situ emission by \citet{delburgo}. However, \citet{mattila07} and \citet{witt10} have shown that most 
if not all of it originates as scattered light. Detection of radio emission by spinning very small grains
by \citet{vidal} and of molecular hydrogen IR lines by \citet{ingalls} emphasize the
growing interest for LDN~1780 as a laboratory for ISM physics. 
For the distance of LDN~1780 we have adopted the weighted mean of the distances of the LDN~134/183/1780 
group of clouds, $d = 110\pm10$\,pc, as determined  by \citealt{schlafly14} with optical photometry of 
stars from the Pan-STARRS-1 data release in the area  $l\sim-1\degr -  +11\degr, b\sim36\degr - 38\degr$.

LDN~1642 has been extensively studied by optical, infared, mm- and cm- molecular and H\,{\small I} 21-cm lines
(see for example \citealt{sandell}, \citealt{lehtinen04},\, and references therein).
Over most of the area its extinction ($A_V \la 4$\ mag) characterizes LDN~1642 as translucent, but 
it has also an opaque core with $A_V > 15$\,mag. And associated with the core there are two newly 
formed stars \citep{sandell87}, a rare occurrence for such a high-latitude ($b \sim -36\degr$) cloud. 
Its distance is $d = 124^{+11 }_{-14 }$ pc, as determined with  Pan-STARRS-1 data \citep{green}.

The Draco nebula has been detected as an intermediate velocity H\,{\small I} 21-cm cloud { \citep{goerigk}} and
has been studied also via molecular line emission and optical surface brightness \citep{mebold, witt08}.
It is at a substantial distance of $\sim800$\,pc \citep{gladders,penprase}, making it an object belonging
to the inner Galactic halo. FUV fluorescence emission of the Draco area by \citet{park09} and 
of a foreground dust filament close to the Draco main cloud by \citet{sujatha10} have been studied. 
In this paper we are mainly concerned with the brightest part, 'The Head' of Draco.

Because of their different distances from the Galactic plane, LDN~1780, LDN~1642 ($z \approx 60$ pc) 
and Draco nebula ($z \approx 400$ pc) are exposed to different ISRF. Moreover, their different
directions as seen from the Sun cause a different scattering geometry because of the 
anisotropic scattering function of dust. It is also possible that the dust scattering properties
in the intermediate velocity Draco cloud differ from those in the local clouds.

\begin{table*}
\caption{Some basic properties of the sample of dark nebulae and the additional UV sources of 
illumination for LDN~1780. The last column gives the reference for distance and.
$A_V$\,\,range refers to the range of observed positions in 
the present study. }
\label{table:0}
\centering 
\begin{tabular}{lccrrcccc} 
\hline
Object      & R.A.& Dec.& $l\,\,\,$ & $b\,\,\,$ & $d\,\,\,$[pc]  &$A_V$\,\,range[mag] &$A_V$\,\,max[mag] & Ref.\\
    
\hline
LDN~1642             &04:36      &-14:12   &210.9&-36.5&$124^{+11 }_{-14 }$& 0 -- 4    &$>10$      &  1     \\
LDN~1780             &15:40      &-07:15   &359.0&37.0 &$110\pm10$        &0.4 -- 3.5 &$\sim4$    &  1     \\
LBN~406 (Draco nebula, 'Head')&16:48      &59:55    &90.0 &39.0 &$\sim$800         & 0 -- 1.3  & $\sim2$   &  2     \\
Sco\,OB2             &16:15      &-24:12   &351.4&19.0 & $144\pm3$        & 0.11 -- 1.76&         &  3     \\
$\zeta$\,Oph         &16:37      &-10:34   &6.3  &23.6 & $112\pm3$        & 0.90      &           &  4     \\     
\hline
\end{tabular}

\tablefoot{(1) \citet{schlafly14}; (2) \citet{penprase,gladders};
(3) \citet{dezeeuw,gordon94}; (4) \citet{vanleeuwen07,liszt} 
}
\end{table*}

\section{Observations and archival data}

We have observed LDN~1780 and Draco clouds with the same photometric equipment at the Calar Alto Observatory 
using the same methods and calibration. LDN~1642 was observed at ESO/La Silla using very similar methods.
The homogeneity of the observational material thus enables their comparison 
essentially free of instrumentally based differences. In addition, we have made wide field CCD imaging with 
the University of Bochum VYSOS6 telescope of LDN~1780 in the $BVRi$ bands and of  LDN~1642 in the $i$ band.
We have also used archival $GALEX$ near and far UV data for LDN~1780 and the Draco nebula, and SPEAR/FIMS
far UV data for  LDN~1780.

\subsection{Intermediate band surface photometry of LDN~1780 and Draco nebula}
Observations  of the diffuse sky brightness in the areas of L~1780 and
Draco nebula were carried out using the 2.2-m and 1.23-m telescopes of the Calar Alto 
Observatory. Each telescope was equipped with a photoelectric photometer, with 
photomultiplier of type RCA 31034A-02 and an identical set of filters. The following six
filters were used, with central wavelengths and, in parentheses, the half widths given in \AA\,: 
Strömgren $u$ 3500 (300) and $b$ 4670 (200) and Omega Optical Inc. interference filters 3920 (90), 4030 (90), 
5250 (250), and 5800 (120), and in addition for L1780 only, the filters 7100 (150), and 8200 (150).
The response curves of the filters as well as other details of the photometers are found in
\citet{leinert95}. The observations were made in the nights 1989-05-02, 05-05, 05-06, 
1990-06-21, 06-22, 06-23, and 1991-06-16. 

In order to eliminate the influence of the airglow time variations we used a two-telescope technique.
The 1.23-m telescope was used as a monitor and was pointed towards a fixed position (R.A., Dec.) in the dark
nebula area during the whole measuring cycle and its integration  times in each filter were
synchronized with those of the 2.2-m within $\sim$1 s. The 2.2-m telescope was used to measure several 
individual positions within or outside the dark cloud in an area of $\sim$1 -- 2 deg in size. 
This method was possible because the ratio of the intensities as measured at the two telescopes remained
very stable during a whole night despite the temporal variations of the airglow. A simple linear
interpolation between two ``standard position'' measurements was justified by the stability of the
ratio values.

In Table \ref{table:1} we give the coordinates of the observed positions in and around LDN~1780 and the
Draco nebula. Also given are the 250 $\mu$m  or the 200 $\mu$m intensities as measured by the 
{\em Herschel Space Observatory} 
\citep{pilbratt} or with the ISOPHOT instrument \citep{lemke} aboard $ISO$ \citep{kessler} as well as 
estimates for visual extinction. The standard position, the ON and the OFF positions are indicated for each cloud.
A standard position was measured typically once or twice an hour. The 1.23-m measurements were used 
to calculate ``interpolated standard position'' values for exactly the same times when the 
``other positions'' were measured at the 2.2-m telescope.
The surface brightnesses of all the other positions, both the OFF positions outside the cloud and the
ON positions within the cloud, were measured relative to the standard position. The differential 
surface brightness of the standard position was then determined as the excess over the mean of
the OFF position values. The other ON position values were referred to the standard position 
value.  In Table \ref{table:2} we give the differences ON -- OFF relative to the 
mean of the OFF-cloud values. 

The diameter of the focal plane circular aperture for the 2.2-m photometer was 118.4 arcsec. 
The effective aperture solid angle was determined by mapping the response with a star placed at 
a grid of positions within the aperture. It was found to be 6.47\,$10^{-4}$ deg$^2$ or  
1.97\,$10^{-7}$ sterad, corresponding to an aperture correction factor of 0.76. The diameter
of the aperture used at the 1.23-m telescope was 10.5 arc min. No aperture correction factor was 
required because it was used as monitor only. 
The calibration and the determination of atmospheric extinction coefficients were done by observing 
in each night about ten spectrophotometric standard stars from the list of \citet{massey}. The 
extinction coefficients are given in Table~2 of \citet{leinert95}. 
The surface brightness values listed in Table~\ref{table:2}  have been corrected to outside 
the atmosphere using these extinction coefficients and are expressed in units of \cgs\,.
 
Because of the small extent of regions covered ($\la$2~deg) and the symmetrical distribution
of the OFF positions around the ON positions, no special reduction procedures
were applied to correct for the (small) differential atmospheric diffuse light or zodiacal light effects.
We have estimated for the 2.2-m telescope the contribution by faint stars in the measuring aperture and 
the instrumental straylight from off-axis stars. Measurements of the off-axis straylight were made using
Vega. Then, using the Nomad star catalogue compilation \citep{zacharias04, zacharias05} we have estimated 
the straylight for each ON and OFF position. Only small corrections of $\la$1\, \cgs had to be applied to
the ON - OFF differences. The values listed in Table 3 have been corrected for these effects.

\begin{figure}
    \hspace{-1cm}
    \vspace{-0cm}
%    \center
%   \resizebox{\hsize}{!}
            {\includegraphics[width=12cm,angle=-90]{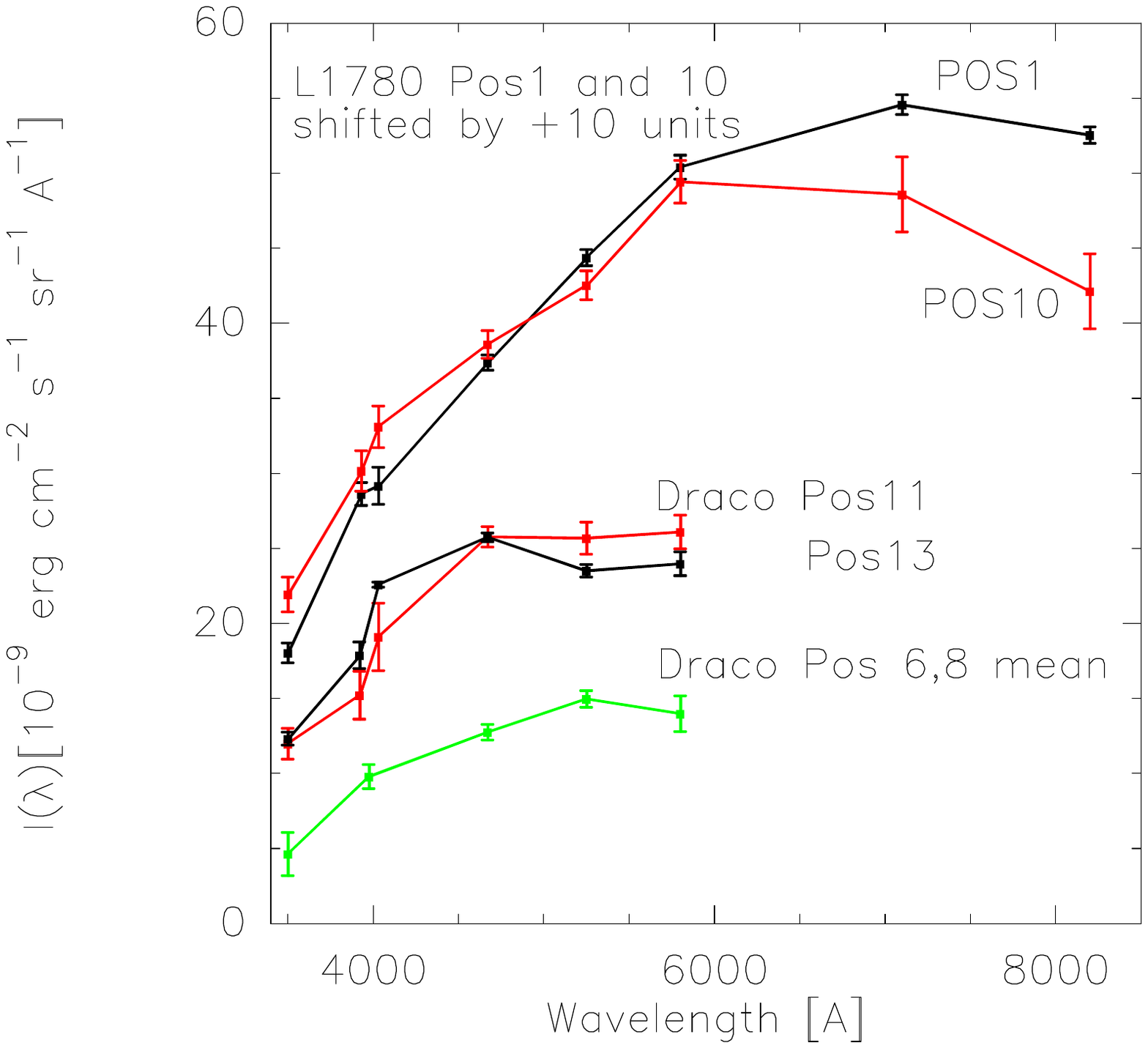}}
    \hspace{-0cm}
    \vspace{-1cm} 
     \caption{ Differential spectral energy distributions  $I$(nebula) { minus} $I$(sky) for selected 
positions in LDN~1780 and the Draco nebula.  The observed values are from Calar Alto photometry as 
described in Section~2.1 and given in Table~\ref{table:2}.  
              }
         \label{FigSEDs}
   \end{figure}

Spectral energy distributions (SED) of selected positions in LDN~1780 and the Draco nebula are shown 
in Fig.~\ref{FigSEDs}. { We emphasize that these SEDs represent the differential $I$(nebula) minus $I$(sky)
values which have resulted directly from the observations; no correction for the diffuse galactic light
has been applied here (see Section 3.4 for the interpretation.)}
 
\begin{table}
\caption{Positions in Draco and LDN~1780: Calar Alto intermediate-band photometry and VYSOS
surface brightness imaging.}
\label{table:1}
\centering 
\begin{tabular}{lllrrr} 
\hline
Pos  &  R.A.(J2000)& Dec.(J2000)& $I_{250}$ & $I_{200}$ & $A_V$ \\
  & hh:mm:ss.s  &  dd:mm:ss  & \multicolumn{2}{c}{MJy sr$^{-1}$}& mag \\  
\hline
 & \multicolumn{5}{l}{Draco nebula ON, Calar Alto photometry positions}\\
1 & 16:44:21.9  & +60:11:36 &     21.4 & &1.11 \\
2 & 16:46:55.5  & +60:17:36 &     18.7 & &0.96 \\
3 & 16:47:10.0  & +60:13:23 &     14.5 & &0.75 \\
6 & 16:48:33.3  & +59:56:53 &     10.6 & &0.55 \\
8 & 16:48:47.7  & +59:54:51 &     12.2 & &0.64 \\
9 & 16:48:59.6  & +59:56:55 &     17.4 & &0.90 \\
11& 16:49:14.7  & +59:55:06 &     24.4 & &1.27 \\
12& 16:49:22.9  & +59:47:01 &      8.3 & &0.43 \\
$13^*$& 16:49:23.7&+59:50:54&     20.6 & &1.07 \\
 & \multicolumn{5}{l}{Draco nebula OFF, Calar Alto photometry positions}\\
21& 16:38:22.7  & +60:18:43 &      1.2 & &0.06 \\
25& 16:41:13.2  & +59:35:44 &      1.2 & &0.06 \\
26& 16:45:08.9  & +59:23:47 &      1.3 & &0.07 \\
28& 16:52:30.8  & +60:21:52 &      1.4 & &0.07 \\
29& 16:54:55.8  & +60:20:16 &          & & \\
31& 16:56:03.6  & +59:47:50 &          & & \\
  &             &           &          & & \\
  & \multicolumn{5}{l}{LDN~1780, Calar Alto photometry positions}\\
$1^*$& 15:39:40.2    & -07:11:32 & &71.4& 3.00\\
10  & 15:40:33.8     & -07:12:35 & &45.6& 1.84\\

33  & 15:39:12.7& -06:48:48 & 8.9  & &0.48\\
48  & 15:39:27.1& -07:36:34 & 8.8  & &0.48\\
  &             &           &      & &    \\
   \multicolumn{6}{l}{Areas set to zero intensity in LDN~1780 VYSOS images}\\
  &15:39       &-08:12      &7.4   & &0.43 \\
  &15:37       &-07:12      &7.4   & &0.43 \\
\hline
\end{tabular}
\tablefoot{* Standard position}

\end{table}

\begin{sidewaystable}
\caption{Calar Alto intermediate-band photometric results: measured surface brightness values in units of \cgs\,.}
\label{table:2}
%\centering 
\begin{tabular}{rrrrrrrrrrrrrrrrrr } 
\hline\hline
 Pos& N &\multicolumn{2}{c}{3500 \AA\,}&\multicolumn{2}{c}{3920 \AA\,}&\multicolumn{2}{c}{4030 \AA\,}&\multicolumn{2}{c}{4670 \AA\,}&
                   \multicolumn{2}{c}{5250 \AA\,}&\multicolumn{2}{c}{5800 \AA\,}&\multicolumn{2}{c}{7100  \AA\,}&\multicolumn{2}{c}{8200  \AA\,}\\     
\hline
 &   &\multicolumn{11}{c}{Draco nebula}\\
  1& 2   &  4.17 & 1.74 &   6.17 & 1.62  &  6.33 & 0.78 &  17.24 & 3.05 &  14.28 & 2.14 &  17.57&  2.32 &&&&\\ 
  2& 2   &  9.60 & 2.71 &  14.59 & 1.33  & 17.30 & 1.55 &  21.40 & 0.90 &  18.97 & 0.90 &  18.80&  1.94 &&&&\\ 
  3& 2   & 17.70 & 1.93 &  20.77 & 1.65  & 28.16 & 2.03 &  26.46 & 0.62 &  23.01 & 1.10 &  21.05&  1.14 &&&&\\ 
  6& 3   &  5.16 & 2.62 &   8.98 & 2.43  &  6.96 & 1.08 &  11.60 & 1.04 &  11.99 & 1.29 &  11.92&  6.48 &&&&\\ 
  8& 2   &  4.34 & 1.72 &  13.37 & 1.46  &  9.09 & 2.50 &  13.07 & 0.60 &  15.60 & 0.62 &  14.01&  1.21 &&&&\\ 
  9& 2   & 10.52 & 2.20 &  13.55 & 1.56  & 15.32 & 3.53 &  18.54 & 1.50 &  18.53 & 1.24 &  19.32&  1.81 &&&&\\ 
 11& 2   & 11.94 & 1.03 &  15.17 & 1.59  & 19.06 & 2.25 &  25.72 & 0.68 &  25.64 & 1.07 &  26.05&  1.12 &&&&\\ 
 12& 3   &  3.52 & 3.46 &  10.85 & 2.30  & 14.84 & 5.09 &  19.34 & 2.22 &  11.60 & 3.87 &  14.32&  3.86 &&&&\\ 
 13& 4   & 12.27 & 0.43 &  17.83 & 0.89  & 22.54 & 0.17 &  25.70 & 0.29 &  23.46 & 0.42 &  23.93&  0.79 &&&&\\ 
  &\multicolumn{11}{c}{ }\\
 & &  &\multicolumn{11}{c}{LDN~1780}\\
1& 1 &  7.99 & 0.66  & 18.58 & 0.76 &  19.12 & 1.24 &  27.34 & 0.51 &  34.33 & 0.53 &  40.36&  0.80 &  44.52 & 0.66 &  42.50 & 0.55 \\  
 10& 1 & 11.89 & 1.16  & 20.12 & 1.35 &  23.06 & 1.39 &  28.55 & 0.92 &  32.49 & 0.97 &  39.39&  1.42 & 36.6 & 2 &  30.5 & 2 \\ 
\hline
\end{tabular}
\end{sidewaystable}

%                                             Two column Figure 
%-------------------------------------------------------------
%
%-------------------------------------------------------------

\subsection{Intermediate band surface photometry of LDN~1642}

Intermediate band surface photometry of selected positions in the area of LDN\,1642 has 
been carried out in five filters, centred at 3500 \AA\, $(u)$, 3840 \AA\,, 4160 \AA\,, 4700 \AA\, $(b)$ and 
5550  \AA\, $(y)$, using the ESO 1-m and 50-cm telescopes at La Silla \citep{mat90,mat96}. 
The observations were made differentially, relative to a standard position in the centre of the cloud. 
Subsequently, the zero level was set by fitting in  Dec, RA 
coordinates a plane through the darkest positions well outside the bright cloud area.  
 The 50-cm telescope monitored the airglow variations. 
A preliminary set of these data has already been used by \citet{laureijs} to derive the
dust albedo over the wavelength range 3500 to 5500 \AA\,.

\subsection{Wide-field imaging of LDN~1780 and LDN~1642  }

We have obtained wide-field ($2.7 \times 2.7$ deg) imaging of LDN~1780
using the University of Bochum 15 cm VYSOS6 refractor (now one of the RoBoTT twin 
telescopes\footnote{https://www.astro.ruhr-uni-bochum.de/Astrophysik/V6\_neues.html}) located 
next to Cerro Armazones, northern Chile \citep{haas12}. It was equipped with a 4 k $\times$ 4 k 
Alta U16M CCD-Camera of Apogee  (pixel size $9 \times 9 \mu$m, corresponding to $2.37 \times 2.37$ arcsec).
The broad band $BVRi$ filters had the following central wavelengths and half widths: Astrodon-Schuler $BVR$: 
$B$ 4400 \AA\, (1000 \AA\,), $V$ 5400 \AA\, (1000 \AA\,), $R$ 6200  \AA\, (1500 \AA\,); $i$(Pan-STARSS)  7480 \AA\, (1250 \AA\,).

The observations for LDN~1780 were carried out over eight photometric 
nights between 2009 May 21 - 26 (new moon) and July 19 - 23 (hours with no moon).
The total accumulated exposure times were 2.25 h in $B$ and 1.25 h in $V, R$, and $i$. 
Imaging of LDN~1642 in the $i$ band was carried out in 68 nights between 2009 January 16 and April 12 
as part of a variable star monitoring programme. Stacking all the frames resulted in an image with
5.6 h total integration time.
The frames were reduced for dark current and bias, and the flat field correction (using dome flats)
was done in the standard way  \citep{haas12}.
Calibration was performed using the standard star fields LSE259, LTT6248 and SA107 \citep{landolt}. 
Atmospheric extinction was corrected by standard methods as explained in \citet{haas12}.

Wide field photometry with CCD detectors has the obvious advantage over the single-channel
photoelectric photometry that all pixels in the field are observed simultaneously. However, flat fielding
represents a problem especially in a situation, as discussed in the present paper, where small
surface brightness differences are to be discerned from a large foreground sky signal. In the 
single-channel approach there are no flat-fielding problems since the whole instrumental setup, that is
telescope optics, instrumental effects (like internal straylight) and detector geometry and 
sensitivity are the same for all observed positions. 

 We have rescaled the VYSOS $BVRi$ broad band photometric maps for LDN~1780 using the results of the intermediate
band results. 
We have determined from the VYSOS maps the differences 
$\Delta I({\rm Pos1}) =  I({\rm Pos1}) - I({\rm Pos33,48})$, 
and $\Delta I({\rm Pos10}) =  I({\rm Pos10}) - I({\rm Pos33,48})$ 
corresponding to the same ON and OFF positions used for the Calar Alto photometry.
Synthetic $BVRi$ intensities were calculated from the intermediate-band photometry following 
the recipes in \citet{leinert95}. The intensity ratios, Calar Alto/VYSOS, were 1.0., 0.97, and 1.10
for $B, V$ and $R$, respectively. For the $i$ band a larger scaling factor of 1.46 was found.
For the surface brightness determination 
the stars were removed from the images and the removed pixels were replaced by average background
values in the surrounding area. Some residuals of the brightest stars remained 
but they cover only a negligible area of the maps (see the $B$-band image of LDN~1780 in 
Fig.~\ref{L1780FUV_NUV_smooth} and the $i$-band image of LDN~1642 in Fig.~\ref{L1642I}).

\subsection{$GALEX$ FUV and NUV imaging}

The {\em Galaxy Evolution Explorer (GALEX)} \citep{martin05} has covered a large fraction of the sky with 
images of 1\fdg25 diameter and 5'' to 7'' resolution in two bands, far ultraviolet (FUV) 1350--1750~\AA\,
and near ultraviolet (NUV) 1750--2850~\AA\,.
We have downloaded the images covering the LDN~1780 and LBN~406 areas from the {\em GALEX} 
archive\footnote{http://galex.stsci.edu/gr6/?page=start} where an all-sky map of the UV diffuse 
background as presented by \citet{murthy14} is available at the address 
{\tt https://archive.stsci.edu/prepds/uv-bkgd/}. 
The intensity unit of the archival images is counts s$^{-1}$ cm$^{-2}$ sr$^{-1}$\AA\,$^{-1}$ 
and the pixel size is $2'\times2'$. The conversion from photon units to cgs units is:
1 photon\,cm$^{-2}$s$^{-1}$sr$^{-1}$\AA$^{-1}$ = 1.29 $10^{-11}$\, and 
0.864 $10^{-11}$\,erg~cm$^{-2}$s$^{-1}$sr$^{-1}$\AA$^{-1}$ for FUV and NUV, respectively. 

\subsubsection{LDN~1780}

\begin{figure*}
    \center
{\includegraphics[width=16.5cm,angle=0]{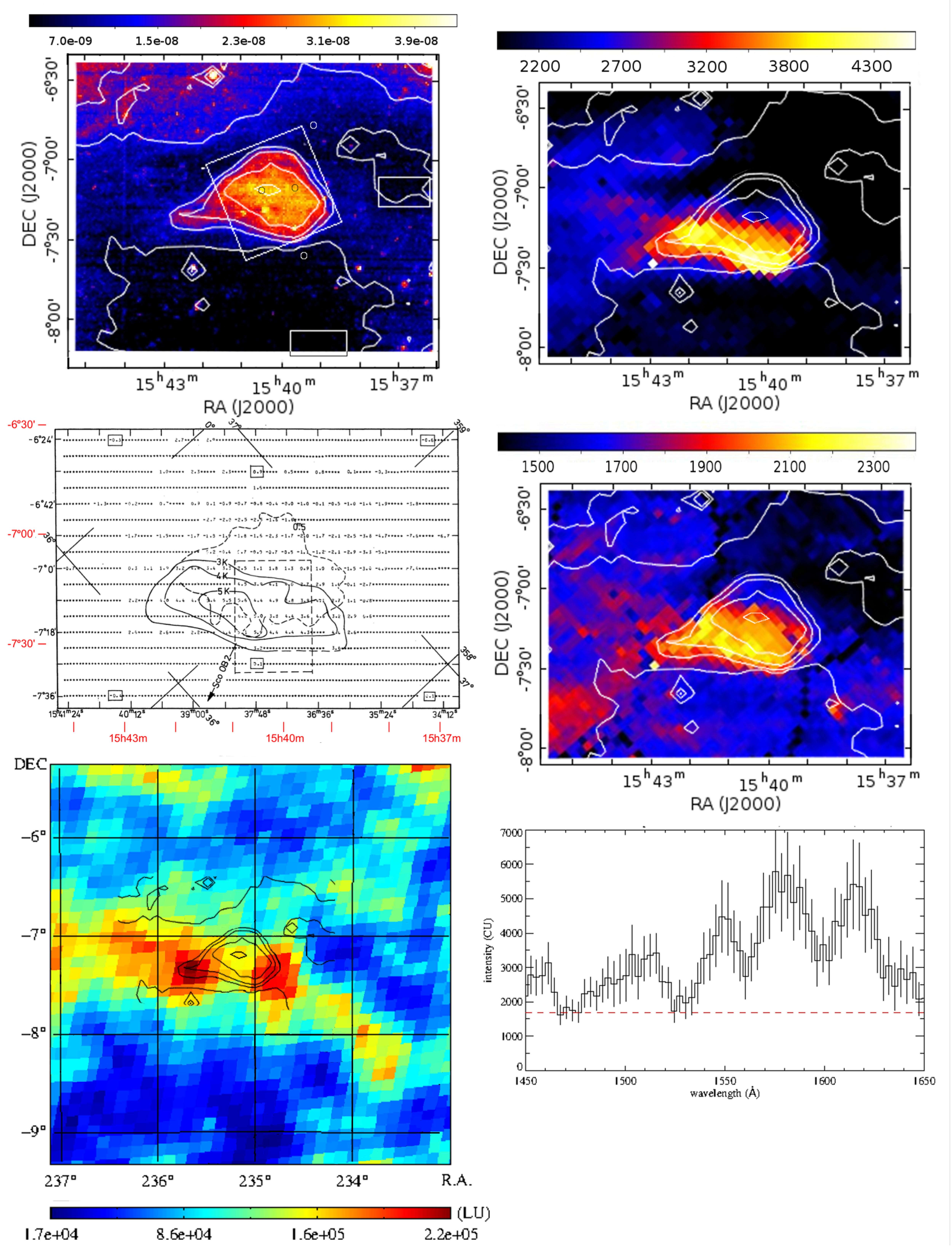}}
\caption{ {\em LDN~1780 data.} {\bf (upper left)} $B$ band VYSOS image with
Calar Alto photometry positions shown as circles, the ISOPHOT  200 $\mu$m map area 
{ shown as rectangle} and the two surface brightness 'zero' 
areas as rectangles; contours are at 30.0, 23.75, 17.5, 11.25 and 5.0\,\cgs\,. 
{\bf (upper right)} $GALEX$ FUV map; {\bf (middle right)} $GALEX$  NUV  map; 
{\bf (middle left)} Map of H\,{\small I} 21-cm excess emission (line area), adopted from 
\citet{mattila+sandell} Fig. 5; {\bf (bottom left)} SPEAR/FIMS  H$_2$ fluorescence emission map; 
 {\bf (bottom right)} SPEAR/FIMS spectrum at R.A. = 15h 42m (235\fdg5), Dec = -7\fdg4 (J2000),
the brightest  H$_2$ spot  in the cloud.
In the $GALEX$  images some artefacts of the 1\fdg25 field edges are seen as black arcs; 
the contours of  the $B$ band image  are overplotted { in the  $GALEX$ and SPEAR/FIMS} for reference. 
Coordinates for the H\,{\small I} 21-cm map are R.A. and Dec. (B1950.0); approximate J2000 
coordinates have been added in red. The $A_B$ = 0\fm5 extinction contour is shown as a dashed line.
{ For  the colour bars of the $GALEX$  images and for the SPEAR/FIMS spectrum the unit 
is ph\,s$^{-1}$cm$^{-2}$sr$^{-1}$\AA$^{-1\,}$}; for the colour bar of the SPEAR/FIMS H$_2$ image 
the unit is ph\,s$^{-1}$cm$^{-2}$sr$^{-1}$ (=LU, line unit). 
}
         \label{L1780FUV_NUV_smooth}
   \end{figure*}

\begin{figure}
    \hspace{0cm}
    \center
%   \resizebox{\hsize}{!}
            {\includegraphics[width=9cm,angle=0]{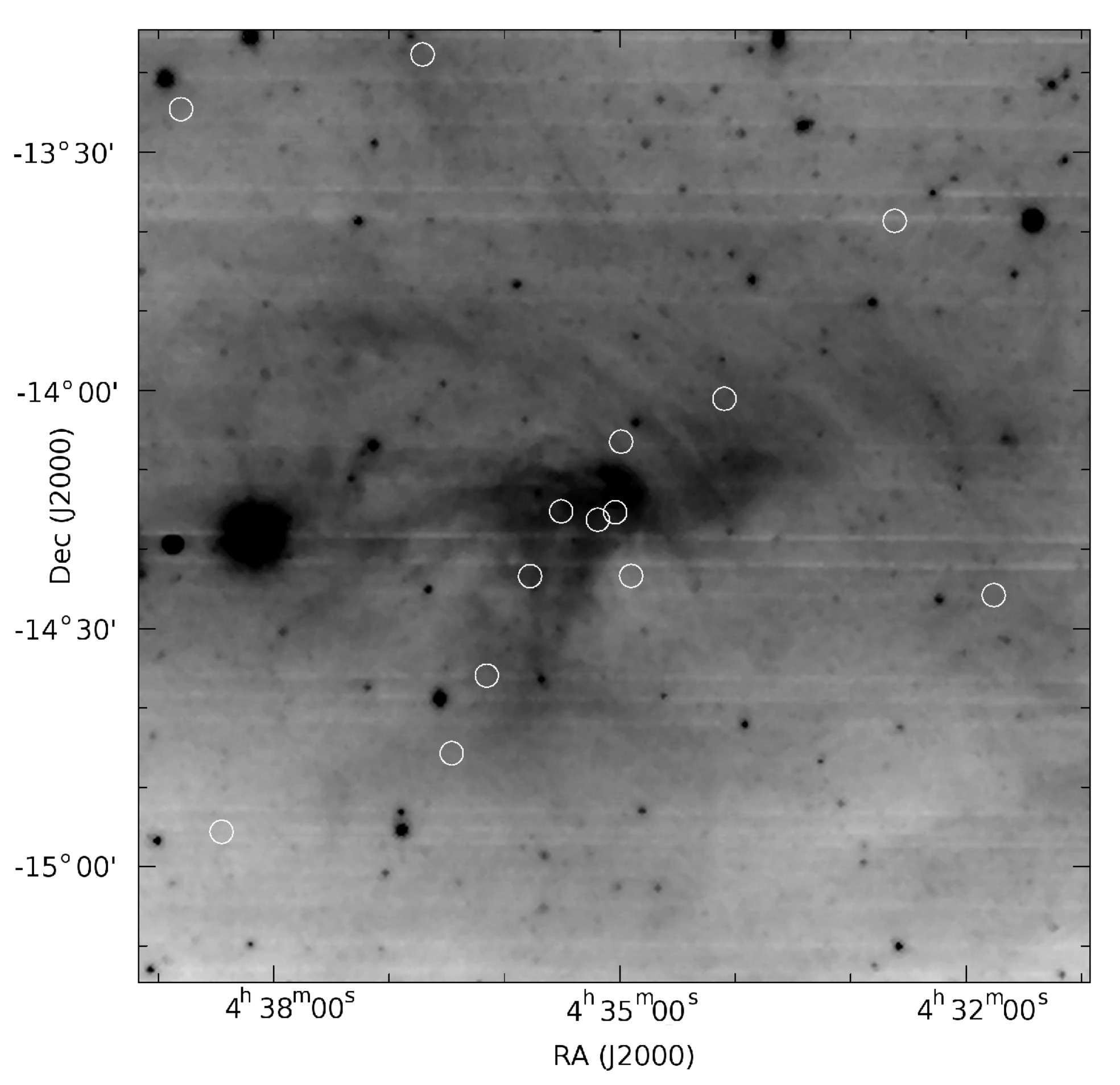}}
    \hspace{-0cm} 
     \caption{VYSOS $i$ band image of LDN~1642; most stars have been removed. 
The La Silla photometry positions that are within this image area are shown as circles.}
         \label{L1642I}
   \end{figure}

$GALEX$ archival FUV and NUV images covering an area of $\sim2.2\degr \times 1.6\degr$ around LDN~1780
are shown in Fig.~\ref{L1780FUV_NUV_smooth}. In order to avoid showing the empty pixels in the \citet{murthy14}
 data base a pixel size of $3' \times 3'$  has been chosen.

\subsubsection{Draco nebula}

\begin{figure*}
%   \resizebox{\hsize}{!}
\vspace{-0.0cm}
\hspace{-0.5cm}
            {\includegraphics[width=19cm,angle=0]{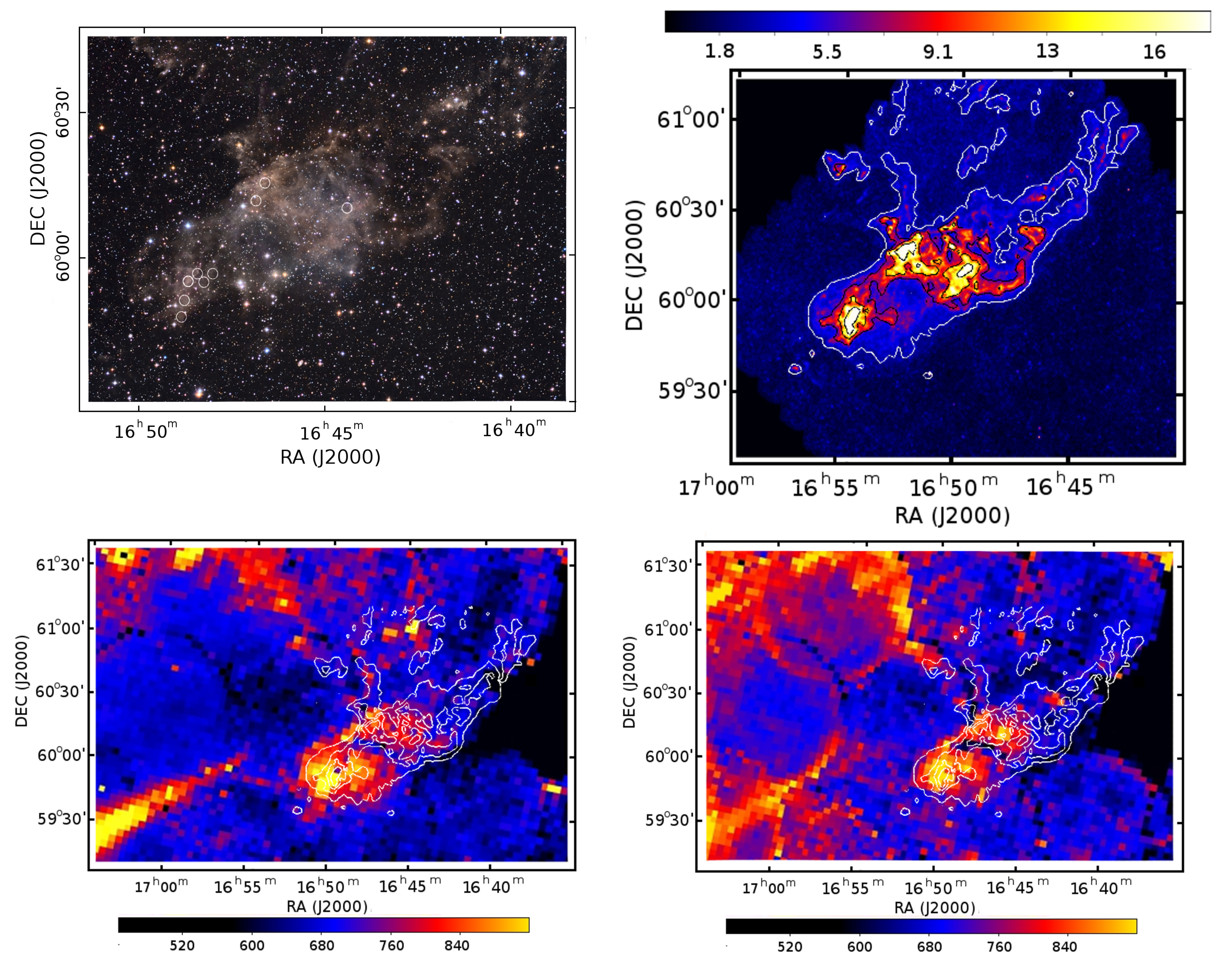}}

      \caption{{\em Draco nebula images.} {\bf(upper left)} broad band optical; 
{\bf(upper right)} $Herschel$ 250 $\mu$m; {\bf(bottom left)} $GALEX$ FUV; 
{\bf (bottom right)}  $GALEX$ NUV. The 250 $\mu$m contours { at 3, 6.6, 11, and 17} MJy\,sr$^{-1}$ 
are shown and are also overplotted on the FUV and NUV images. They correspond to $A_V$ values of 
$\sim$0.15, 0.36, 0.57, and 0.78 mag. The unit in the FUV and NUV colour bars is 
ph\,s$^{-1}$cm$^{-2}$sr$^{-1}$\AA\,$^{-1}$. Pixel size for FUV and NUV is 3'. 
At Eastern edge of the UV images the tip of the \citet{sujatha10} foreground filament is seen. 
The black area on the Western edge has no FUV and NUV data. Some artefacts
of the $GALEX$ 1\fdg25 field edges remain. The optical image was provided by Jim Thommes.} 
\vspace{-0cm}
        \label{Draco_GALEX}
   \end{figure*}

$GALEX$ archival FUV and NUV images of an area of $\sim2.5\degr \times 2.0\degr$ covering the main part (LBN~406) 
of the Draco nebula
is shown in Fig.~\ref{Draco_GALEX}. As for Fig.~\ref{L1780FUV_NUV_smooth} the data of \citet{murthy14} have been
smoothed to a pixel size of $3' \times 3'$. For reference, also a {\em Herschel} 250 $\mu$m 
and a broad band optical image (private communication, Jim Thommes 2018) of the same area are shown.

\subsection{SPEAR/FIMS FUV spectral imaging of LDN~1780}
Archival data of SPEAR/FIMS \citep{edelstein06a,edelstein06b} aboard the 
Korean satellite $STSAT-1$ were utilized to study the distribution of  H$_2$ fluorescence emission in the 
area of LDN~1780 (see Fig~\ref{L1780FUV_NUV_smooth}). The long-wavelength channel (L band; 1350--1750 \AA\,, 
$7\fdg4 \times 4\farcm3$ field of view) was used to cover a spectral region with intense H$_2$  fluorescence emission. 
The spectral and spatial resolution of the instrument are 
$\lambda/\Delta\lambda \sim550$ and $\sim 5\arcmin$, respectively. We pixelated the
photon data using  HEALPix scheme \citep{gorski05} with a resolution parameter Nside=512 corresponding 
to a pixel size of $\sim6\farcm9$. After removing point sources from the image, we constructed the FUV continuum 
and  H$_2$  fluorescence emission maps. They were convolved with a FWHM=0\fdg2. For the  H$_2$ map
the continuum was first subtracted from each spectrum. The SPEAR/FIMS FUV continuum map (see Section 4.4 and Fig.~15) 
is similar to the $GALEX$ FUV map even though, due to its lower spatial resolution, it shows a smaller 
peak intensity and the map details are more diffuse in comparison with $GALEX$. In the  H$_2$ map shown in 
 Fig.~\ref{L1780FUV_NUV_smooth} the contours of the $B$ band surface brightness map are overplotted. 
As an example we show the spectrum at the H$_2$  peak position at (R.A., Dec.) = (235\fdg5, -7\fdg4), 
SE of the LDN~1780 cloud centre. The characteristic H$_2$ fluorescence emission features are seen in the spectrum.   
We will analyse the properties of the H$_2$ emission in Section 4.4.

\subsection{Determination of visual extinction}

\subsubsection{LDN~1780}
Visual extinction is the natural parameter to be used in connection with modelling the scattered light. 
For LDN~1780 the extinction has been determined using 
near-IR colours of background stars. In order to reach a better { statistical} precision than 
reached via the directly measured  $A_V$ values we have used as proxy the 200 $\mu$m emission as mapped 
with ISOPHOT \citep{lemke} aboard $ISO$ \citep{kessler}. 

The methods used for the determination of extinction and for the ISOPHOT 200 $\mu$m mapping  
have been described in detail in  \citet{ridderstad}. In short, the optimized multi-band NICER 
technique of \citet{lombardi11} was applied to the 2MASS data. 
The values used for the extinction-to-colour-excess ratio were $A_{V}/E(J-H) = 8.86$ and 
$A_{V}/E(H-K_{\rm _S}) = 15.98$; they correspond to $R_V = A_{V}/E(B-V) = 3.1$ \citep{mathis}.
The extinction values were averaged using a Gaussian with FWHM = 3.0\arcmin.
The typical total error was 0.47 mag in $A_V$, including  the variance of the intrinsic colours 
$(J-H)_0$ and $(H-K)_0$ and the photometric error.
At the resolution of 3.0\arcmin\, the value of maximum extinction is $\sim$3.4 mag.

The extinction values as derived from 2MASS photometry are relative to a low-extinction reference field
leaving, however, the extinction zero point somewhat uncertain.
From three recent extinction surveys based on different methods we have found the following  
``absolute'' extinction ($A_V$) estimates for the VYSOS zero intensity level areas (see Table 1):
$A_V$ = 0.34 mag \citep{schlafly} (FIR emission), 0.54 mag \citep{schlafly14} (Pan-STARRS), and 
0.40 mag \citep{Planck2013XI} (FIR and sub-mm).
They are consistent with a mean value of $A_V = 0.43\pm 0.05$ \,mag.

Over the mapped area of $39\times39$ arcmin$^2$ the 200 $\mu$m intensity correlates well with 
$A_V$; the relation is represented by the linear fit
\begin{equation}
A_V = (0.0451\pm0.0007) \times I(200 \mu{\rm m}) - 0.22\pm0.02 \, {\rm mag}. 
\end{equation}

For positions that are outside  the $39\times39$ arcmin$^2$ area 
we have used a {\em Herschel} SPIRE  250 $\mu$m map\footnote{hspirepsw447\_30pxmp\_1539\_m0706\_1476933801545.fits,\\
centre RA = 234.76159 deg, DEC = -7.1134 deg (J2000),\\
 map size $\sim 1\fdg8\times 1\fdg8$, start date 210-18-04T11:00:13}   
available in the  {\em Herschel} archive\footnote{http://archives.esac.esa.int/hsa/whsa/}. 
A tight correlation (with an  rms = 0.65 MJy\,sr$^{-1}$) 
was found between $I(200\mu$m) and $I(250\mu$m) over the ISOPHOT map area
\begin{equation}
I(200\mu{\rm m})  = (0.856\pm0.006) \times I(250\mu{\rm m}) +7.9\pm0.17. 
\end{equation}
Together with equation (1) this then enables us to derive $A_V$ values also from $I(250\mu$m):

\begin{equation}
A_V = 0.0387 \times I(250 \mu{\rm m}) + 0.14\, {\rm mag}.
\end{equation}
The  $A_V$ values are given in Table \ref{table:1}.

\subsubsection{LDN~1642}

We have used the {\sc nicer} method \citep{lombardi11} to derive extinctions from the 2MASS $JHK$ 
colour excesses of stars in $\O$$6\arcmin$ areas corresponding to our photometry  positions
in and around LDN~1642 (\citealt{mattila17}, Appendix C). 
However, because of the relatively low numbers of 2MASS stars at the high 
galactic latitude of L\,1642 these extinction values had large statistical uncertainties of  
ca. $\pm0.2-0.3$~mag. The high precision  achieved for $I(200\,\mu$m) values (ca. $\pm0.5$~MJy\,sr$^{-1}$) 
allows a better precision  of  $\sim \pm0.03$\,mag to be reached for $A_{\rm V}$, especially at low extinctions. 
Therefore, visual extinctions for our low-to-moderate extinction positions, $A_V\la1$mag, 
were determined using ISOPHOT 200~$\mu$m observations.
An area of $\sim$1.2~$\sq\degr$ around L\,1642 was mapped using ISOPHOT 
\citep{lehtinen04} and, in addition, our surface photometry positions
were observed also in the {absolute photometry mode} at 200~$\mu$m.

A fit to $I(200\,\mu$m) vs $A_{\rm V}$ at $I(200\,\mu$m) $<$30 MJy\,sr$^{-1}$ gave the slope of 
$19.0\pm2.5$~MJy\,sr$^{-1}$mag$^{-1}$. The zero point of $I(200\,\mu$m)\, was corrected for a
zodiacal emission (ZE) of $0.8\pm0.2$~MJy\,sr$^{-1}$ and a Cosmic Infared Background of 
$1.1\pm0.3$~MJy\,sr$^{-1}$ \citep{hauser}. The ZE at the time of our 200~$\mu$m observations 
(1998-03-19/20, longitude difference $\lambda - \lambda_{\sun}=63\fdg7)$
was estimated using a 270 K black-body fit to the ZE intensities at 100, 140 and 
240~$\mu$m. They were interpolated from the weekly DIRBE Sky and Zodi Atlas (DSZA)
\footnote{https://lambda.gsfc.nasa.gov/product/.../dirbe\_dsza\_data\_get.cfm}
maps based on the \citet{kelsall} interplanetary dust distribution model. 
The extinction values were then calculated from \\
$A_{V} = (I(200\,\mu$m) - 1.9~MJy\,sr$^{-1}$)/19~MJy\,sr$^{-1}$mag$^{-1}$.\\
We estimate their errors to be ca. $\pm0.05$~mag.

\subsubsection{Draco nebula}
The large distance and small angular size of the Draco core area  
make it impracticable to determine its extinction distribution with the 2MASS $JHK$ colour 
excess method. The extinction mapping by \citet{schlafly14} using Pan-STARRS data
reaches far enough (4.5 kpc) but with an insufficient resolution (14' $\times$ 14') for our purpose.   
Instead, we have used the far-infrared emission at 250 $\mu$m as a proxy for the visual extinction.

We have determined the $A_V$ vs $I(250 \mu$m) relationship in two steps:
{\bf (1)}  \citet{miville} have mapped the Draco nebula at 250 $\mu$m  using the SPIRE 
instrument \citep{griffin} aboard the {\em Herschel Space Observatory} \citep{pilbratt}.
A plot of $I(250 \mu$m), 
versus the optical depth at 353 GHz, derived from the {\em Planck Surveyor} all-sky survey 
\citep{Planck2013XI}, gave the relationship  
$I(250\mu$m) =$6.37\,10 ^5 \times \tau_{353{\rm GHz}} - 1.46.$ 
The intercept was removed from the $I(250\mu$m) map to set its zero level.

\noindent {\bf (2)} The second step consists of determining the connection between $A_V$ and  
$\tau_{353{\rm GHz}}$.
For the diffuse dust at high galactic latitudes \citet{Planck2013XI} have found
a tight relationship between the reddening $E(B-V)$ and $\tau_{353{\rm GHz}}$:
 $E(B-V) =(1.49 \pm 0.039\,10^4)\times \tau_{353{\rm GHz}}$ 
which for $R = A_V/E(B-V) = 3.1$ corresponds to $A_V = 4.62\,10^4\, \tau_{353{\rm GHz}}$.
On the other hand, 
for three molecular clouds, Orion A, Orion B and Perseus,  the extinction $A_K$, 
derived from 2MASS $JHK$ photometry has been correlated with $\tau_{353{\rm GHz}}$ by
\citet{lombardi14, zari} and the results are presented in Table \ref{table:3} in the form\\
$A_K/0.112 = A_{V} = \gamma \times  \tau_{353{\rm GHz}} + \delta.$  

The slope $\gamma$ is seen to be consistently smaller in the molecular clouds, 
$\gamma = 2.20 - 3.51\,10^4 $, 
as compared to the diffuse high-latitude dust with  $\gamma = 4.62\,10^4$\,.
This is understood because of the higher far-IR absorption cross section of dust grains 
in molecular clouds vs diffuse dust. Although the dust in the Draco nebula may be different
from the 'local-velocity' molecular clouds we prefer to adopt for  $\gamma$ the value
$3.30\,10^4$, representative of the translucent parts of the Orion B and Perseus clouds, 
rather than the diffuse-dust value. Using this  $\gamma$ we end up for Draco nebula with the
relationship: $A_{V} = 0.052 \times I(250 \mu{\rm m})$.

This is also consistent with the relationship found in the previous Subsection for LDN~1780,
The relationship for Draco 'with molecular dust' is, as expected, in between the cases for
'diffuse dust' and for the somewhat more opaque LDN~1780.

\begin{table}
\caption{NIR and optical extinction vs thermal dust emission for four molecular 
clouds and for the high-latitude diffuse medium. The relationship is expressed as 
$A_V = \gamma \times  \tau_{353{\rm GHz}} + \delta$ and as $A_{V} = \gamma ' \times I(250 \mu{\rm m}) + \delta '$
}
\label{table:3}
\centering 
\begin{tabular}{lcccl} 
\hline
Object      & Extinction   & $\gamma$ & $\gamma$' &Ref.\\
            & range ($A_V$)&  $10^4$  &        &       \\
\hline
Orion A                    &$\la 5$   & 2.20   &  & 1 \\
Orion B                    &$\la 6$   & 3.09   &  & 1 \\
Perseus                    &$\la 6$   & 3.51   &  & 2 \\
Draco nebula                &$\la 1.5$ &        &  &   \\
\quad w. 'diffuse dust' $^{*}$   &          & 4.62   & 0.072 & 3\\
\quad w. 'molecular dust'$^{**}$ &          & 3.30   & 0.052 & 4\\
LDN~1780                  &$\la 4$   &        & 0.039 & 5 \\   
\hline
\end{tabular}
\tablefoot{(1) References: \citet{lombardi14}; (2) \citet{zari};  (3)\citet{Planck2013XI};
(4) \citet{miville}; (5) This Paper}
\tablefoot{* Using the  $\gamma$ value as derived for diffuse dust by \citet{Planck2013XI}
from from $E(B-V) =1.49\,10 ^4 \tau_{353{\rm GHz}}$ and $A_V/E(B-V)=3.1$}
\tablefoot{** Using the mean of the  $\gamma$ values of Orion B and Perseus}

\end{table}

\subsection{Optical and UV surface brightness vs visual extinction}

\subsubsection{LDN~1780}
\begin{figure}
%   \resizebox{\hsize}{!}
\vspace{0cm}
\hspace{-1cm} 
            {\includegraphics[width=10.8cm,angle=0]{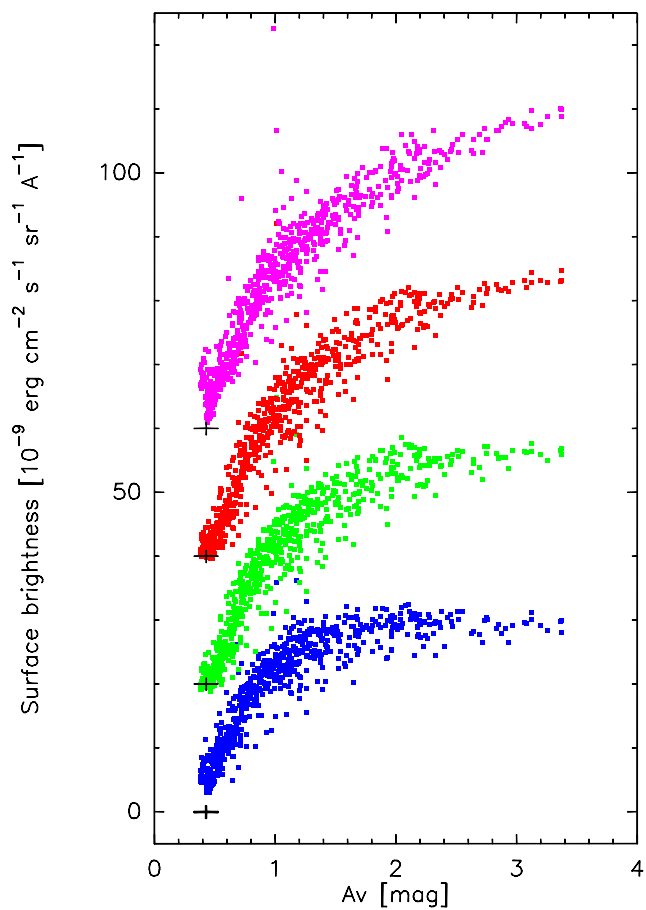}}
\vspace{-1cm}
      \caption{VYSOS $BVRi$ surface brightness vs visual extinction in LDN~1780. %VYSOS broad band results in $BVRi$.
The relationships are shown from bottom to top in different colous: $B$ (blue dots), $V$ (green, shifted by +20 units), 
$R$ (red, shifted by +40 units), and $i$ (magenta, shifted by +60 units). The black crosses at  $A_V = 0.43$ mag
indicate the zero level for each surface brightness band.
 }
 \vspace{-0cm}
        \label{VYSOSplots}
   \end{figure}

The $BVRi$ surface brightnesses from VYSOS imaging of LDN~1780 are shown as function 
of visual extinction $A_V$ in Fig.~\ref{VYSOSplots}. 
The surface brightnesses have been extracted for the 39' $\times$ 39' area covered by
the ISOPHOT 200 $\mu$m map. 

The zero point for the surface brightness is set in the minimum intensity areas 
of the VYSOS images (see Table \ref{table:1} and Fig. \ref{L1780FUV_NUV_smooth}). 
In these areas the mean extinction is  $A_V = 0.43$\,mag; the zero points are indicated with black crosses in 
Fig.~\ref{VYSOSplots} . In the $V$, $R$ and $i$ bands the zero level is reached within the  39' $\times$ 39' 
area of the ISOPHOT 200 $\mu$m map, but in the $B$ band there is an offset of $\sim 4$ \cgs.
 
The $B$ band diagram is characterized by a linear part at  $A_V \la 1.3$ mag followed by a partial and then a full 
saturation which sets in when the extinction at the wavelength in question reaches a value of  
$A_{\lambda} \approx 2.5$ mag. At  still higher extinctions the surface brightness starts to decrease, a phenomenon
that is already weakly evident at $A_V \ga 3$. The behaviour is
analogous in the other bands. Because of decreasing extinction from $B$ to $i$ the turn-over is expected to occur, 
instead of at $A_V \sim1.5$ mag corresponding to  $A_B \sim2.0$ mag ($B$), at $\sim2.0$ mag ($V$), $\sim2.5$ mag ($R$), 
and $\sim3.2$ mag ($i$). This is also seen to be roughly the case.

    \begin{figure}
%   \resizebox{\hsize}{!}
\vspace{1cm}
\hspace{-0cm} 
            {\includegraphics[width=9.0cm,angle=0]{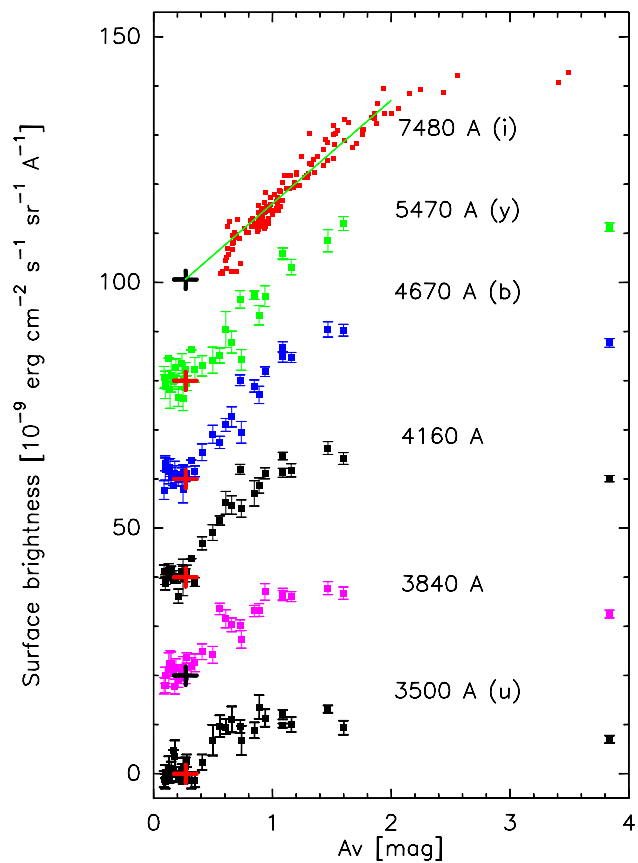}}
\vspace{-0cm}
      \caption{Surface brightness vs visual extinction in LDN~1642.
The relationships are shown from bottom to top for: 3500 \AA\, ($u$) (black dots), 
3840 \AA\,(magenta, shifted by +20 units), 4160 \AA\, (black dots, +40 units),
 4670 \AA\, ($b$) (blue dots, +60 units), 5470  \AA\, (green dots, +80 units) and
7478\,\AA\, ($i$) (red dots). The crosses at  
$A_V = 0.27$ mag indicate the zero level for each surface brightness band. }
 \vspace{-0cm}
        \label{L1642plots}
   \end{figure}

    \begin{figure}
%   \resizebox{\hsize}{!}
\vspace{0cm}
\hspace{-1.5cm} 
            {\includegraphics[width=10.8cm,angle=0]{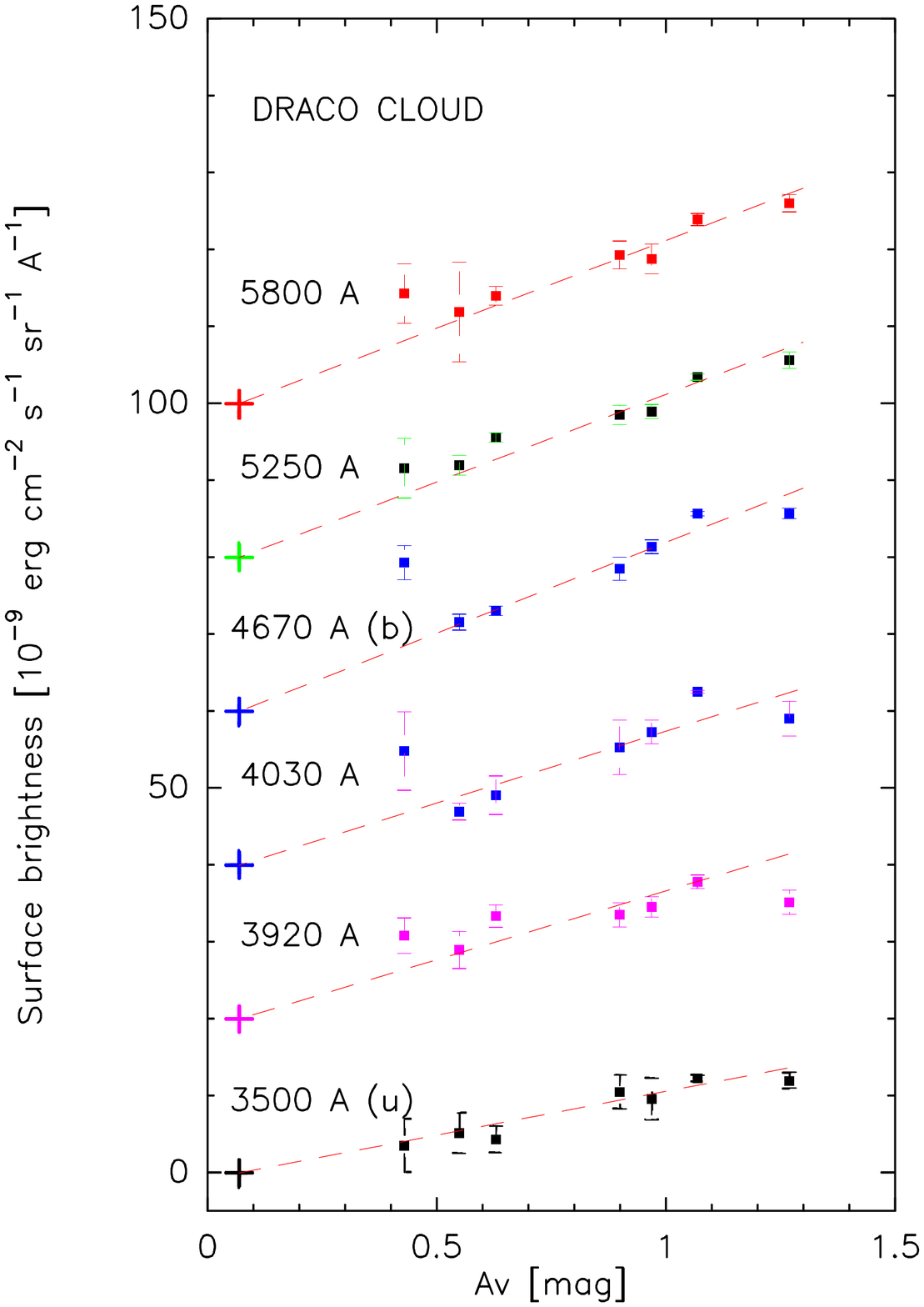}}
\vspace{-1cm}
      \caption{Surface brightness vs visual extinction in the Draco nebula.
The relationships are shown from bottom to top in different colours: 3500 \AA\, ($u$) (black dots), 
3920 \AA\,(magenta, shifted by +20 units), 4030 \AA\, (blue dots, +40 units),
 4670 \AA\, ($b$) (blue dots, +60 units), 5250  \AA\, (black dots, +80 units) and 
 5800  \AA\, (red dots, +100 units). The crosses at $A_V = 0.07$ mag indicate the zero level for 
each surface brightness band. To guide the eye approximate linear fits are shown as dashed lines.
Data for positions 1 and 3 (see Table \ref{table:2}) deviate by more than $3\sigma$ in most colours 
and are not shown in this plot. }
 \vspace{-0cm}
        \label{Dracoplots}
   \end{figure}

The NUV and FUV surface brightnesses are plotted against the visual extinction in  Fig.~\ref{GALEXvsAV}.
For each raster position of the $A_V$ map, 
corresponding to the ISOPHOT 200~$\mu$m map positions, the mean value of the {\em GALEX} pixels within 
1.8' in $l$ and $b$ was taken. The zero level was set in the same area that was 
used also for the $BVRi$ bands.%  R.A. = 15:37 Dec = 7:12 (J2000).

The data points are shown with red symbols for the southern and black symbols for the northern half, 
with the division line at Dec.(J2000) = -7\fdg20 corresponding to the extinction maximum. Data in the 
Eastern ``tail'' area 
of the cloud at R.A.(J2000) > 15h 41m have not been included. These diagrams demonstrate quantitatively the dichotomy 
as seen in the UV images in Fig.~\ref{L1780FUV_NUV_smooth}. In the southern half the intensity
is seen to increase almost linearly up to  $A_V \sim 1$ (corresponding to 
$A_{\rm FUV} \approx A_{\rm NUV} \approx 2.7$ mag and, after a quick turn-over, to decrease almost linearly all 
the way to the cloud core where $A_V \sim 3.5$\,mag, corresponding to $A_{\rm FUV} \approx A_{\rm NUV} \approx 9.5$ mag. 
While the  $I_{\rm NUV}$ vs  $A_V$ curve is qualitatively similar to the $I_{B}$ vs  $A_V$  curve 
the case of the FUV band is clearly different. This cannot be explained with different extinction because it is 
practically the same for the NUV and FUV bands. 

The northern half the cloud (black symbols in Fig.~\ref{GALEXvsAV}) shows a very different   
$I_{\rm NUV}$ vs  $A_V$ and  $I_{\rm FUV}$ vs  $A_V$ relationship. All points lie below those for the
southern half. Most points accumulate around a lower boundary which increases roughly linearly from the cloud's 
(northern) edge to the centre. In the NUV curve there is a weak indication of a bump at  $A_V\sim1$ mag,
reminiscent of the pronounced turn-over bump on the southern half. 
The north-south asymmetry is clearly connected to the asymmetrical illumination of LDN~1780 in the ultaviolet. 
Its southern half is exposed to the UV radiation from the galactic plane and from the nearby Scorpius OB2 
association.% (see Fig.~\ref{L1780FUV_NUV_smooth} for the orientation).

While the $BVRi$ and NUV surface brightnesses can be understood in terms of scattered starlight 
the FUV observations require, in addition, another explanation. We will discuss in Section 4.4
its modelling with molecular hydrogen fluorescence emission.

\subsubsection{LDN~1642}

At small optical depths with $A_{\lambda}\la1$~mag, 
the relationship is closely linear. At intermediate opacity positions with 
$A_{\lambda}\sim$1--2~mag the scattered light has its maximum value. 
For still larger optical depths the scattered light intensity
decreases because of extinction and multiple scattering and absorption losses (see Fig.~\ref{L1642plots}).

\subsubsection{Draco nebula}

The maximum extinction in the Draco nebula is $A_V\sim1.3$~mag.   In all optical bands the surface brightness vs  $A_V$ 
relationship appears linear over this range (see Fig.~\ref{Dracoplots}). The NUV and FUV surface brightness
vs   $A_V$ plots are shown in Fig.~\ref{Draco_GALEXvsAV}.

\subsection{H\,{\small I} 21-cm excess emission vs optical and UV surface brightness  in LDN~1780}

The distribution of the H\,{\small I} 21-cm excess emission, adopted from Fig.~5 of \citet{mattila+sandell} is shown 
in the middle panel of Fig.~\ref{L1780FUV_NUV_smooth}. The 21-cm excess emission is seen to be concentrated to 
the southern half of the cloud and its distribution is similar to the FUV distribution.
Also shown in  Fig.~\ref{L1780FUV_NUV_smooth} is  a map of the  H$_2$ fluorescence emission in 
the 1350 -- 1750 \AA\, wavelength range of {SPEAR/FIMS} and, as an example, 
the spectrum at R.A.= 15h 41.6m Dec =-7\fdg4 (J2000), SE of the cloud centre. 
The similarity of the distributions of FUV,  H$_2$ fluorescence and  21-cm excess emission is obvious
and it will be further analysed and discussed  
in Section~4.4.

   \begin{figure}
   \vspace{-0.3cm}
    \hspace{-1cm}
%    \centers
%   \resizebox{\hsize}{!}
            {\includegraphics[width=11.0cm,angle=0]{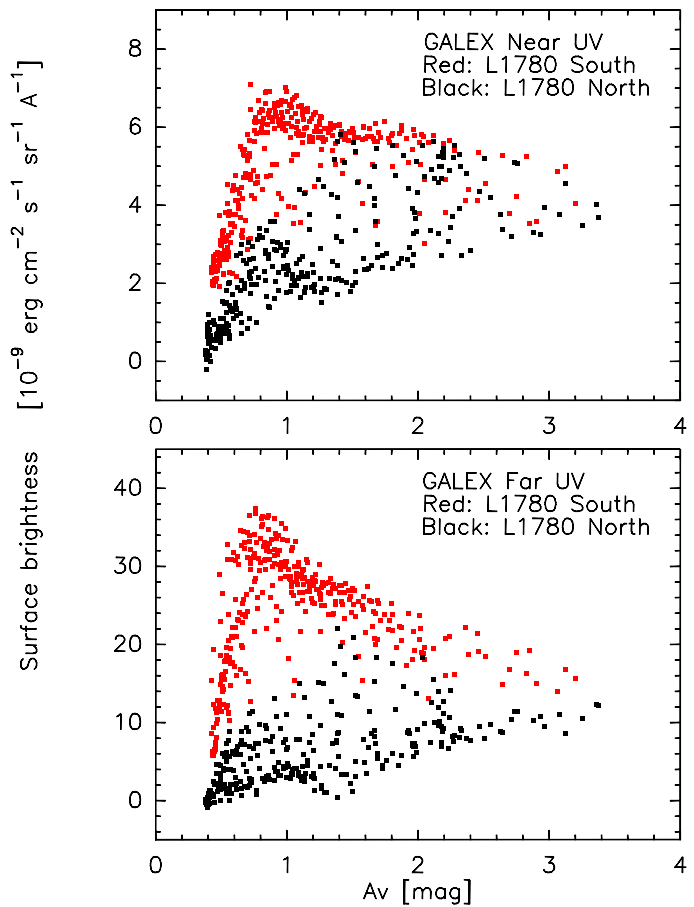}}
    \vspace{-1cm} 
     \caption{{\em GALEX} NUV {\bf (upper)} and FUV {\bf (lower)} surface brightness in LDN~1780 plotted 
against visual extinction. The red dots are for the southern (Dec. $<$-7\degr12\arcmin), the black dots for
the northern (Dec. $\ge$-7\degr12\arcmin) part of the cloud (R.A. < 15h 41.2m for both).}
         \label{GALEXvsAV}
   \end{figure}

   \begin{figure}
   \vspace{-0.3cm}
   \hspace{-1cm}
%    \center
%   \resizebox{\hsize}{!}
            {\includegraphics[width=11.0cm,angle=0]{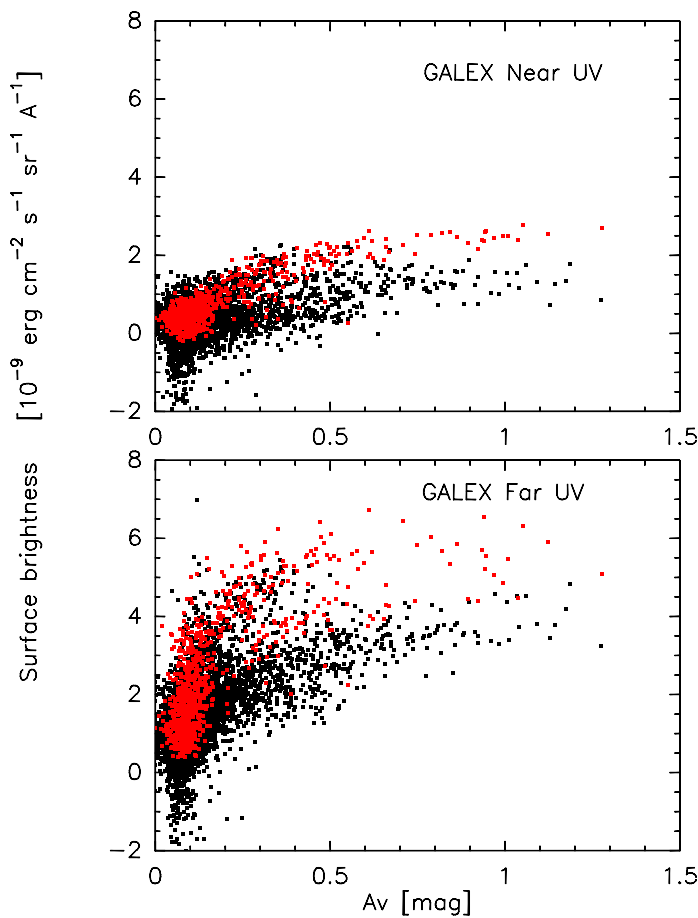}}
    \vspace{-1cm} 
     \caption{{\em GALEX} NUV {\bf (upper)} and FUV {\bf (lower)} surface brightness in Draco 
nebula plotted against visual extinction. Red points are for the ``Head'' at 
$l = 88.8 - 90.2$ deg, $b = 38.0 - 38.5$ deg (see Fig.~\ref{Draco_GALEX}) and black points 
for the ``Tail'' part of the nebula.}
         \label{Draco_GALEXvsAV}
   \end{figure}

\section{Modelling and analysis}

The intensity of the impinging Galactic light, as seen by a virtual 
observer at the cloud's location, is a function of the galactic coordinates, 
$I_{\rm GAL} =I_{\rm GAL}(l,b)$. The impinging radiation is multiply scattered by the dust 
grains in the cloud and the intensity of the scattered light towards the observer is given by
\begin{equation}
 I_{\rm sca}(l_0,b_0) = \sum_{l=0\degr}^{360\degr} \>\> \sum_{b=-90\degr}^{90\degr} \>\>  
I_{\rm GAL}(l,b) \>  S(\theta)\> \cos b\> \Delta l\>  \Delta b. 
\end{equation}
 For the Galactic light we have, in the optical, made use of the 
$Pioneer 10/11$ Imaging Photopolarimeter data (see Section 3.1.1). In the far and near  UV bands 
data from the $S2/68$ experiment aboard the $TD$-1 satellite are used (see Section 3.1.2).

$S(\theta)$ is the scattering function of the cloud;  $\theta$ is the scattering angle, that is
the angle between the direction  $l, b$ of the impinging radiation and the direction
of the cloud, $l_0,b_0$, as seen from the observer's viewing position. $S(\theta)$ is 
a function of the scattering properties of the grains (albedo $a$ and scattering asymmetry
parameter $g$) and the optical thickness $\tau$ of the cloud, $S(\theta) = S(\theta, a,g,\tau)$. 
We have calculated it for a set of $a, g$, and $\tau$ values using a Monte Carlo 
radiative transfer code, see Section  3.2.

\subsection{Milky Way surface brightness}

For the determination of the dust albedo the intensity and distribution of the incident interstellar 
radiation field is equally important as that of the observed scattered light from dust. We utilize 
the empirical results available for the ISRF at the Sun's location at optical and UV wavelengths. 
Before we can apply them to the target clouds of the present study two effects have to be taken into account: 
(1) the observed clouds, even if relatively nearby, are not in the Galactic plane but at a substantial $z$-distance 
of 50 to 400 pc; in the UV there may also be substantial ISRF variations caused by individual bright stars or
clusters; (2) the empirical ISRF is not available for the same wavelength bands as our surface brightness
observations. To account for these effects we have used a model for
the star and dust distribution in the Solar Neighbourhood     

\subsubsection{Optical}
 The Imaging Photopolarimeter instruments aboard the Pioneer 10 and 11 spacecraft got
a clean measurement of the Milky Way $B$ and $R$ band surface brightness distributions, 
$I_{\rm GAL}(l,b)$, during their interpanetary cruise beyond the zodiacal dust cloud at 
heliocentric distance of 3 -- 5 A.U. \citep{weinberg74, toller}. 
All-sky maps have been published by \citet{gordon}; see also 
{\tt http://www.stsci.edu/\~\,kgordon/\\
\~\,pioneer\_ipp/Pioneer\_10\_11\_IPP.html}.
We have adopted from this web page the data files ``{\tt Reformated data with all stars}'' which,
according to the {\tt README} file, ``give the measurements *including* all the stars'', 
that is the stars brighter than $6\fm5$  subtracted by \citet{toller} have been returned. 
We have divided the sky into $\Delta l \times \Delta b $ = $10\degr \times 10\degr$ pixels
centred at $l = 5,15,25, ... ,165,175$ deg and $b = -85, -75, ... ,75,85$ deg. The Pioneer measures 
were ascribed to the pixels according to their centre coordinates and the mean value for 
each  $10\degr \times 10\degr$ pixel was formed.
The hole in the Pioneer data centred at $l \approx 225\degr, b \approx 55\degr$ was filled by 
using the mean values at the same galactic latitude on both sides of the hole.

The average values of the Galactic light, $\overline{I_{\rm GAL}(l,b)}$, over the celestial 
sphere *including* all the stars are 114 
 and 152  S$_{10}(V)_{\rm G2V}$ in $B$ and $R$ band, respectively. 
These values are $\sim 12 - 13\%$
higher than the average values without the $m < 6\fm5$ stars,  101 and 
136 S$_{10}(V)_{\rm G2V}$ in $B$ and $R$, respectively \citep{toller}. 
In the cgs units used in this paper the averages over the sky  
*including* all the stars, are 136 and 151 \cgs\, in  $B$ and $R$, respectively (for a compilation, see
Table \ref{table:7}). 

\begin{table}
\caption{Milky Way mean surface brightness $\overline{I_{\rm GAL}}$ as seen from a vantage point at 
$z = 0$\, pc. The unit for columns (3) and (4) is  S$_{10}(V)_{\rm G2V}$ as used in the Pioneer 
photometry data. The values in units of \cgs\, (= cgs) in the last column (5) have been 
calculated using the conversion factors from Table 2 of \citet{gordon}, 
1~S$_{10}(V)_{G2V} \> \widehat{=} \> 1.192$\,\cgs\, in $B$ and 
1~S$_{10}(V)_{\rm G2V} \> \widehat{=} \> 0.992$\,\cgs\, in $R$.}
\centering 
\begin{tabular}{lcccc} 

           &           & \multicolumn{3}{c}{ $\overline{I_{\rm GAL}}$} \\
           &           &  \multicolumn{3}{c}{----------------------------------------- } \\
%           &           & \multicolumn{3}{c}{----------------------} \\
Band       & $\lambda$ &$m > 6\fm5$ stars & \multicolumn{2}{c}{all stars included} \\
           &            &    included   &    \multicolumn{2}{c}{ } \\
            &  \AA      &S$_{10}(V)_{\rm G2V}$  &S$_{10}(V)_{\rm G2V}$ &cgs\\
\hline
B\_Pioneer  & 4370      & 101                & 114 & 136 \\
R\_Pioneer  & 6441      & 136                & 152 & 151 \\
            &           &                    &     &     \\
$TD$-1 ISL  & 1565      &                    &     & 88 \\
            & 2365      &                    &     & 36 \\
$TD$-1 ISL+DGL  &  1565 &                    &     & 110 \\

\hline
\end{tabular}
\label{table:7}
 
\end{table}

\begin{table}
\caption{The Milky Way mean surface brightness as seen from a vantage point at $z = 50$\, pc and 
$z = 400$\, pc above 
the Galactic plane relative to the surface brightness at $z = 0$\,. The three columns refer to the 
sky areas at high positive and negative latitudes, $b > 25\degr$ and $b < -25\degr$, respectively, 
and to the all-sky mean surface brightness. }
%\label{table:4}
\centering 
\begin{tabular}{llccc} 

             & Band     &$b > 25\degr$&$b < -25\degr$& all-sky\\
\hline
$I_{\rm z=50 pc}/I_{\rm z=0}$& & && \\
             &  B\_Pioneer&0.76 &1.22 &1.02 \\
             &  R\_Pioneer&0.84 &1.16 &1.28 \\
             &  FUV 1565  &0.52 &1.42 &0.72 \\ 
             &  NUV 2365  &0.56 &1.40 &0.74 \\

$I_{\rm z=400 pc}/I_{\rm z=0}$& & & &\\
             &  B\_Pioneer&0.19 &1.62 &0.93 \\
             &  R\_Pioneer&0.26 &1.65 &1.28 \\
             &  FUV 1565  &0.002&1.5  &0.18 \\
             &  NUV 2365  &0.019&1.5  &0.20 \\

\hline
\end{tabular}
\label{table:4}
%\tablefoot{* } 
\end{table}

\begin{table}
\caption{Intensity ratios of Milky Way surface brightness in different filter bands relative to  
$B\_$Pioneer and $R\_$Pioneer filter bands. Values are given for two vantage points at a distance 
of $z = 50$\,pc and 400 pc above the Galactic plane. The values refer to the mean over the sky.
In case of the $u$,  3840/3920,  4030/4160 and $b$ bands there is a dependence on galactic latitude
(positive/negative) by up to $\pm20\%$; it has been included into the modelling. }
\centering 
\begin{tabular}{lcc} 

 & \multicolumn{2}{c}{$I({\rm Band})/I(B\_$Pioneer)} \\
   Band       & \multicolumn{1}{c}{$z = 50$ pc} & \multicolumn{1}{c}{$z = 400$ pc } \\
\hline

 $B$  & 1.07 &1.10 \\
 3500 ($u$)  & 0.65 &0.58 \\
 3840/3920 & 0.87 &0.78 \\
 4030/4160 & 1.10 &1.04 \\
 4670 ($b$)  & 1.14 &1.21 \\
 & & \\
  & \multicolumn{2}{c}{$I({\rm Band})/I(R\_$Pioneer)} \\
   Band       & \multicolumn{1}{c}{$z = 50$ pc} & \multicolumn{1}{c}{$z = 400$ pc } \\
\hline
 $V$  &1.09 & 1.04  \\
 $R$  &1.03 & 1.02  \\
 $I$  &0.92 & 0.92  \\
5250       &1.08 & 1.05 \\
5470 ($y$)   &1.10 & 1.08 \\
5800       &1.10 & 1.08 \\
7100       &0.93 & 0.94 \\
8200       &0.82 & 0.87 \\

\hline
\end{tabular}
\label{table:5}
\end{table}

The Pioneer data set of Milky Way photometry is ideal because it is free of
the atmospheric and interplanetary foreground components and also because it does 
include the Diffuse Galactic Light (DGL) which, besides the ISL, contributes substantially 
($\sim$ 10 -- 35\% of the ISL) to the Galactic background light \citep{toller}. 

Before utilizing the Pioneer  ${I_{\rm GAL}(l,b)}$
data for our observations we need to pay attention to two modifications: 
(1) our observed clouds are away from the Galactic plane; 
(2) the VYSOS $B$ and $R$ filter bands differ somewhat from the Pioneer ones. Furthermore, for other filter 
bands, including VYSOS $V$ and $i$, no corresponding, directly observed Milky Way surface brightness
maps exist. They can be derived from the Pioneer $B$ and $R$ band maps by applying colour transformation
coefficients based on Milky Way spectrum models.
 
 To estimate these corrections we use the results of a Solar Neighbourhood Milky Way model as presented 
in \citet{lehtinen13}. 
Empirical $z$-distributions for stars of different 
spectral types and dust have been used. The spectral library of \citet{pickles} was used 
to synthesize the ISL spectrum. The spectrum of the DGL was assumed to be a copy of the ISL 
spectrum. Based on the model we give in Table~\ref{table:4} the 
correction factors to be applied to the Milky Way surface brightnesses when
the cloud is at a distance of $z$ = 50 pc or $z$ = 400 pc above the
Galactic plane. Three different numbers are given: one is for the mean surface brightness over
the sky (``all-sky''); the other two are for the 
high galactic latitudes, $b > 25\degr$ and $b < -25\degr$, assuming that the cloud is above the plane.

The synthetic starlight spectra for  $z$ = 50 pc and $z$ = 400 pc are used to calculate the 
transformation coefficients from Pioneer $B$ and $R$ bands to VYSOS $BVRi$, Str\"omgren $uby$ 
and the custom-made intermediate bands, see Table~\ref{table:5}.

\subsubsection{Ultraviolet}

The UV astronomy experiment $S2/68$ \citep{boxenberg73} aboard the ESRO $TD$-1 satellite produced a star catalogue 
\citep{thompson78} which was used by \citet{gondha80} and \citet{gondha90} to derive the ISL
in the four channels around 1565  \AA\, ($\Delta\lambda =300$ \AA\,), 1965  \AA\, (330  \AA\,), 2365  \AA\, 
(330 \AA\, ), and 2740  \AA\, (300  \AA\,). Stars brighter than $1\, 10^{-12}$ erg\,s$^{-1}$cm$^{-2}$\AA$^{-1}$ 
were individually included and for the fainter stars a small correction ($\la10\%$)  was added. 
The tables of \citet{gondha90} give the summed-up values in $10\degr\times10\degr$ pixels over the 
whole sky. The mean intensity over the sky is 88 and 36 \cgs\, at 1565  \AA\, and 2365 \AA\,, 
respectively (Table\,\ref{table:7}).

The contribution of the diffuse component to the background radiation was estimated using 
the diffuse background results of \citet{seon11} based on the SPEAR/FIMS sky survey in the FUV band at 
1370 -- 1710 \AA\,. The mean diffuse background intensities, $I_{\rm FUV}^{\rm diffuse}$,
for the different galactic latitude slots, separately for north and south, were read from their Fig. 9 and added 
to the $TD$-1 1565 \AA\, ISL values. To estimate the NUV values at 2365 \AA\, we used the scaled FUV values, 
$I_{\rm NUV}^{\rm diffuse}= 0.476\,I_{\rm FUV}^{\rm diffuse}$, where the coefficient has been estimated 
from the plot as presented in Fig.\,9 of \citet{murthy14}. The resulting total mean Milky Way 
brightnesses (ISL + DGL) over the sky are  110 and 47 \cgs\, at 1565  \AA\, and 2365  \AA\,, respectively.
They are by 25\% and 30\% higher than the $TD$-1 starlight-only intensities (see Table \ref{table:7}).  
Also, \citet{witt97} have estimated that for an albedo $a=0.5$ the scattered light increases the
FUV background intensities by typically 25\,\%.

Because of their substantial $z$ distances the radiation field at the positions of LDN\,1780, LDN\,1642 
and the Draco nebula differs from the one at  $z$ = 0.  As in the case of the optical radiation field we have
used the  the results of the  Solar Neighbourhood Milky Way model of \citet{lehtinen13}
to estimate the corrections to be applied at the different galactic latitudes. Characteristic values are given
in  Table~\ref{table:4} for positive and negative latitudes and for the mean over the sky. Because of the 
narrow $z$--distribution of the OB stars and the stronger extinction the radiation field in the NUV 
and especially in the FUV is much more strongly affected by the  $z$-distance than in the optical.  

\subsubsection{Enhanced UV radiation field for LDN~1780} 

LDN~1780 is located close to the directions of the Sco\,OB2 association and the bright O9.2\,IV star 
$\zeta$\,Oph. With the heliocentric distances of Sco\,OB2 ($d_{\sun} = 144\pm3$~pc, \citealt{dezeeuw}), $\zeta$\,Oph 
($d_{\sun} = 112\pm3$~pc, Simbad, \citealt{vanleeuwen07}) and LDN~1780 ($d_{\sun} = 110\pm10$~pc, see Section 1) 
 the distances of  Sco\,OB2 and $\zeta$\,Oph from LDN~1780 are $d_{\rm L1780}=54$~pc and  $d_{\rm L1780}=28$~pc, 
respectively. The flux which LDN~1780 receives from Sco\,OB2 is $(d_{\rm L1780}/d_{\sun})^{-2} = $ 7.2 times and 
from $\zeta$\, Oph $\sim$15.5 times as large as that received from these sources at the Sun's position.
This causes an asymmetrical and enhanced illumination of LDN~1780 especially from Sco\,OB2,
as already suggested by \citet{mattila+sandell} and further substantiated by \citet{laureijs95}.

In order to estimate the FUV flux from Sco\,OB2 we have made use of the results of \citet{gordon94}.
Using NRL's FUV Cameras experiment on STS-39 they imaged in two bandpasses with $\lambda_{eff} = 1362$~\AA\,
and 1769 \AA\, an area of $\sim20\degr$ encompassing most of the association's OB stars and a huge reflection 
nebula around them. Using $IUE$ archives and the $TD$-1 catalogue \citep{thompson78} they found the total stellar 
fluxes in these two bands to be 61.8 and 29.0\,$10^{-9}$\,erg~cm$^{-2}$s$^{-1}$\AA$^{-1}$. The contribution
by the reflection nebula was found to be very substantial, amounting to 58 -- 92\,\% and 68 -- 92\,\% of
direct starlight; the nebular contribution depended  on the adopted background sky level. 
Assuming intermediate values of the nebular contributions of 75 and 80\,\% 
and applying the distance correction factor $(d_{\rm L1780}/d_{\sun})^{-2}$ = 7.2
we find for the total fluxes, expressed as surface brightness contribution over the sky, the values
of 62.0 and 29.3\,\cgs\,. 

For the $GALEX$ FUV band we have adopted the mean 
of these two values:  $I$(1540~\AA\,) = 45.6\,\cgs\,. For the $GALEX$ NUV band we have adopted the value 
$I(2320~\AA\,) = 0.48 \times I(1540~\AA\,)$ = 21.9\,\cgs\, where the coefficient 0.48 corresponds to the 
intensity ratio of the $GALEX$ NUV and FUV diffuse backgrounds as found by \citet{murthy14};
 this may be somewhat overestimated because a bigger part of the diffuse radiation from  Sco\,OB2 
and $\zeta$\,Oph than that of the general DGL may be due to H$_2$ fluorescence. We have 
assumed that the foreground extinction in the direction of Sco\,OB2 as seen from LDN~1780 is the same as 
 that seen from the Earth, that is the extinction of the Sco\,OB stars is assumed to be
mainly caused  by dust within the association itself. 

$\zeta$\,Oph has in the $TD$-1 catalogue \citep{thompson78} the fluxes of $3.303\pm0.011$ and 
$1.292\pm0.004$\,\cgs\, at 1565 and 2365 \AA\,, respectively. It has a colour excess of $E(B-V)=0.32$ mag
corresponding to $A_V=0.90$ mag, of which\, $\la$ half is thought to originate in the diffuse ISM between the star
and the Sun \citep{liszt}. Because the line of sight between $\zeta$\,Oph and LDN~1780 is located at high
altitude, $z > 44$~pc, we assume that the total extinction of $\zeta$\,Oph as seen from  LDN~1780
is  $A_V=0.45$ and that half of the UV light removed from the star's line of sight is returned via
scattered light from the large reflection nebula surrounding it \citep{choi}. With these assumptions and
applying the distance correction factor of $(d_{\rm L1780}/d_{\sun})^{-2}$ = 15.5 we find that
 $\zeta$\,Oph contributes by 20.8 and 9.2 \cgs\, to the illumination of LDN~1780. This is 
almost half as much as the contribution by  Sco\,OB2. 

With the addition of the contributions of Sco\,OB2 and $\zeta$\,Oph
the illumination for LDN~1780 is increased to 181 and 93 \cgs\, at 1565 and 2365~\AA\,, respectively.
These values are 1.57 and 1.65 times as large as the general Milky Way ISL+DGL surface brightness 
for an ``observer'' at $z = 50$ pc. For comparison, and in good agreement with this result, the direct 
summing-up of the radiation contributions by individual Tycho stars 
(for the method see \citealt{sujatha}) results  in a radiation field 
that is $\sim1.7$ and $\sim1.4$  times as large as that near the Sun in these two $GALEX$ bands 
(J. Murthy, private communication, May 2012). As seen from L\,1780 Sco\,OB2 and $\zeta$\,Oph are
seen in the directions $(l, b) = (338\degr, -21\degr)$ and $(34\degr, -48\degr)$, respectively. Their
contributions are added to the corresponding pixels in the FUV and NUV Milky Way maps
used for the L\,1780 illumination. 
Given the distances as adopted above the scattering angle 
of radiation reaching us via LDN~1780 from Sco\,OB2 and $\zeta$\,Oph is $\sim60\degr$ and  
$\sim70\degr$, respectively. The direction towards Sco\,OB2 as projected on the sky
is indicated by the arrow in the H{\small I} 21-cm map in Fig.\,\ref{L1780FUV_NUV_smooth}.

\subsection{Monte Carlo RT modelling of scattered light}

We have performed a multiple scattering calculation %of the scattered light
using the Monte Carlo Method as described by \citet{mattila70b}. The cloud is modelled as a homogeneous
sphere of optical thickness (diameter) $\tau$. The scattering function of a single grain is assumed
to have the form as introduced by \citet{henyey}, characterized by the single scattering albedo $a$ and
asymmetry parameter $g = \langle{\rm cos}\theta\rangle$. To obtain the scattering function of the cloud, $S(\theta)$, 
that shall be used in Eq.~(4), photons are
shot into the cloud from a fixed direction. The photons leaving the cloud are classified as function
of $\theta$, the angle between the initial and the final flight direction of the photon.

It should be noticed that $S(\theta)$ refers to the scattered light from the central spot of the cloud disk, 
the area of the spot is 1/10\,th of the total disk area. In order to model the scattered light of the
cloud as function of optical depth, $I_{sca}(\tau)$, we have used spherical model clouds with different
diameters $\tau$. While this approximation does not correspond to the geometrical intuition, it still
approximately takes into account the relevant optical depth and the varying amount of illumination at the 
off-centre cloud positions.

\subsection{Effects of foreground and background dust}

\subsubsection{LDN~1780}

It has been seen above in Section 2.4.1 and in Table \ref{table:1} that at the OFF (= sky) positions of LDN\,1780
there is an extinction of $A_V\approx 0.43$~mag. { From line-of-sight extinction data, 
available for example in \citet{green},  it is not possible to determine where the LDN~1780 cloud 
is located relative to this extended dust stratum: in front, behind or in between. The extinction
in the surrounding sky area rises at  $110\pm10$\,pc, that is closely the same distance 
where the complex of high-latitude dark nebulae LDN~134, LDN~169, LDN~183 and LDN~1778/1780 is located.
All these clouds are seen to be embedded in a large ($\sim10\degr$) dust envelope which 
is well seen in the $IRAS 100\,\mu$m and $Planck$  images as shown for example in \citet{laureijs95} 
and \citet{Planck2013XI}. We consider it most plausible that LDN\,1780 is located about half way between 
the front and back side of the envelope.}  

Then, a layer with  $A_V\approx0.215$~mag would be located in front and a layer with  $A_V\approx0.215$~mag
behind of LDN\,1780. Scattered light intensity from the layer in front of LDN\,1780 amounts
to half, and from the layer behind of it to the other half of the total intensity at 
the OFF positions:\\ 
$ I^{\rm foreground} =  I^{\rm background} = \frac{1}{2} I_{\rm OFF}$. \\    
{ However, because of the distance uncertainty we cannot exclude the alternative that LDN\,1780
is located in front of the diffuse dust envelope in which case\\
$ I^{\rm foreground} = 0$; \, $I^{\rm background} =  I_{\rm OFF}$. \\
This case will be considered in Section~4.3 in addition to our standard case with equal fore- and background
dust opacities.} 

The value of $I_{\rm OFF}$ corresponding to the {total line-of-sight} extinction at the OFF positions 
is determined by fitting a straight line to the low-extinction section 
of the $I_{\rm sca}$ vs $A_V$ curves, as shown in Fig.~\ref{VYSOSplots}, and extrapolating it 
to  $A_V$ = 0. The values are given in Table~\ref{table:6}.

We have handled the effects of the diffuse dust stratum towards L~1780 as follows:
{\bf (1)} The Galactic light $I_{\rm GAL}(l,b)$ impinging at the boundary of the envelope is 
attenuated by factor $\sim$e$^{-\tau_0}$ where $\tau_0$ corresponds to $\frac{1}{2} A_V$(OFF)= 0.215 mag. 
Because of scattering by the dust particles, the envelope adds a diffuse radiation field that 
compensates  part of the attenuation. The combined effect has been tabulated for  
isotropic incident radiation impinging on a plane parallel layer by \citet{whitworth}.
The attenuation can be represented in terms of an effective optical depth, $\tau_0({\rm eff})$. 
For $a = 0.60,\, g = 0.5 - 0.9$ we obtain from Table 1 of  Whitworth $\tau_0({\rm eff}) = 0.55 \cdot \tau_0$ 
which results in the attenuation factors e$^{-\tau_0({\rm eff})}$ as given in Table~\ref{table:6}. 
{\bf (2)} In our model half of the scattered light of the envelope is coming from beyond LDN\,1780.
It is attenuated when passing through LDN\,1780. Thus, we add to the model ON -- OFF signal 
a component $\Delta I = I^{\rm background}\times (e^{-\tau}-1)$. 
{\bf (3)} Finally, the scattered light from the LDN\,1780 cloud is attenuated by the dust layer in front 
of it, assumed to have an optical depth of $\tau_0$ corresponding to $\frac{1}{2} A_V$(OFF) = 0.215 mag. 
The attenuation factors e$^{-\tau_0}$ are given in  Table~\ref{table:6}. 

The observed surface brightness differences ON -- OFF\, are thus modelled according to the expression:

\begin{equation}
 \Delta  I^{\rm obs}_{\rm ON-OFF}(\tau) = [I_{\rm sca}(\tau) + I^{\rm background}\times(e^{-\tau}-1)]e^{-\tau_0} .
\end{equation}

\begin{table}
\caption{Parameters related to the ISRF attenuation, foreground extinction, and the off--cloud DGL models 
for fitting the LDN~1780 and LDN~1642 surface brightness observations.  $I_{\rm OFF}$ is in units of \cgs\,,
$A^{\lambda}_{\rm OFF}$ in magnitudes.}
\label{table:6}
\centering 
\begin{tabular}{lllll} 
Band    & $I_{\rm OFF}$& $\frac{1}{2} A^{\lambda}_{\rm OFF}$& e$^{-\tau_0{\rm(eff)}}$ & e$^{-\tau_0}$ (*)\\
\hline
 LDN~1780        &         &     &     &     \\
FUV   &20.5      &0.57 &0.75 & 0.59\\
NUV   &12.0      &0.61 &0.73 & 0.57\\     
$u$   &8.9$^{\#}$&0.34 &0.84 & 0.73\\
$B$         &12.5     &0.29 &0.87 & 0.77\\
$V$         &14.7     &0.215&0.90 & 0.82\\
$R$         &15.2     &0.17 &0.92 & 0.86\\
$i$         &14.0     &0.13 &0.94 & 0.88\\

                 &         &     &     &     \\
 LDN~1642        &         &     &     &     \\
$u$   &1.7      &0.21 & 0.90 &0.82\\
3840 \AA\,       &2.3      &0.20 & 0.90 &0.83\\
4160 \AA\,       &3.6      &0.19 & 0.91 &0.84\\
$b$   &3.4      &0.16 & 0.92 &0.86\\
$y$   &3.0      &0.135& 0.93 &0.88\\
\hline
\end{tabular}
\tablefoot{* $\tau_0 = \frac{1}{2} A^{\lambda}_{\rm OFF}/1.086$; \,\,\,  $\tau_0{\rm(eff)}= 0.55\, \tau_0$ 
  \newline
$\#$ estimated from $I_{\rm OFF}$($B$)}
%\label{table:6}
\end{table}

\subsubsection{LDN~1642}

For LDN~1642 we treat the influence of the foreground and background dust extinction and scattered light
in the same way as for LDN~1780. The assumption that the cloud is embedded in a diffuse dust stratum 
is based on the finding that its distance, $d = 124^{+11 }_{-14 }$ \citep{schlafly14}, coincides with that 
of the inner wall of the Local Bubble in its direction. This has been accurately determined by 
\citet{wyman} using interstellar  Na\,{\small I} D  absorption line measurements of $Hipparcos$
stars. The OFF area extinction is $A_V = 0.27$~mag, substantially less than that of LDN~1780. 
The quantities as used for the corrections,  $I_{\rm OFF}$,  e$^{-\tau_0{\rm(eff)}}$  and e$^{-\tau_0}$,
are given in the second part of Table \ref{table:6}.% for the filter bands of the LDN~1642 observations.

\subsubsection{Draco nebula}

For the Draco nebula the extinction at the OFF positions is small,  
$A_V\approx 0.065$~mag (see Section 2.4.2 and\,Table\,\ref{table:1}) and it is likely caused by dust
in the foreground. Thus, an additional treatment with a background dust layer is avoided.
Also, the attenuation by the foreground dust layer remains small enough to be neglected in view
of the other error sources.

%\newpage

\begin{figure}
\hspace{-0cm}
            {\includegraphics[width=10cm,angle=-0]{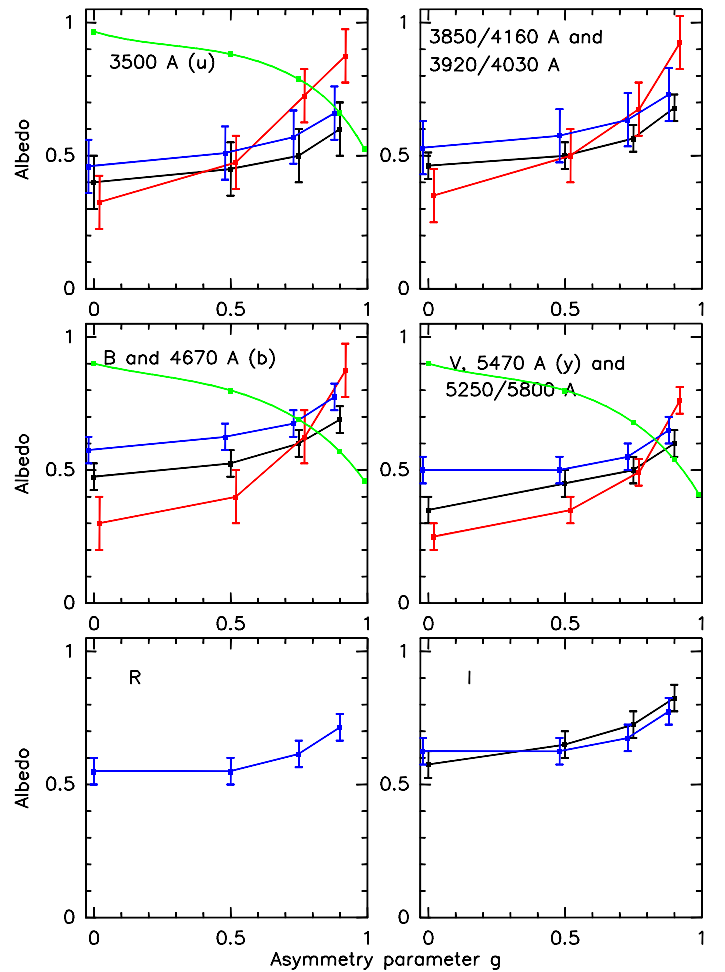}}
\vspace{-0cm}
\caption{Albedo values vs asymmetry parameter $g$  derived from modelling as shown in Fig. \ref{Isca_models}. 
Results are shown for LDN~1780 (blue symbols),  LDN~1642 (black) and Draco nebula (red). For reference
also the results for the Coalsack are shown in the $UBV$ bands (green lines) \citep{mattila70b}.
}
             \hspace{-.0cm}
        \label{albedo_vs_g}
   \end{figure}

\subsection{Model fitting of the spectral energy distributions}

The observed SEDs of LDN~1780 and Draco nebula as shown in Fig.~\ref{FigSEDs} shall be 
compared with model spectra of the integrated starlight.
In the scattering process the ISL spectrum, while preserving its spectral features, experiences a broad-band
reddening (or bluening) that depends on the optical thickness of the cloud and the scattering properties of the grains.
Because of the more qualitative aims of the SED comparison we have not made a full self-supporting RT modelling
according to Eq.~(4). Rather, { following the the analysis of \citet{witt08} } the scattered 
light spectra of the ON-cloud positions with $A_V \sim 1 - 3$\,mag have been { approximated using the ansatz}
\begin{eqnarray}
\lefteqn{I^{\rm sca}_{\lambda}({\rm cloud}) = {} }  \nonumber  \\
 &  {} = [C_V I^{\rm ISL}_{\lambda} {\rm e}^{-\tau_{\lambda}^{\rm eff}} (1 - {\rm e}^{-\tau_{\lambda}({\rm cloud})})
+ I^{\rm sca}_{\lambda}({\rm bg}) {\rm e}^{-\tau_{\lambda}({\rm cloud})}] {\rm e}^{-\tau_{\lambda}({\rm fg})}
\end{eqnarray}
{ where $I^{\rm ISL}_{\lambda}$ represents the model SED for the ISL; $C_{V}$ is a scaling factor used to adjust the model SED 
intensity level at 5500\,\AA\, to the observed spectrum; $\tau_{\lambda}$(cloud) is the line-of-sight optical depth through the cloud 
at the observed position; and $\tau_{\lambda}$(fg) is the optical depth of the foreground dust layer. 
The term $1 - {\rm exp}[{-\tau_{\lambda}({\rm cloud})}]$ accounts for the dependence on the finite and 
wavelength-dependent optical path length along the line of sight. 
It is a good approximation for moderate optical depths, $\tau_{\lambda}$(cloud) $\la 0.5$.

For larger cloud thicknesses one has to account for the reddening that the impinging ISL may experience before 
reaching the line-of-sight column.  This is the purpose of the term  ${\rm exp}(-\tau_{\lambda}^{\rm eff})$ where 
$\tau_{\lambda}^{\rm eff}$ can be thought to consist of two components, $\tau_{\lambda}^{\rm eff}$(envelope) 
and $\tau_{\lambda}^{\rm eff}$(cloud); it can encorporate also any additional
reddening (or bluening) which may be required to correct for the approximative nature of the extinction  
correction in our ISL SED model. It is assigned as 'effective' optical depth to emphasize that it is 
not directly given by the optical depth of the cloud and its envelope. Therefore, $\tau_{\lambda}^{\rm eff}$  
is in the first place to be understood as a fitting parameter when adjusting  the ISL model SED to 
the observed one. 

The scattered light is coming mainly from the cloud's surface layer with $\tau_{\lambda}\la 2$, facing the 
observer. This can be qualitatively seen from Fig.~5 where the surface brightness saturates when 
$\tau_{\lambda}{\rm (cloud)} \ga 2 $ at the respective wavelength $\lambda$. For $B$ band the saturation 
occurs at  $\tau_{V}{\rm (cloud)} \sim 1.5$ whereas for the  $I$ band it happens only at 
$\tau_{V}{\rm (cloud)} \sim 3.5$. This situation also qualitatively demonstrates that the scattered from
lines-of-sight with $\tau_{V} \ga 1 - 1.5$  can appear redder than the impinging ISL: while the $B$ band
intensity saturates the $V$ to $I$ band intensities still continue to increase.

The scattered light from the background dust layer at $d>d_{\rm cloud}$ passing through the cloud 
is represented by the term $I^{\rm sca}_{\lambda}({\rm bg}){\rm exp}[{-\tau_{\lambda}({\rm cloud})}]$.} 
For the wavelength dependence of $\tau_{\lambda}^{\rm eff}$,  $\tau_{\lambda}$(cloud) and $\tau_{\lambda}$(fg) 
the interstellar extinction law of \citet{cardelli89} with $R_V = 3.1$ was adopted.

At the sky positions, with $A_V \la 0.5$\,mag, the scattered light from the background dust layer at  
$d>d_{\rm cloud}$ can be modelled as 
\begin{equation}
I^{sca}_{\lambda}({\rm bg}) = I^{sca}_V({\rm bg})
\frac{I^{\rm ISL}_{\lambda}(1 - e^{-\tau_{\lambda}({\rm bg})})}{I^{ISL}_V(1 - e^{-\tau_{V}({\rm bg})})}e^{-\tau_{\lambda}({\rm fg})}
 \end{equation}
where $\tau_{V}$(bg) and $\tau_{V}$(fg) are the optical depths of the background ($d>d_{\rm cloud}$ ) and of the 
foreground ($d<d_{\rm cloud}$) dust layer in $V$ band. In this case
the scattered light will be bluer than the impinging ISL, $I^{\rm ISL}_{\lambda}$; this is demonstrated by
the model SEDs for the OFF position backround light shown in the upper and middle panels of Fig.~14.

The observed surface brightness 'cloud minus sky' is then given by
\begin{equation}
\Delta I^{\rm sca}_{\lambda} = I^{\rm sca}_{\lambda}{\rm(cloud)} - I^{\rm sca}_{\lambda}({\rm bg}).
\end{equation}
{ The scattered light intensity of the foreground dust layer ($d<d_{\rm cloud}$) is assumed to
be the same towards the cloud and the sky: it cancels out in the difference.}

We have calculated the ISL spectra, $I^{\rm ISL}_{\lambda}$, using a synthetic model of the Solar neighbourhood 
as described in Section 3.1.1. The calculations were made for different galactic latitudes and for three locations in the Solar 
neighbourhood: at $z$=0, 50 and 400\,pc, the latter two corresponding to the locations of LDN~1780 and Draco nebula.  
For a single scattering phase function of \citet{henyey} with $g = 0.80$, about 70\% of light is scattered 
within an angle of $\theta < 30\degr$. Thus, the illumination for LDN~1780 and Draco nebula from the sky area behind 
the cloud, that is from galactic latitudes of $b = 10\degr - 70\degr$, is heavily weighted. This is, however, counterweighted by
the brighter ISL in other parts of the sky. For the Draco nebula (assumed to be at $z = 400$\,pc) the $B$ band sky 
background at $b \sim 36\degr$ is only $\sim 17\%$ of the mean sky brightness and $\sim 18\%$ of the corresponding value at 
 $b \sim -36\degr$ (see also Table~\ref{table:4}). 

For the purpose of comparing the observed spectra with the ISL model spectra we have adjusted  in Eq.~(6) the 
parameters $C_V$ and $\tau_{\lambda}^{\rm eff}$ in such a way that a best overall agreement was found. 
The  $V$-band sky background intensity, $I^{\rm sca}_V$(bg), has been estimated independently as described 
in Sections~3.3.1 and 3.3.2 (see also Table~\ref{table:6}). For other wavelengths $I^{\rm sca}_{\lambda}$(bg) 
was calculated using Eq.~(7).

\section{Results and discussions}

\subsection{Dust albedo in the optical}

Model fits of scattered light data in LDN~1780 are shown in Fig.~\ref{Isca_models} for
$B$ and $R$ bands and in Fig.~\ref{Isca_modelU} for the $u$ band.
Models are shown for four different scattering parameters, $g$\, = 0, 0.50, 0.75 and 0.90, 
and for a range of albedos as indicated in the figures. 
Corresponding figures are utilized for LDN~1642 and the Draco nebula.

As can be seen from the figures the observational data enable, for each $g$ value, a determination 
of the albedo with a good precision, typically $\pm 0.05$. 
While at low extinctions, $A_V \la 1$~mag,  $I_{\rm sca}$ is linearly proportional 
to albedo, the dependence is substantially steeper at intermediate and high extinction values,
$A_V = 1.5 - 4$~mag. This is caused by the increasing contribution of multiple scattering with increasing
optical depth.

The albedo values as function of the asymmetry parameter $g$ are shown in Fig.~\ref{albedo_vs_g} 
for six different wavelength bands. Two or more filter bands are grouped together in the cases  
3850/3920/4030/4160 \AA\,, $B$/$b$ and $V$/$y$/5250/5800\AA\,. 

In the bands between 3500 -- 5500\,\AA\, all three clouds have been observed. They show different $a$ vs $g$ 
dependences: while LDN~1642 and LDN~1780 have a moderately rising albedo from  $g=0$ to  
$g=0.90$ the curve for Draco nebula is a much steeper one, ranging from  $a\sim$0.25
to  $\sim$0.75 for $V/y$ and from $a\sim$0.3 to  $\sim$0.9 for the $B/b$ band. This behaviour is caused by the
anisotropy of the impinging Galactic ISRF: if $g=0$ the cloud scatters with roughly equal weight the light from
all directions, while for $g=0.75 - 0.90$ the illumination comes with higher weight from the dimmer sky area
 behind the cloud at high galactic latitudes, $|b| > 25\degr - 40\degr$. With increasing $z$-distance from 
$\sim$50 -- 60 pc for LDN~1780 and LDN~1642 to $\sim$400 pc for Draco the ISRF an-isotropy is strongly enhanced (see 
Table~\ref{table:4}). On the other hand, when a cloud is projected against the bright areas near the galactic 
equator its surface brightness shows an opposite trend: it decreases with increasing value of $g$. 
To qualitatively demonstrate this effect we have plotted in Fig.~\ref{albedo_vs_g} also the $UBV$ band results 
for Coalsack ($l, b \approx 301\degr, -1\degr$) according to \citet{mattila70b}. 
 
\begin{figure}
\vspace{-2cm}
\hspace{-0.2cm}
            {\includegraphics[width=9cm,angle=-0]{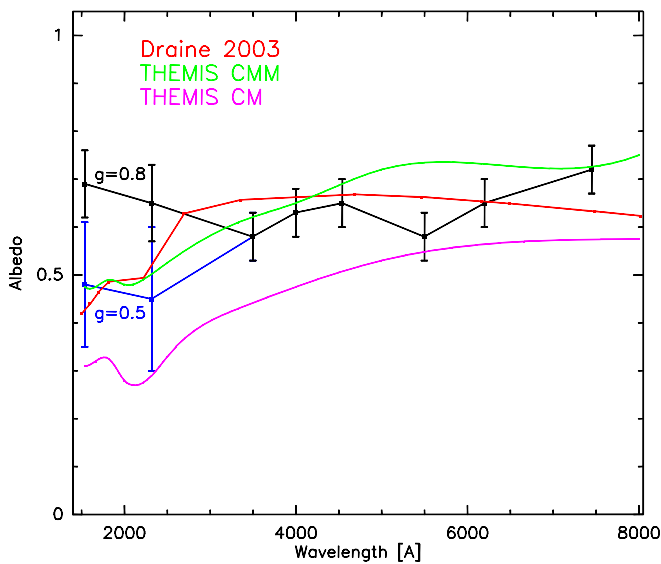}}
\vspace{-1.5cm}
\caption{Our derived dust albedos according to results shown in 
Figs.~\ref{albedo_vs_g} and \ref{Galex_a_vs_g} and Table~\ref{table:9}.  For the black dots 
with error bars the g-parameter value of 0.80 has been adopted for all bands. For FUV and NUV
albedos also the values for $g = 0.50$ are shown as blue point with error bars.
 For FUV ($\lambda 1540$\,\AA\,)
and NUV ($\lambda 2320$\,\AA\,) the mean of LDN~1780 and Draco nebula is shown, with the Draco value at 
the lower and LDN~1780 value at the upper end of the error bar. 
Dust model albedo values are shown according to \citet{draine03} 
(red curve), THEMIS CM (magenta) and THEMIS CMM (green) \citep{jones}.  
}
\hspace{-0cm}
\label{albedo_vs_lambda}
\end{figure}

If we assume that the scattering properties of dust are the same in LDN~1642, LDN~1780 and Draco nebula
then the crossing area of the three loci in the $a,g$ plane provides a determination of both parameters. For the
$B/b$ and $V/y$ bands this results in the values  $g \sim0.80, \,a \sim0.70$ and  $g \sim 0.80, \,a \sim 0.60$, 
respectively. For the $u$ and 3850/3920/4030/4160 \AA\, bands somewhat smaller g-parameter values of 
$\sim0.60$ and $\sim0.65$
are indicated, resulting in albedo values of 0.50 and 0.60, respectively. If, on the other hand, $g = 0.80$ is adopted
also for these bands, then slightly larger albedos of 0.58 and 0.63 are obtained from the LDN~1780 and LDN~1642
data. Because of the larger $z$-distance of Draco larger correction factors have been applied, especially for the
$u$ and  3850 -- 4030 \AA\, bands (see Tables~\ref{table:4} and \ref{table:5}). In view of these uncertainties
the uniform value of $g = 0.80$ appears preferable also for these bands. We give in Table~\ref{table:9} and 
Fig.~\ref{albedo_vs_lambda} the resulting albedo values when $g =0.80$ is adopted.

\begin{table}
\caption{Albedo values $a$ for asymmetry parameter value $g = 0.80$. Values of
 $g$ as determined independently at each wavelength are given in column 4. In the last column the
 $g$ value as favoured by our observations is given. }
%\label{table:9}
\centering 
\begin{tabular}{lcccccc} 
Band  &$\lambda[\AA\,]$  & $a$ for $g=0.8$& $g$ observed\\
\hline
$u$  &3500 &  $0.58\pm0.05$ &0.60$\pm0.1$\\
3850/3920/4030/4160 &4000  & $0.63\pm0.05$ &0.65$\pm0.1$ \\
$B/b$  & 4540 &  $0.65\pm0.05$ & 0.80$\pm0.1$\\  
$V/y/5250/5800$ &5500 &  $0.58\pm0.05$ &0.80$\pm0.1$ \\ 
$R$ &6200 &  $0.65\pm0.05$ & -- \\   
$i$  & 7450 & $0.72\pm0.05$ & -- \\
\hline
\end{tabular}
\label{table:9}
%\tablefoot{* } 
\end{table}

\subsection{Ultraviolet dust albedo} 

The ultraviolet surface brightness distribution in LDN~1780 differs from that in 
the optical (Fig.~2). It is, especially in the FUV band, much stronger in the southern half
of the nebula (Dec.\,$<-7\degr15\arcmin$) as compared with positions with the same $A_V$
in the northern half. For a quantitative presentation see Fig.~8. A somewhat similar dichotomy,
though much less pronounced, is observed also between the ``Head'' and the ``Tail'' parts of the
Draco nebula (see Figs.~4 and 9).
This behaviour can be understood for the Draco nebula (both FUV and NUV) and LDN 1780 (NUV, but 
not FUV as shown below) in terms of scattered light, with enhanced illumination from 
the hemisphere facing the galactic plane and with strong intra-cloud shadowing by the dust in UV.

\begin{figure}
\vspace{-0cm}
\hspace{-1.2cm}
            {\includegraphics[width=11.5cm,angle=-0]{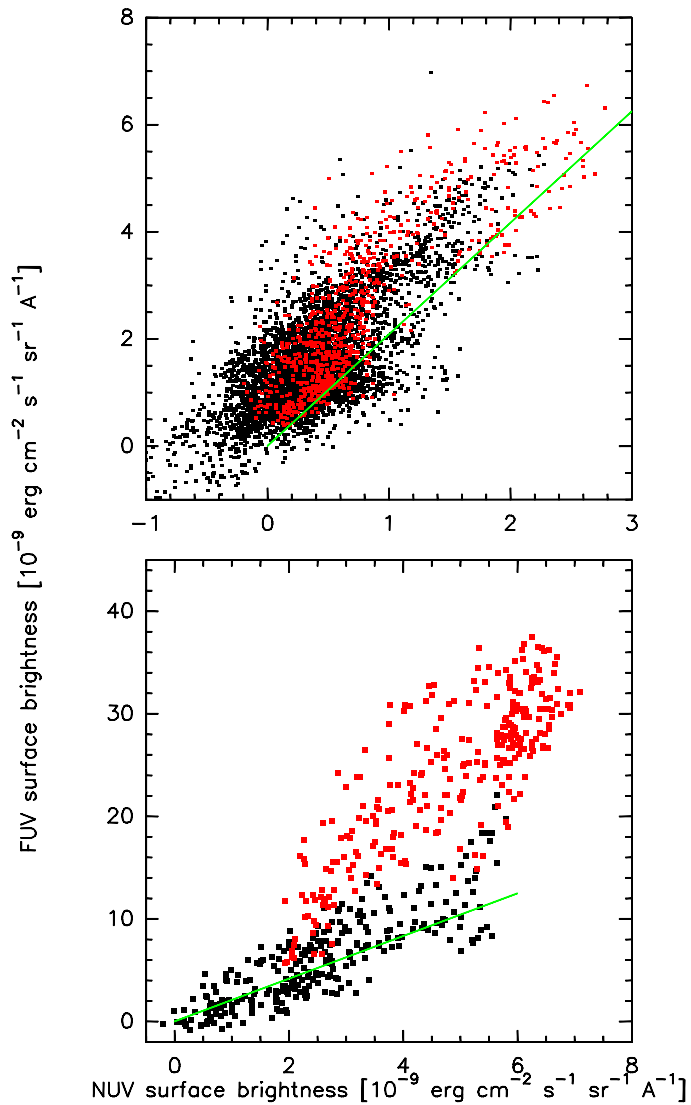}}
\vspace{-0cm}  
\caption{FUV vs NUV surface brightness in LDN~1780 (lower) and Draco nebula (upper). 
The southern part of LDN~1780 and the ``Head'' of Draco are indicated with red and the 
northern part of LDN~1780 and ``other'' parts of Draco with black dots (see Figs.~8 and 9). 
The mean intensity ratio for the all-sky diffuse background radiation according to 
\citet{murthy14} is shown as the green line. 
}
\hspace{-0cm}
\label{FUV_vs_NUV}
\end{figure}

We show in Fig.~\ref{FUV_vs_NUV} a plot of FUV  ($\lambda = 1344 - 1786$\,\AA\,,) vs NUV
($\lambda = 1771 - 2831$\,\AA\,) surface brightness for  LDN~1780 (lower panel) and Draco nebula 
(upper panel). The positions in the southern part of LDN~1780 and in the ``Head'' of Draco 
are indicated with red points; for choice of areas see explanations for Figs.~8 and 9. 
While the NUV intensity is likely due to scattering only, the FUV band contains, potentially, 
also fluorescent emission bands of  H$_2$ and spectral lines of atomic species. In Draco
both the ``Head'' (red) and the ``Tail'' points (black) are seen to obey the same relationship
$I_{\rm FUV} = k\, I_{\rm NUV} + b$. The slope indicated with the green line corresponds to the
mean $I_{\rm NUV}/I_{\rm FUV}$ ratio of 0.48 found by \citet{murthy14} for the all-sky
diffuse background radiation. For LDN~1780 the ``shadow-side'' points (black) are seen to follow 
this same mean $I_{\rm NUV}/I_{\rm FUV}$ relationship. For the ``bright-side'', however, the 
points (red) show a different slope, and relative to the mean $I_{\rm NUV}/I_{\rm FUV}$ value there
is a FUV excess of up to $\sim25$\,\cgs\,; or, the ratio  $I_{\rm FUV}/I_{\rm NUV}$ is up to 
$\sim3.5$ times as large as that on the ``shadow side''. We interpret 
this behaviour as evidence for H$_2$ fluorescence plus possible atomic emission lines and will
analyse it in Section 4.4. 
(For a similar argument applied to a foreground dust filament in Draco see \citealt{sujatha10}.)

For a strongly forward throwing scattering function, $g\ge 0.7$, the starlight
``from behind'' dominates the illumination even for high-latitude clouds  
in the optical bands, 3500 to 8000 \AA\,. For the NUV (2320 \AA\,) and FUV (1540 \AA\,) 
bands the situation is different: here the background starlight at high latitudes,
and especially when the cloud is at high altitude ($z\ga200$~pc), becomes so weak that
light from the galactic disk dominates also for $g\ge 0.7$.

The NUV observations of LDN~1780 can be reasonably well understood in terms of scattered light, see 
Fig.~\ref{L1780_Galex_models}. At small extinctions, $A_V \la 1.5$\,mag,  there is a large systematic 
difference between the ``bright side'' (red) and ``dark side'' (black) branches of points. However, 
in the cloud centre, corresponding to $A_V \approx 2.5 - 3$\,mag, both branches converge. It is just
here that our simple Monte Carlo simulation is also expected to be valid. Reasonable albedo estimates
can thus be obtained for the $g$ values of 0, 0.50 and perhaps also for 0.75, whereas the models 
for $g = 0.90$ do not reproduce the observations for any albedo.  

\begin{figure}
\vspace{0.5cm}
\hspace{-1cm}
           {\includegraphics[width=10.5cm,angle=-0]{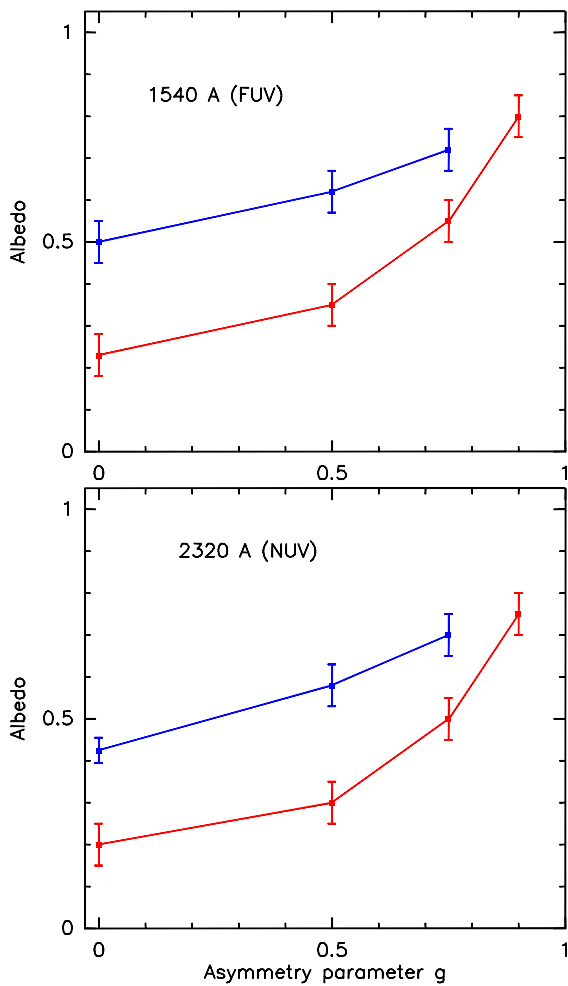}}
 \caption{Albedo values vs asymmetry parameter $g$  derived from modelling as shown in 
Figs. \ref{L1780_Galex_models} and \ref{Draco_Galex_models}. 
Results are shown for LDN~1780 (blue symbols), and Draco nebula (red). 
}
\label{Galex_a_vs_g}
\end{figure}

In the Draco nebula the NUV intensity at $A_V\approx1$\,mag is by a factor of $\sim2 - 3$
lower than that for LDN~1780 at the same $A_V$.  The model fits reproduce well the observed intensity 
vs $A_V$ behaviour and, although the observational scatter is large, enable determination of albedo values 
to a precision of $\sim\pm0.1$ for all values of $g$ from 0 to 0.9. The NUV albedo values for LDN~1780 
and Draco are shown in the lower panel of Fig.~\ref{Galex_a_vs_g} as function of  $g$.

Also the FUV observations of the Draco nebula can be satisfactorily interpreted in terms of the scattered
light model curves for $g = 0 - 0.9$ . Although the observational scatter at faint surface brightness levels is
large it still allows the determination of the albedo to a precision of $\sim \pm 0.1$ or better. Also in this 
case most weight is put to the positions at the highest extinction range at   $A_V \ga 1$\,mag where our Monte 
Carlo modelling corresponds best to the cloud illumination geometry. The positions in the cloud ``Head''
(red points in Fig.~\ref{Draco_Galex_models}) are considered to better correspond to our ISRF  
model and they are preferred over the ``Tail'' points. 

In the case of FUV observations of LDN~1780 we consider that the points (red) at $A_V \la 2$ on the 
``bright side'' are strongly
affected by H$_2$ fluorescence emission and cannot be used for albedo estimates. Thus, the observed
points at  $A_V \ga 2.5$\,mag, where the ``bright'' and ``shadow'' branches converge, are again
preferred for the albedo determination. For $g = 0$ and 0.5 reasonably good fits are possible and
a useful estimate appears possible for  $g = 0.75$ but not for  $g = 0.9$. The  FUV albedo 
values are shown in the upper panel of Fig.~\ref{Galex_a_vs_g} for the three values of $g$. 

 In Fig.~\ref{albedo_vs_lambda} the NUV and FUV albedo values are shown with black symbols for  
the $g$-parameter value of 0.80 which in the optical appears reasonably well justified. However, as has 
been pointed out above, the observations in NUV cannot be well fitted with models for $g > 0.75$. 
And the FUV observations, especially for LDN~1780, are not well-fitted for  $g > 0.5$. Therefore, 
we show in Fig.~\ref{albedo_vs_lambda} the FUV and NUV albedo values also for $g = 0.5$ (blue symbols).

The FUV ``bright side'' intensity peak of LDN~1780 at  $A_V \approx 0.9$\,mag cannot be 
satisfactorily fitted with any combination of the scattering parameters $a$ and $g$ (see Fig~\ref{L1780_Galex_models}). 
Even the best-fitting model curves for $g = 0$,  $a \sim 0.6$ leave an excess of $\sim 15$\,\cgs\, 
or $\sim 40$\,\% of the total signal unexplained.

\subsubsection{Dust scattering properties: comparison with dust models}

Two model concepts for grains are at present mostly being discussed and used 
for predicting observational phenomena of interstellar dust:
(1) Based on the pioneering work of \citet{hoyle} grain models consisting of a mixture of distinct 
populations of bare graphite and silicate particles, each one having its own size distribution; 
this concept has lead to the popular MRN dust model of \citet{mathis77} and its modified versions 
presented  in \citet{draine84} and \citet{weingartner}; 
(2) the so-called core-mantle grains which have graphite or silicate cores covered by ice or 
 ``organic'' amorphous carbon mantles were introduced by \citet{greenberg86}, 
 \citet{duley87}, \citet{jones88} and others and are the basis of the dust modelling framework
THEMIS\footnote{The Heterogeneous dust Evolution Model for Interstellar Solids} \citep{jones} . 

Besides the wavelength dependence of extinction, the scattering properties of the grains, the albedo $a$ and the 
asymmetry parameter $g$, are important in putting constraints on the grain models.
Scattering and the determination of $a$ and $g$ have been studied in a variety of dusty environments;
for a review see for example \citet{gordon04}. 
  
In Fig.~\ref{albedo_vs_lambda} we compare our observed albedo values with the predictions of three dust models: 
(1) the albedo values as presented in \citet{draine03} for the \citet{weingartner} model (WD for short) 
for the ``Milky Way dust'' and $R_V$ = 3.1 are shown as red line;
(2) albedos for two models in the THEMIS family, CM (for core-mantle, magenta line) and CMM 
(for core-mantle/mantle, green line) are shown according to \citet{jones} and \citet{ysard}. 
In the optical, $\lambda = 3500 - 7500$\,\AA\,, our observed albedo values (given for $g = 0.8$) are in good 
agreement both with the WD and THEMIS CMM models but lie, at most of the wavelength 
points, substantially above the THEMIS CM model values. 
In the ultraviolet (FUV and NUV) we give the albedo values  
for two asymmetry parameter values, $g =0.8$ and $g =0.5$. The values for $g = 0.5$ are seen to be in good
agreement with both the WD and THEMIS CMM model whereas the values for  $g = 0.8$ disagree with both.

Among the THEMIS family the CM particles represent the ``basic'' form, thought to prevail in the diffuse medium.
Moving towards the denser phases the CMM grains have more evolved mantles of amorphous carbon and C:H compounds.
The physical environment of translucent dark clouds and cloud envelopes appears to be the appropriate 
environment where CMM type grains prevail. The more evolved THEMIS AMM and AMMI grains, giving albedo
values in close agreement with CMM grains and our observations, are thought to exist mainly in the denser
core regions of molecular clouds.

Based on our results for the bands 4000\,\AA\,, $B$ and $V$ the asymmetry parameter value $g = 0.8$ was adopted 
as a reasonable estimate for the whole optical wavelength band.  There is also a wide agreement 
from studies in different environments and among most authors on a large value of $g$ ($\sim 0.75 - 0.9$) 
in the optical \citep{gordon04}. 
However, in the ultraviolet bands, $\lambda = 1350 - 1750$\,\AA\, (FUV) and $\lambda = 1750 - 2850$\,\AA\, (NUV), 
our observations 
favoured a smaller value,  $g \la 0.5$ against  $g = 0.8$. A trend towards a smaller $g$ values in the ultraviolet
was indicated also by our ``best-fit'' values  $g = 0.65\pm0.1$ at 4000\,\AA\, and   $g = 0.60\pm0.1$ at 3500\,\AA\,
(see Table~\ref{table:9}). 

For the WD model the value of $g$ \citep{draine03} increases from $\sim$0.45 at 7500\,\AA\, to  $\sim$0.65 at 
1500\,\AA\,. In the ultraviolet, $\lambda \le $ 4000\,\AA\, this is compatible with our preference for intermediate 
$g$ values but it is in contradiction with the observational values at the wavelengths,  $\lambda \ge  $ 4500\,\AA\,. 
The same problem occurs and is even more pronounced  w.r.t. the THEMIS CMM grains \citep{ysard}: 
it predicts a value of $g$ increasing from $\sim$0.45 at 7500\,\AA\, to  $\sim$0.8 at 1500\,\AA\,.

While it is out of the scope of the present paper to review and discuss the previous $a$ and $g$ determinations
(see for example the compilation in \citealt{gordon04}\footnote{see also the updated compilation in Karl D. Gordon's web page 
http://www.stsci.edu/~kgordon/Dust/Scat\_Param/scat\_data.html}) we note that in their recent study
 \citet{togi} find results closely similar to ours for the albedo in optical. Their results are based on $UBVRI$ photometry
of the translucent outer parts of the dense Taurus core Barnard~207 (LDN~1489). Their albedo values were
calculated assuming that $g = 0.9$. In the $u$, 4000\,\AA\, and $B$ bands there is a perfect agreement with our values,
but in the $V, R, I$ bands their albedo values are by $\sim$0.10 -- 0.16 larger than ours. A substantial part of this
could result from the different   $g$ values adopted: had we used  $g = 0.9$ instead of  $g = 0.8$ our albedo 
values would have increased by $\sim$0.05 -- 0.10 (see Fig.~\ref{albedo_vs_g}).

\subsection{Comparison of the SEDs of the ISL and scattered light}

 While the aim of Sections 4.1 and 4.2 was to model the absolute intensity of scattered light 
and thereby derive the dust albedo, this section has more qualitative goals: 
(1) to study the connection between the spectral energy distributions (SED) of the dark 
nebula and that of the ISL; 
(2) to see if such a connection could enable to determine the location of the nebula, for example its distance; and 
(3) to find if the nebular surface brightness can be understood solely as scattered starlight or 
whether other components, such the Extended Red Emission (ERE), are needed.

\subsubsection{LDN~1780}

{ For Pos1 and Pos10 in LDN~1780 the green lines in the upper and middle panels of Fig.~\ref{SED+ISRF} show the 
model spectra $I_{\lambda}^{\rm sca}$(cloud) and $I_{\lambda}^{\rm sca}$(bg) for the case that the ISL spectral shape 
is given by the starlight at  $b = 36\degr$, representative of the high-latitude sky in the direction 
of the nebula (see Eqs. 6 and 7). Their difference, $\Delta I^{\rm sca}_{\lambda}$, 
is shown as the red lines; it is to be compared with the observed differential SEDs shown as 
black squares with error bars.  The ISL spectrum for $b = 36\degr$ (magenta line)
is shown with normalization to the  $\Delta I^{\rm sca}_{\lambda}$  values at 5500 \AA\,. 

In the upper left and middle left panels of Fig.~\ref{SED+ISRF} the $I_{\lambda}^{\rm sca}$(bg) spectra 
are for the case that half of the total diffuse galactic sky background 
light at OFF positions, $I_V^{\rm sca}$(tot) = 15\,\cgs\,, corresponding to $A_V = 0.43$\,mag, 
originates at distances $d > d_{\rm cloud}$; the other half at  $d < d_{\rm cloud}$ has been assumed to be the same
for the direction of the cloud and the sky positions and has thus canceled out in the difference (see Section
3.4). In the upper right and middle right panels of Fig.~\ref{SED+ISRF} the cloud is assumed to be in front of 
all widely distributed
dust and, therefore, the total diffuse galactic sky backgrond light at OFF positions is ascribed to
the backround component $I_{\lambda}^{\rm sca}$(bg), relevant for the ON - OFF subtraction.

In each of the four cases the model ON - OFF spectrum has been fitted to the observed SED by adjusting
the parameters $C_V$ and $\tau_V^{\rm eff}$ in Equation (6). The best-fitting values of 
$\tau_V^{\rm eff}$ for the case of $I_V^{\rm sca}$(bg) = 7.5\,\cgs\, were 0.70 and 0.80 
for Pos1 and Pos10, respectively; for $I_V^{\rm sca}$(bg) = 15\,\cgs\, the $\tau_V^{\rm eff}$  values
were somewhat smaller, 0.3 and 0.5 for Pos1 and Pos10, respectively.
}

From comparison of the observed with the model spectra the following conclusions can be drawn: 
{ (1) Good model fits are obtained for both background choices adopted above, $I_V^{\rm sca}$(bg) = 
7.5 or 15 \,\cgs\,.\,  A larger $I_V^{\rm sca}$(bg) and $\tau_{V}^{\rm eff}$ influence the differential spectrum,
 $\Delta I^{\rm sca}_{\lambda}$, in the same direction: making it redder. 
Thus, somewhat larger $\tau_{V}^{\rm eff}$ values result for our favoured case
where the LDN~1780 cloud is embedded halfway within the surrounding diffuse dust envelope. We note that 
the $\tau_{V}^{\rm eff}$ value obtained for the higher opacity Pos1 ($A_V = 3.0$\,mag) is almost the same as that for Pos10
with the somewhat lower opacity ($A_V = 1.84$\,mag). This is as expected in the situation where the scattered
light comes mostly from a $\tau_{V} \la 2$ surface layer facing the observer (see Sect. 3.4).
(2) No additional components except scattered light are needed for a good fit. }

\citet{chlewicki} argued that the peak  at $\sim$6500\,\AA\, in the SED of LDN~1780 surface brightness \citep{mattila79} 
could not be explained by scattered light with the 'standard' grain models, for example \citet{mathis77}. They suggested 
an explanation in terms of optical fluorescence emission from grains or molecular species. 
In a number of papers LDN~1780 was subsequently presented
as a relatively rare example of ERE occuring in a translucent dark nebula in low UV radiation density environment 
exhibiting a high ERE quantum yield \citep{gordon, smith02, witt04}. 

Our multi-band photometry of LDN~1780 indicates, in agreement with 
\citet{mattila79}, the presence of a broad maximum peaking at  $\sim$6000 -- 7000\,\AA\,. 
We argue, however, that our modelling in terms of scattered light only
is capable of explaining this broad intensity maximum.
  
The fall on the long-wavelength side is jointly caused by the genuine form of the ISL's SED with declining 
intensity towards the longer wavelengths and the decrease with  $\lambda$ of the line-of-sight optical depth 
through the nebula, influencing the intensity via the term $1 - {\rm exp}[{-\tau_{\lambda}({\rm cloud})}]$.

{ The rise on the short-wavelength side of the peak, from 
$\sim$4500 to 6000 \AA\,, results from the combination of two effects: 
(1) the intensity of the diffuse galactic scattered light at the OFF (=sky) positions, from beyond the 
distance of LDN~1780, $I_{\lambda}^{\rm sca}$(bg), is a monotonically rising function towards the shorter 
wavelengths; its effect is thus to make the ON - OFF signal,  $\Delta I^{\rm sca}_{\lambda}$, redder;
(2) reddening of the impinging ISL occurs when passing through the large-scale diffuse dust envelope 
in which LDN~1780 is embedded, and subsequently also within the nebula itself; 
these two effects are represented in Eq.~(6) by the term ${\rm exp}({-\tau_{\lambda}^{\rm eff}})$.}

For lines of sight with intermediate optical depths, $\tau_{\rm V} \sim$1 to 3 mag, these two dust 
scattered light and opacity effects give rise to a broad maximum in the range $\sim$6000 -- 7000\,\AA\,.

{ While our differential SEDs of LDN~1780 can well be explained in terms of reddening by dust and 
by the blue SED of the OFF position signal, we cannot exclude that  ERE is present in LDN~1780.
Because far-UV photons with energies 10.5\,eV < E <13.6\,eV are required for the ERE excitation 
\citep{witt04,lai17} the ERE phenomenon is limited to low-extinction areas.  In case of LDN~1780 this means 
that the extended diffuse dust layer covering also the OFF positions may give rise to an ERE signal as strong 
as that in LDN~1780, thus nulling the signal in the ON - OFF SED.
And secondly, Pos1 and Pos10 represent lines of sight strongly shadowed against the far-UV photons; in the
southern part of the nebula where scattered far-UV light and H$_2$ fluorescence emission are observed (see Fig.~2), 
the chances for ERE detection might be better. }

For optically thin lines of sight, $\tau_V \la$0.5 mag, as was the case for the nebulae in the sample of \citet{witt08}, 
a monotonically falling scattered-light SED from $\lambda \sim$4500 to 9000 \AA\, was found. In that case no similar 
confusion of the ERE signal with dust opacity effects did happen.

\subsubsection{The Draco nebula}

\begin{figure*}
\vspace{-3.0cm}
\hspace{-0cm}
           {\includegraphics[width=10cm,angle=-90]{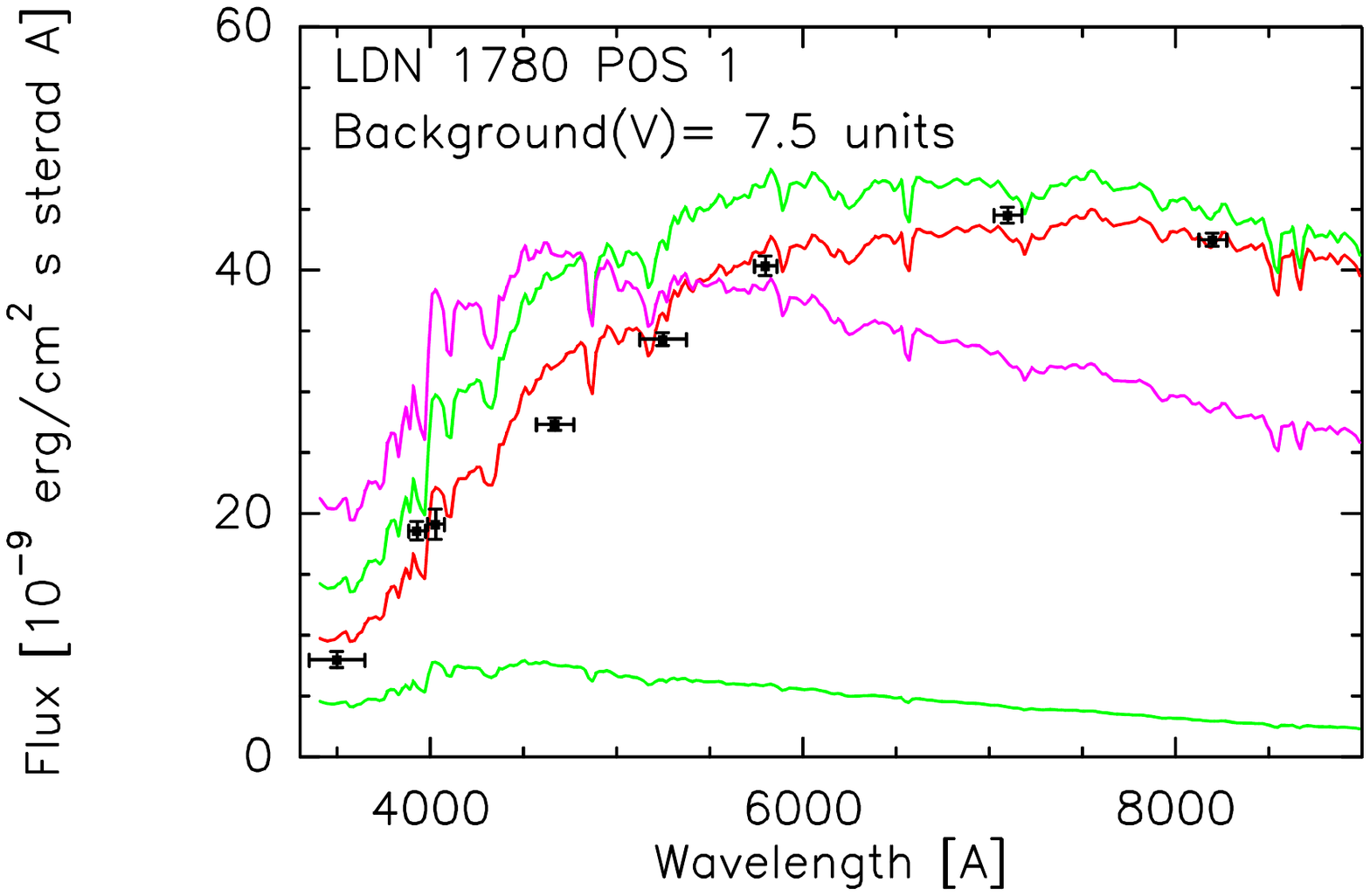}}
\hspace{-5cm}
\vspace{-4cm}
           {\includegraphics[width=10cm,angle=-90]{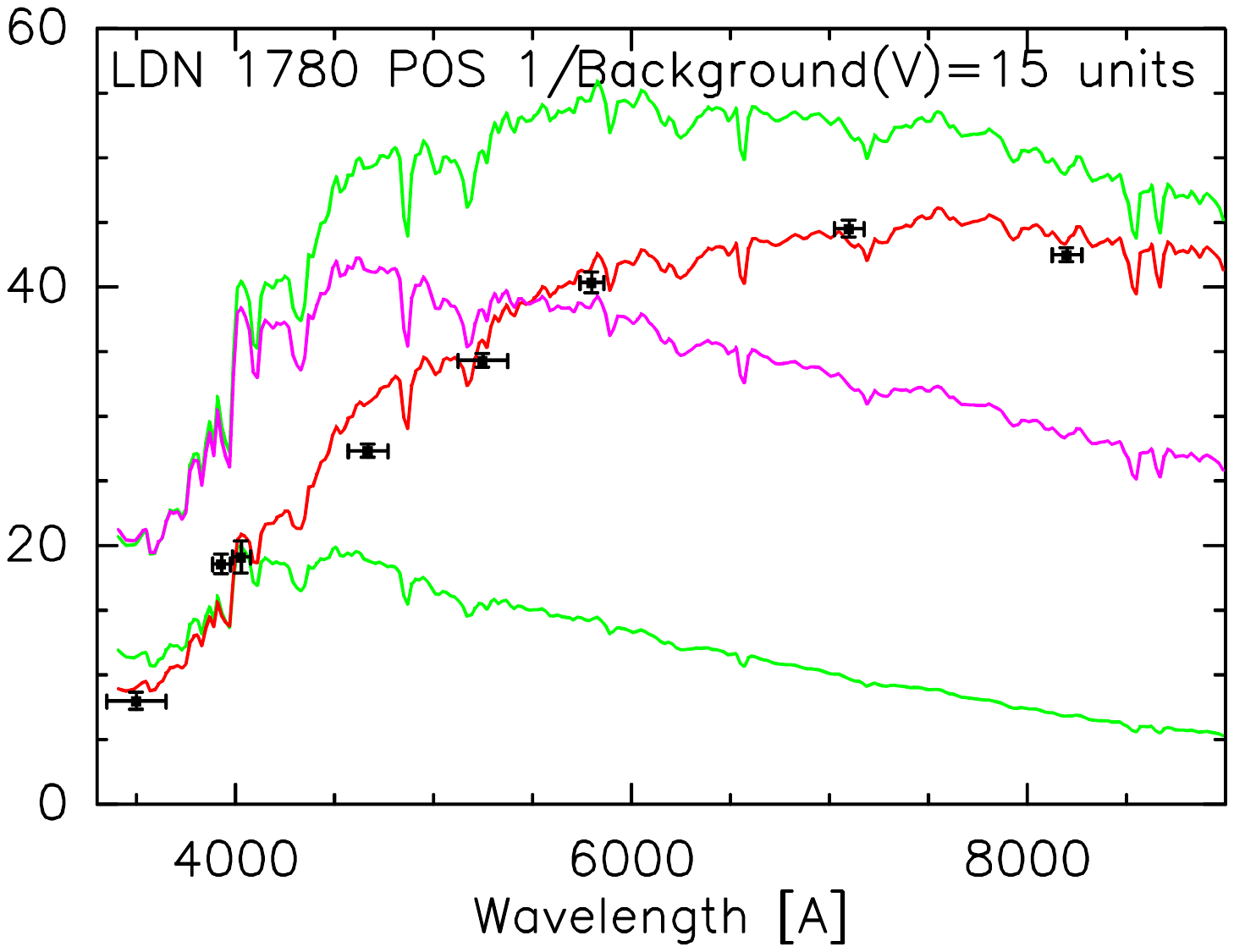}}
\hspace{-5cm}
\vspace{-3cm}
           {\includegraphics[width=10cm,angle=-90]{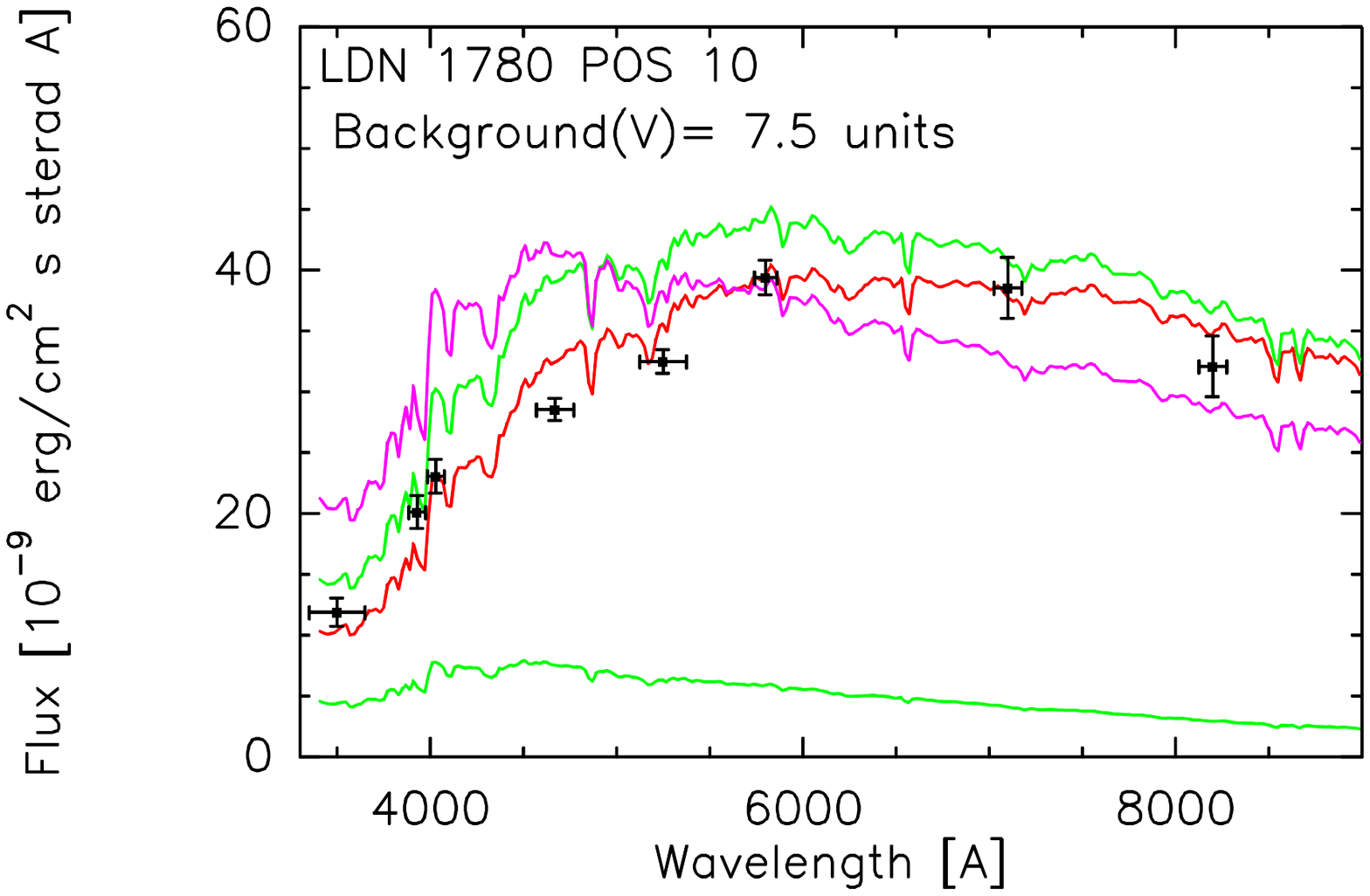}}
\hspace{-5cm}
\vspace{-0cm}
           {\includegraphics[width=10cm,angle=-90]{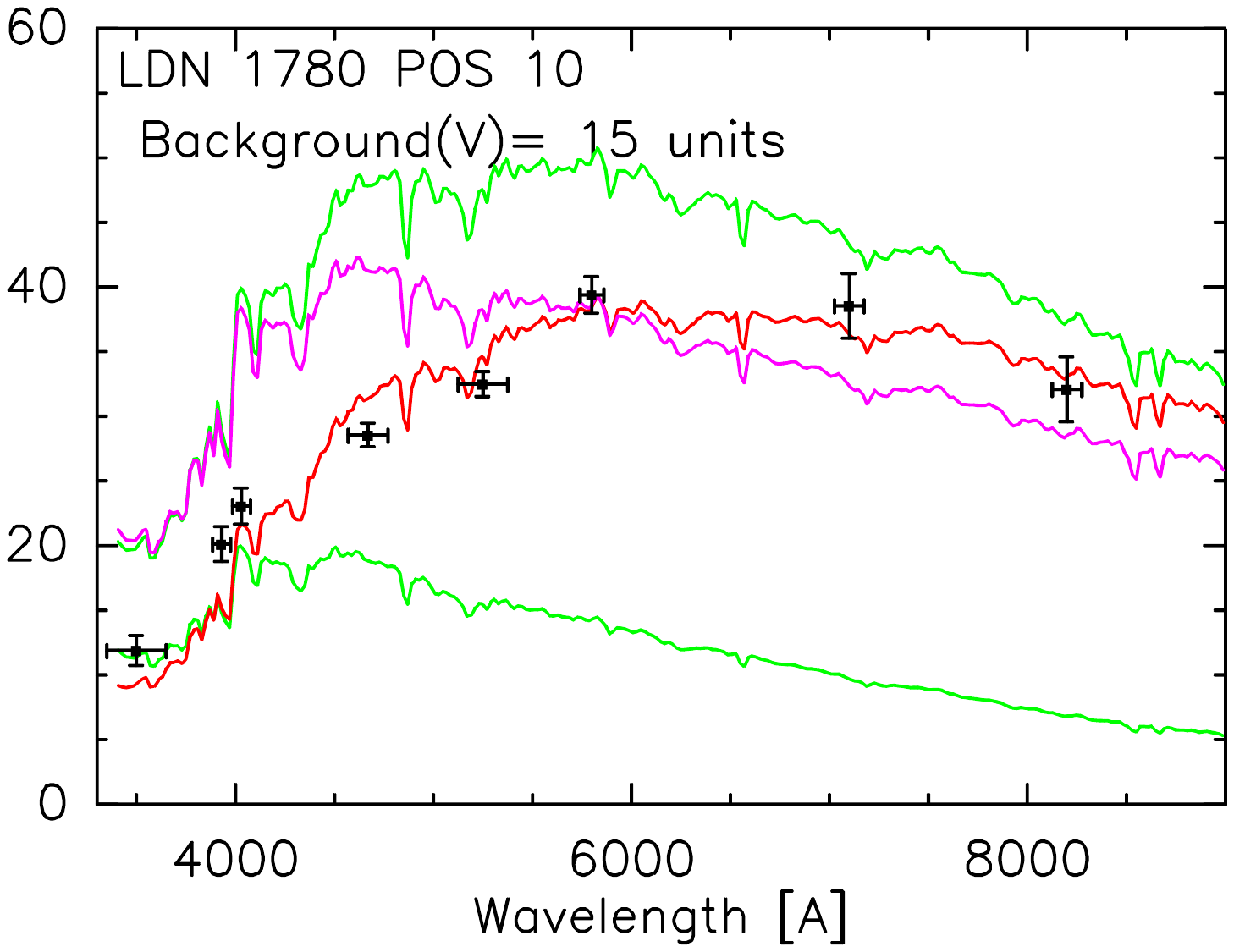}}

\vspace{1cm}
           {\includegraphics[width=10cm,angle=0]{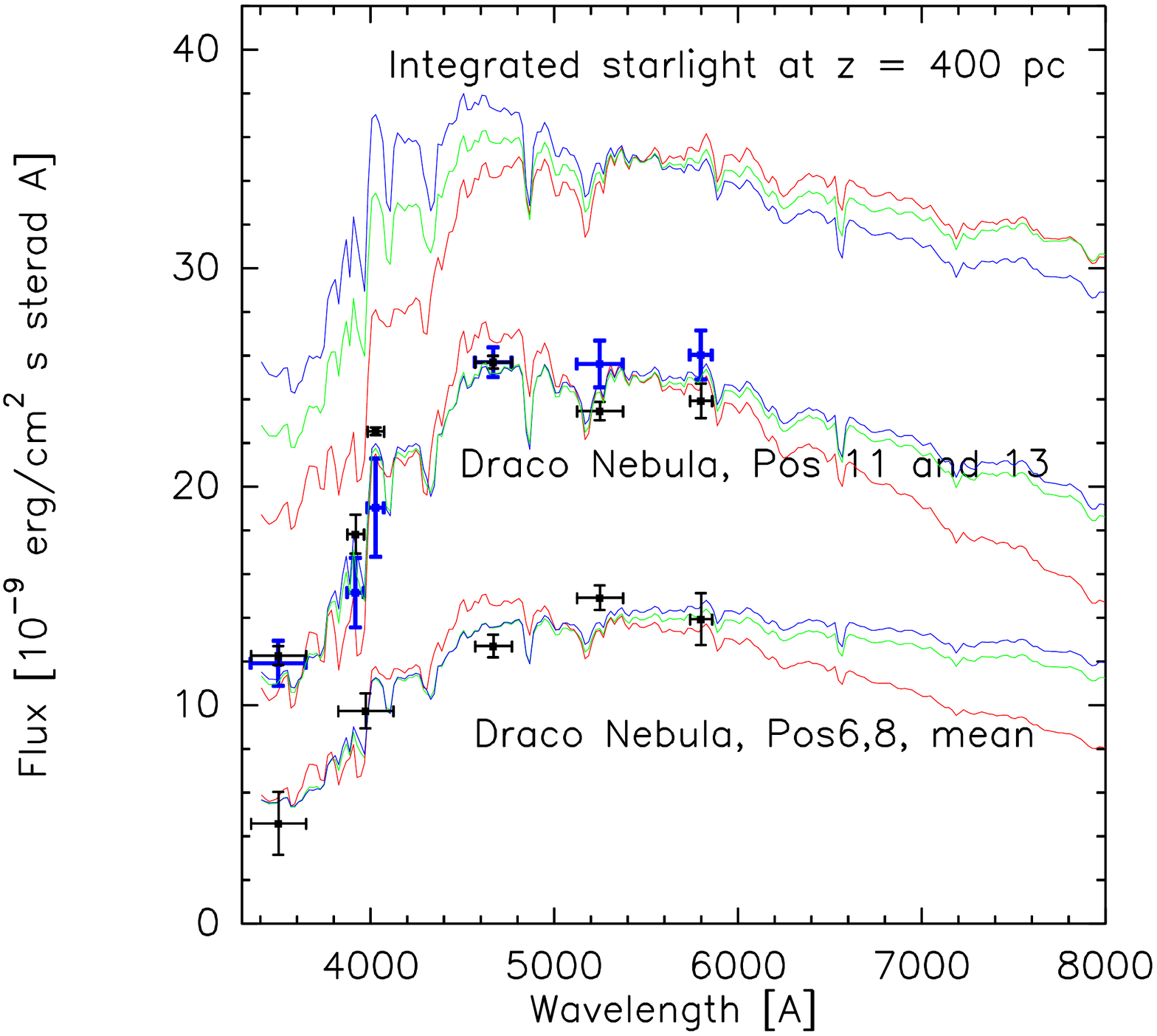}}
%\hspace{0.5cm}
\vspace{-1cm}
 \caption{Spectral energy distributions for Pos 1 ({\bf upper left} and {\bf upper right}) 
and Pos 10 ({\bf middle left} and {\bf middle right}) in LDN~1780, 
and Pos. 11 and 13 and mean of Pos 6 and 8 in Draco nebula ({\bf bottom left}). 
Observations are as in Fig.~\ref{FigSEDs}, model curves are according to starlight models (see text).
}
\label{SED+ISRF}
\end{figure*}

{ The bottom left panel of Fig.~\ref{SED+ISRF} shows the model fitting for the Draco nebula. 
In the upper part, the observed SEDs for the two highest-extinction positions, Pos11 ($A_V = 1\fm27$) and Pos13 
($A_V = 1\fm07 $) are shown together with ISL spectrum models for $z = 400$\,pc; 
in the lower part, the observed SEDs and the  model spectra for the mean of Pos6 and Pos8 are shown. 
In both cases the red lines show models and model fits with ISL at $b = 36\degr$ as impinging radiation.
In order to see how much the scattered light with larger scattering angles, 
$\theta \ga 30\degr$, influences the result we show also the modelling with ISL for $b = -36\degr$ 
(blue lines),
representative of the southern hemisphere high-latitude sky, 
and for mean ISL over the sky (green lines). For Pos11/13 the best-fitting  $\tau_V^{\rm eff}$ values
for $b = 36\degr$, -36\degr and mean-over-sky ISL were 0, 1.3 and 0.6, respectively. 
The corresponding $\tau_V^{\rm eff}$ values for Pos6/8 were 0, 1.55 and 0.8, respectively.
We note that the value of $I_{\lambda}^{\rm sca}$(bg) has been assumed to be negligible (see Sect. 3.3.3).}

Good fits are obtained for both the southern-sky ($b = -36\degr$, blue curve) 
and mean-over-the-sky (green curve) SED models;
however, the ISL spectrum for $b = +36\degr$ is so red that no optimal fit was possible with $\tau_V^{\rm eff} > 0$; 
the fits shown in  Fig.~\ref{SED+ISRF} for the Draco nebula are for $\tau_V^{\rm eff}$ = 0. 
This means that a very strongly forward-peaked scattering function with $g \approx 1$ is not favoured. 

For the location of Draco at $z = 400$ pc there are in the ISL spectrum for the
northern and southern high-latitude sky  distinct spectral differences, 
such as the strength of the g band, the Mg 5170 \AA\, and the Balmer lines, and the size of the 4000 \AA\, 
jump $D4000$. 
In principle, these features could be used to tell where the predominant illumination of Draco nebula is coming from. 
However, the coarse wavelength resolution of the present observations
does not allow such a discrimination except, perhaps, in the case of $D4000$. 
A dominant contribution to Draco's scattered light 
from the northern high-latitude sky would mean, however, that the dust albedo had to be $\ga0.9$ in the $u$ to $B$ bands.
This is not supported by the observed intensity of the scattered light, see Section 4.1 and Fig.~\ref{albedo_vs_g}.    

We conclude that the illumination for Draco is dominated by starlight from the galactic disk, seen from Draco's vantage point at 
$b < 0$. In principle, the scattered light spectrum of a dark nebula could be used to determine its $z$-position and, 
thereby, also its distance from the Sun. The example with Draco demonstrates, however, that this is difficult in practice.

Historically, in the early modelling of optical cirrus clouds at high latitudes by \cite{sandage}  it was suggested 
that their illumination is coming from 'below', that is from the galactic disk. As has been demonstrated in this 
paragraph it is indeed the case for a high-altitude cloud like the Draco nebula at $z = 400$\,pc. However, 
 most of the high-latitude clouds are, like LDN~1642 and LDN~1780, at lower altitudes,  $|z| \la 70$\,pc, and, because of 
the strongly forward-directed scattering function of the grains, their dominant illumination
in optical comes from the high-latitude sky behind them; see also \citet{witt08} for further discussion.

\subsection{H$_2$ fluorescence emission in LDN~1780}
 
\begin{figure*}
\vspace{-0cm}
\hspace{-1.0cm}
            {\includegraphics[width=8.5cm,angle=-0]{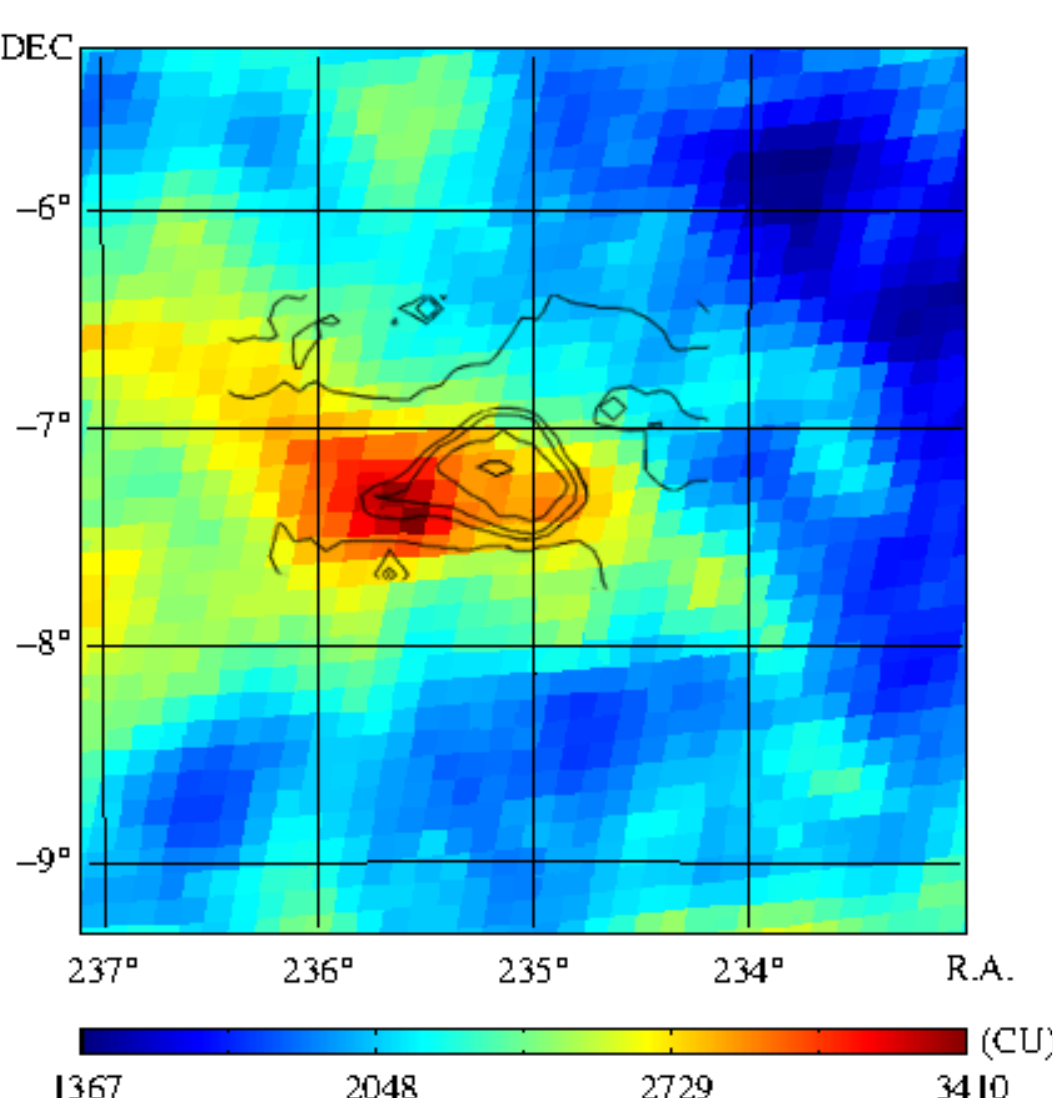}}
\hspace{-0cm} 
\vspace{-7cm}
            {\includegraphics[width=8.5cm,angle=-0]{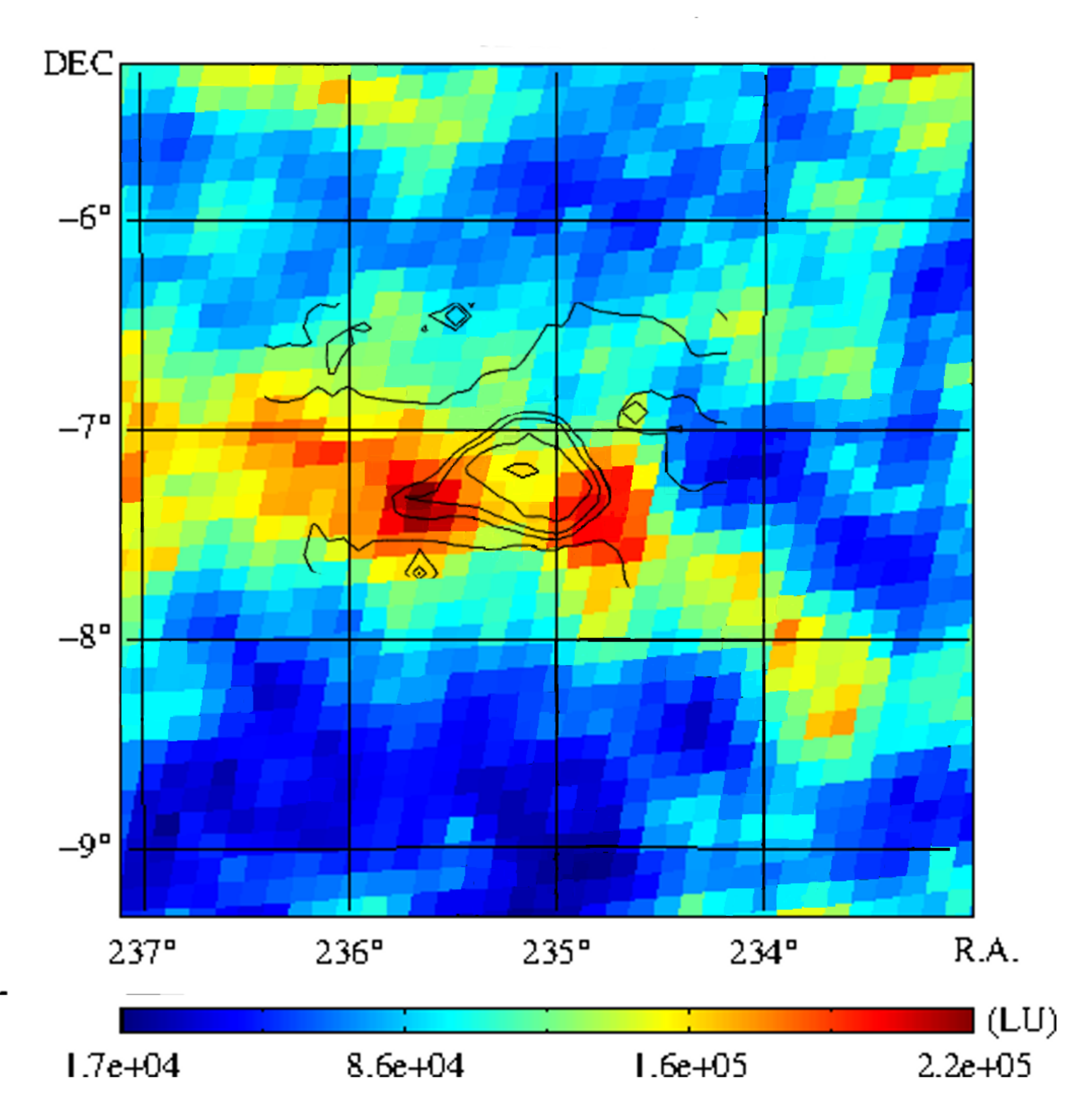}}
\hspace{1cm}
\vspace{-0cm}
            {\includegraphics[width=9.5cm,angle=-0]{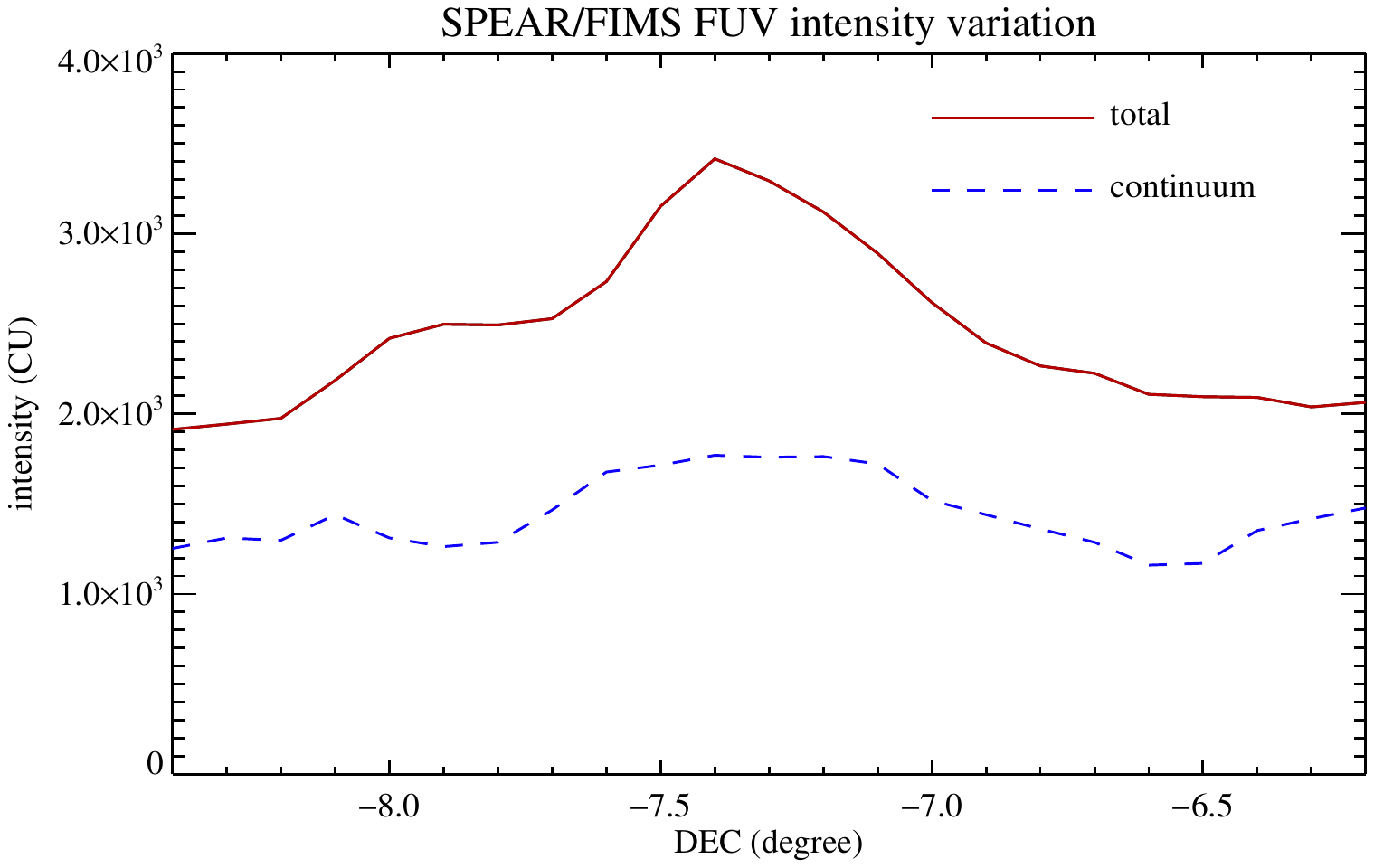}}
\vspace{0cm}
\hspace{1cm}
            {\includegraphics[width=9.5cm,angle=-0]{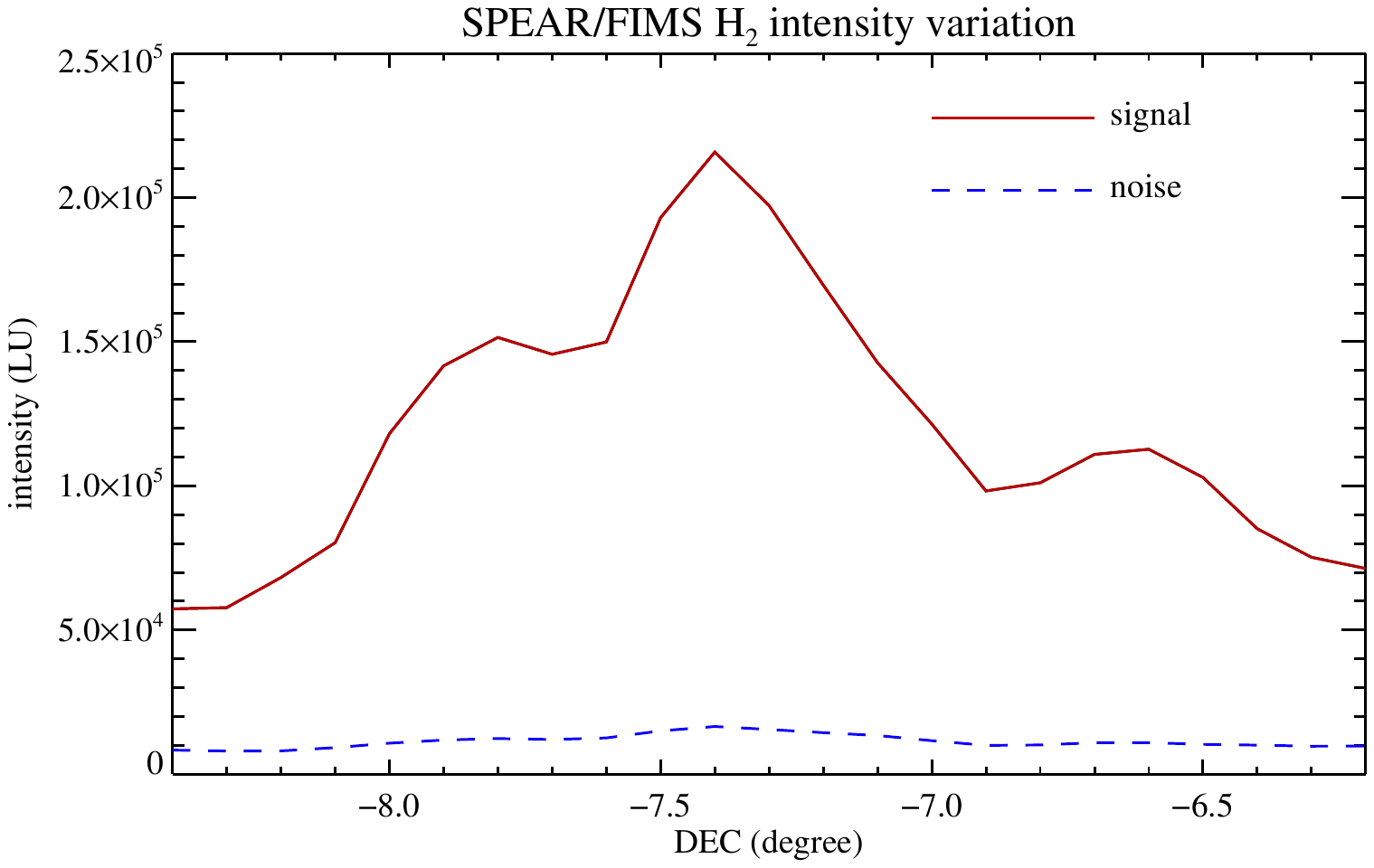}}
\vspace{0cm}
\caption{FUV continuum and H$_2$ fluorescence emission in LDN~1780; 
{\bf upper left:} SPEAR/FIMS FUV continuum map with $B$ band surface brightnes contours overlayed;
{\bf upper right:} the SPEAR/FIMS H$_2$ fluorescence emission map + $B$ band contours; 
{\bf lower left:} total (red line) and continuum (blue dashed line) intensities 
shown as function of the declination along the path from south to north; 
{\bf lower right:} H$_2$ intensity (red line) and its statistical error (blue dashed line) along the 
same path across LDN~1780.
The intensity unit for the total and continuum is  photons cm$^{-2}$s$^{-1}$sr$^{-1}$\AA\,$^{-1}$ (CU; continuum unit) 
and for the line emission photons cm$^{-2}$s$^{-1}$sr$^{-1}$  (LU; line unit).}
\hspace{-0cm}
\label{H2_comb}
\end{figure*}

As has been seen above (Section 4.2 and Figs.~\ref{FUV_vs_NUV} and ~\ref{L1780_Galex_models}) there is in LDN~1780 
an excess of FUV surface brightness that can not be explained in terms of scattered light.  In Section 2.5 and 
Fig.~\ref{L1780FUV_NUV_smooth} we have already presented SPEAR/FIMS data which demonstrate
the presence of  H$_2$ fluorescence emission in the LDN 1780 FUV spectrum $\lambda$ = 1450 -- 1665 \AA\,. 
To further analyse this, we present in  Fig.~\ref{H2_comb} a FUV continuum map (upper left hand panel) and 
an  H$_2$ fluorescence emission map (lower left), both extracted from  the SPEAR/FIMS data archive,
processed to 0\fdg2 resolution. The H$_2$ map has its peak intensity in the SE part of the cloud. 
The upper right hand panel of Fig.~\ref{H2_comb} shows the total and the continuum intensities plotted as function 
of declination from south to north. In order to increase the signal-to-noise ratio, pixels with 
the same declination were merged over the right ascension range from 234\fdg8 to 235\fdg8, and the mean values 
were then convolved in N-S direction with FWHM=0\fdg2.

 The difference between the total and continuum intensities 
can be ascribed to  H$_2$ emission; atomic lines, such as Si~{\small II}* (1533Å) and C~{\small IV} (1550Å), 
are negligible in the LDN~1780 spectra. The lower right hand panel of Fig.~\ref{H2_comb} shows the H$_2$ intensity 
variation across LDN 1780. The peaks of the total FUV and H$_2$ intensities are found at the same 
declination of -7\fdg4 which is close to the southern edge of the cloud. The extinction peak ($A_V \sim 4$\,mag) is at 
Dec.$\approx -7\fdg15$ and the cloud edge ($A_V \sim 0.5$\,mag) at Dec.$\approx -7\fdg45$. 
According to Fig.~\ref{H2_comb} the H$_2$ fluorescence emission is equally or even more strongly 
concentrated to the southern edge of LDN 1780 than the total FUV emission. 

A substantial fraction of the total FUV intensity in LDN 1780 is due to the H$_2$ fluorescent emission. 
At the intensity peak the H$_2$ contribution to the total intensity in the FUV band is
$\sim$ 160\,10$^3$ photons cm$^{-2}$s$^{-1}$sr$^{-1}$ (LU, line units) or $\sim$12.9\,\cgs\,, which is 
$\sim$2/3$^{rds}$ of the total intensity of $\sim$19.4\,\cgs\,. 
This is in good agreement with our previous evidence for a high level of H$_2$ contribution to the total FUV 
intensity (see Figs.~\ref{FUV_vs_NUV} and \ref{L1780_Galex_models}). The total SPEAR/FIMS FUV intensity 
( = continuum + H$_2$ emission) is, however, substantially lower than that for {\em GALEX}. 
This may be to a major part caused by the difference in the spatial resolutions of these two instruments.

\subsubsection{H$_2$ fluorescence emission from the H\,{\small I} -- H$_2$ transition layer in LDN~1780}

When an interstellar cloud is exposed to ISRF the ground state H$_2$ molecules absorb FUV photons 
in the wavelength band  $\lambda = 912 - 1100$\,\AA\, and are exited to the electronic states 
B$^1\Sigma^+_u$ and C$^1\Pi^u$. Some 10\% of the excited  H$_2$ molecules dissociate to H atoms while the remaining 
90\% radiatively de-excite to vibrationally excited states of the ground electronic state X$^1\Sigma^+_g$.
These bound -- bound electronic transitions give rise to H$_2$ fluorescence emission in the Werner and Lyman 
bands (see for example \citealt{williams}).

 H$_2$ molecules are continuously being formed on and released from grain surfaces.
Extinction and absorption of photons by dust grains and  H$_2$ molecules screen the inner parts of the cloud
against UV radiation. Thus, a transition zone consisting of a mixture of H atoms and  H$_2$ molecules is formed and upheld
in the cloud surface layer. In LDN~1780 such a photodissociation region (PDR) is manifested by the presence 
of H$_2$ fluorescence emission and H\,{\small I} 21-cm excess emission in a zone at the southern edge of the cloud.
As has been described in Section 3.1.3 above, there is an enhanced, asymmetrical FUV radiation field
impinging on the cloud preferentially from the southern side.

H{\small I}-to-H$_2$ transition zones are an important ingredient in ISM physics. With LDN~1780 we have a 
good opportunity to study a simple, almost spherically symmetric cloud immersed in a well-known radiation field.
As can be seen from Fig.~\ref{L1780FUV_NUV_smooth} the distributions of the FUV and H$_2$ fluorescence 
emissions in LDN~1780 are qualitatively similar to that of the H\,{\small I} 21-cm excess emission. 
All these distributions show a bright zone along the southern edge of the nebula. Less distinctly, also
the NUV intensity distribution has the same north-south dichotomy. Therefore, part of the FUV asymmetry is
likely caused by the stronger illumination from the southern hemisphere, leading to stronger scattered light
on that side. A substantial part of the FUV emission is, however, due to H$_2$ fluorescence.

\citet{williams} first pointed out the possibility of detecting diffuse ultraviolet fluorescence emission 
from interstellar H$_2$ and showed that its intensity in the wavelength window $\lambda = 1400 - 1700$\,\AA\,
is comparable to that of the scattered light. Following their notification the equilibrium condition
for  H$_2$ formation/destruction is
\begin{equation}
\frac{d \,n({\rm H_2})}{dt} = 0 = kn({\rm H}) \times n({\rm H{\small I}}) - \beta n({\rm H_2}),
\end{equation}
where $n$(H{\small I}), $n$(H$_2$) and  $n$(H) = $n$(H{\small I}) + 2$n$(H$_2$) are the atomic, 
molecular and total hydrogen nucleii 
number densities,  $\beta$ is the H$_2$ destruction rate and $k$  the rate constant for  H$_2$ formation 
= 3\,10$^{-17}$cm$^{-3}$s$^{-1}$. 

In  equilibrium the  H$_2$ fluorescence volume emissivity is proportional to the number of H$_2$ destructions
\begin{equation}
J({\rm fluorescence}) = \beta n({\rm H_2})<h\nu> = kn({\rm H}) \times n({\rm H{\small I}})<h\nu>,
\end{equation}
where $<h\nu>$ is the average photon energy. 
This expression refers to the fluorescence continuum emission which results from H$_2$ dissociations, that is
transitions to the vibrational continuum ($v$'' $>$14) of the X$^{1} \Sigma^+_g$ electronic state. 
To take the transitions to the bound vibrational states ($v$'' $\le$14) approximately into account 
\citet{williams} multiplied this expression by 2. 

We assume that the total density $n({\rm H})$ can be considered to be constant in the transition layer and that the layer can
be considered to be optically thin. Then, integrating  $J({\rm fluorescence})$ along the line of sight through 
the nebula, one finds that the intensity of the fluorescence  emission is proportional to the H{\small I}
column density, $I({\rm fluorescence}) \propto  N({\rm H{\small I}})$. 
This result from the general analysis of \citet{williams} has been confirmed also by the detailed 
model calculations of \citet{neufeld} (see their Figs.~1 and 2).

\begin{figure}
\vspace{0.5cm}
\hspace{-0.5cm}
           {\includegraphics[width=9.5cm,angle=-0]{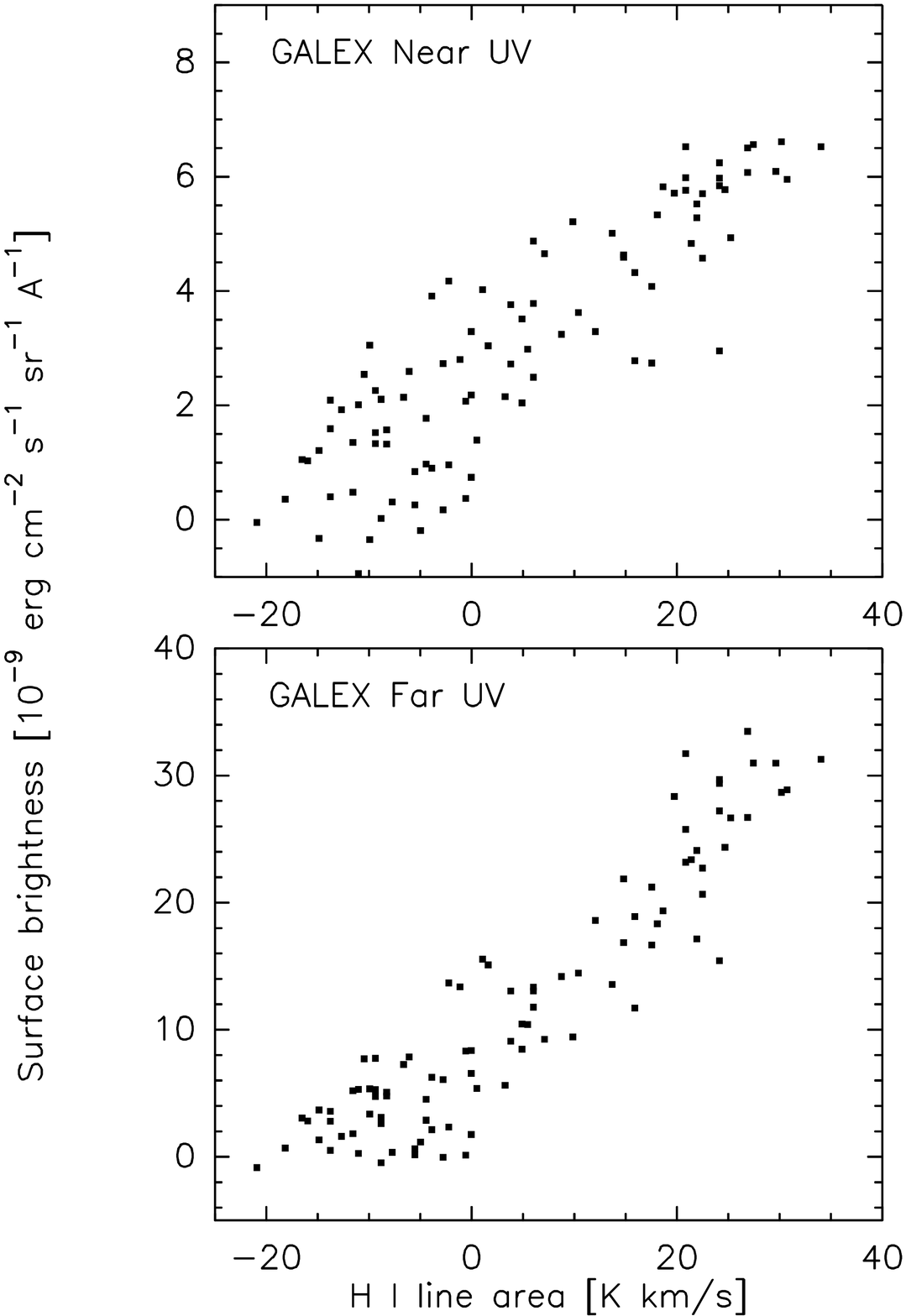}}
 \caption{{\bf (upper panel)} $GALEX$ NUV and {\bf (lower panel)} FUV intensity in LDN~1780 as a function 
of H{\small I} 21-cm excess line area, adopted from Fig.\,5 of \citet{mattila+sandell}.}
\label{GALEXvsHI}
\end{figure}

We show in the upper and lower panel of Fig.~\ref{GALEXvsHI} the $GALEX$ NUV and FUV intensities in LDN~1780 as  
function of H{\small I} 21-cm excess line areas, W(H{\small I}), adopted from Fig.~5 of \citet{mattila+sandell}.
The H{\small I} column density in LDN~1780 is directly proportional to  W(H{\small I}), see
\citet{mattila+sandell}. For each  W(H{\small I}) value, representing the signal within 9\arcmin\, FWHM beam, 
we have averaged the  $GALEX$ values within 6\arcmin\, from the centre. 
The remarkably good correlation of  $I({\rm FUV})$ vs. $W$(H{\small I}) especially at the larger FUV intensities, 
$I({\rm FUV}) \ga 10$ \cgs\,, where the fluorescence emission dominates, can be undestood as observational 
evidence of the linear relationship $I({\rm fluorescence}) \propto  N({\rm H{\small I}})$ 
resulting from the analysis by \citet{williams}.

\subsubsection{Modelling of the H$_2$ fluorescence spectrum}

We have modelled the H$_2$ fluorescence emission spectrum in the FUV  peak-intensity region using the CLOUD code
developed for photodissociation regions assuming plane-parallel geometry  (\citealt{vandishoeck}, 
\citealt{black}; see also \citealt{lee06,lee07,lee08,jo17}). The following three input parameters were adopted and 
kept fixed during the fitting: (1) cloud temperature $T=100$ K; 
(2) the sticking probability and formation efficiency of H$_2$ on dust grains; we adopted $y_{\rm F}$ = 1 as a typical 
value; (3) the strength of incident FUV radiation in the band $\lambda = 912 - 1100$\,\AA\,, $I_{\rm UV}$, in units of the 
average UV radiation field of \citet{draine78}; we adopted the value $I_{\rm UV} = $1.7. It is based, on one side, 
on the direct addition of the radiation of individual stars in the  $TD$-1 and Hipparcos star catalogues 
(\citealt{thompson78}; \citealt{perryman}; see \citealt{jo17} for details) and on the other side on the estimates 
as presented in Sections 3.1.2 and 3.1.3 above. A further parameter which was allowed to vary during the fitting 
process was (4) the total hydrogen density; its initial value was $n$(H) = 1000 cm$^{-3}$ based on 
$N$(H)$ \approx 1\,10^{21}$ cm$^{-2}$, (corresponding to $A_V \approx 0\fm5$) and an estimated line-of-sight 
cloud extent of 0.4 pc at the FUV peak position. 
We generated models with a range of log\,$N$(H$_2$) values from 19.0 to 23.0 with 0.1 steps.
 
\begin{figure*}
\vspace{-10cm}
     {\includegraphics[width=12cm,angle=-0]{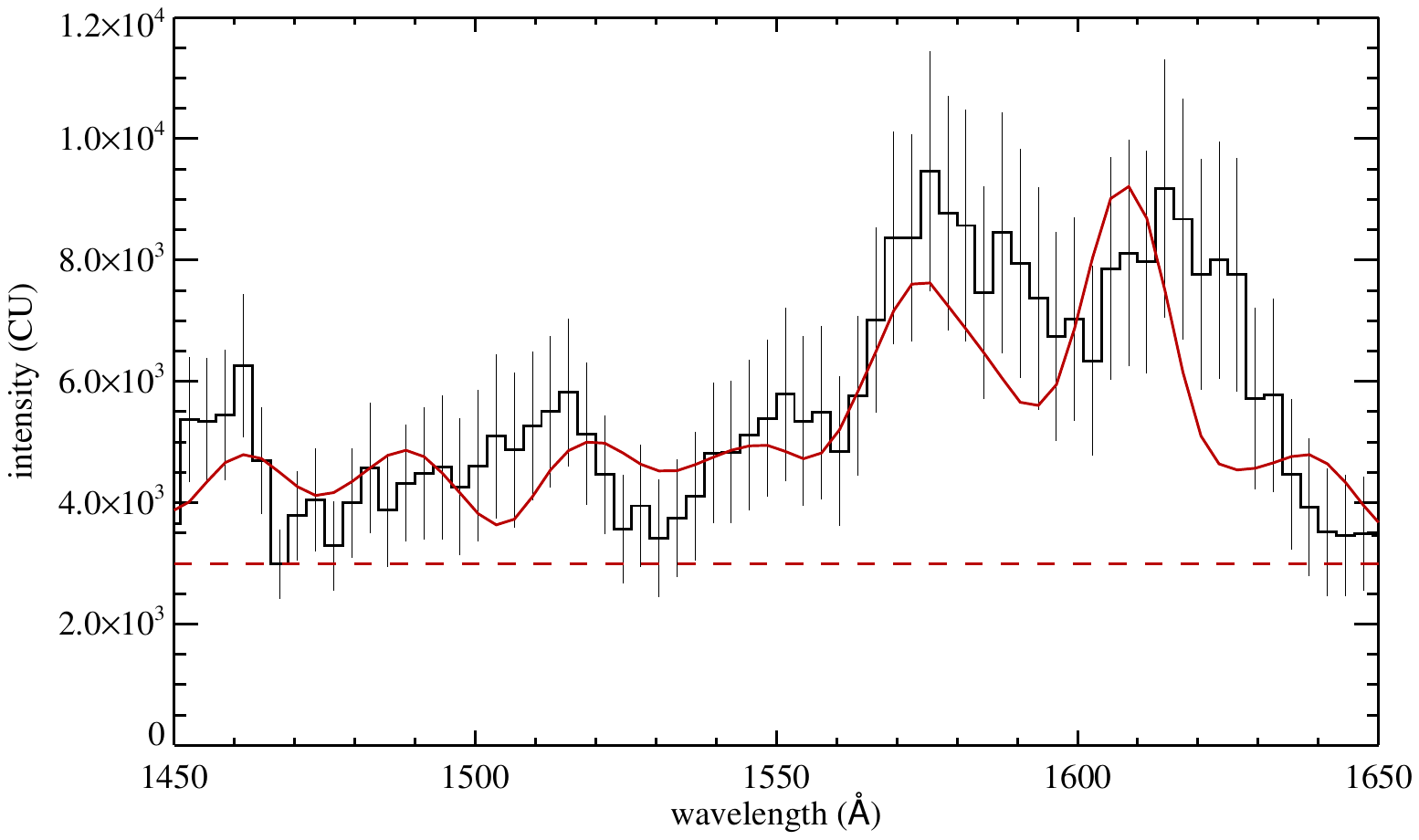}}
\hspace{-3cm}
   {\includegraphics[width=12cm,angle=-0]{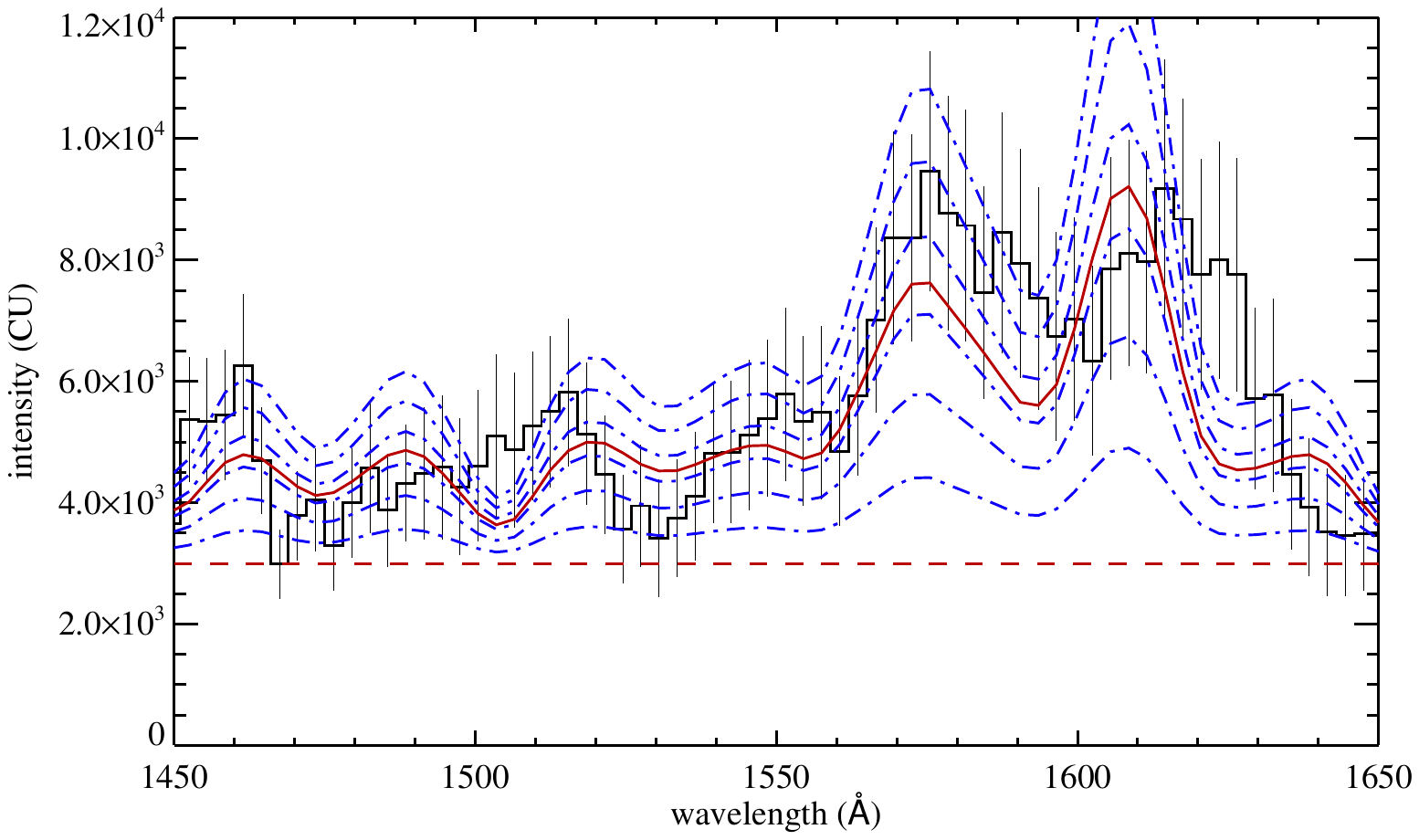}}
\hspace{-3cm}
\caption{{\bf (left)} Extinction-corrected spectrum at the FUV peak region (black histogram) and the best-fit model 
spectrum for $I_{\rm UV}$ = 1.7 (red line). Red dashed line indicates continuum level; 
{\bf (right)} as in the left-hand panel but with models for $I_{\rm UV}$ = 0.5, 1.0, 1.5, 1.7 (red line), 2.0, 
2.5 and 3.0. } 
\label{H2_fit}
\end{figure*}

Before fitting the models we corrected the observed spectrum for the extinction in front of the cloud, 
$A_V = 0\fm215$ (see Table \ref{table:6}); this corresponds to an optical depth of $\tau_{\lambda} = 0.5$ at 
$\lambda = 1550 \AA\,$.  We show in the left panel of Fig.~\ref{H2_fit} the extinction-corrected spectrum 
(black histogram) 
at the FUV peak position (Dec. = -7\fdg4, R.A = 234\fdg8 - 235\fdg8) with the best-fit model spectrum superposed 
(red solid line). The best-fit model corresponds to the value log\,$N$(H$_2$) = 20.89 with reduced $\chi^2$ = 0.976  
and 1-$\sigma$ confidence range of 19.9 - 22.1, or $N$(H$_2$) = 7.8\,$10^{20}$cm$^{-2}$
with  1-$\sigma$ confidence range of (0.8 - 130)\,$10^{20}$cm$^{-2}$.
The 1-$\sigma$ confidence range is wide because of the low signal-to-noise ratio of the observed spectrum and,
as seen from the left panel of Fig.~\ref{H2_fit}, there are substantial deviations between the fitted and 
the observed spectrum.
The total hydrogen column density corresponding to the best-fit $N$(H$_2$) value is $N$(H) = 
2$N$(H$_2$)+ $N$(H\,{\small I}) = 16.2\,$10^{20}$ cm$^{-2}$, the corresponding volume density is  $n$(H) 
= 1.4\,$10^{3}$\,cm$^{-3}$. The atomic hydrogen column density is $N$(H\,{\small I}) = 0.53\,$10^{20}$ cm$^{-2}$
which corresponds to an H\,{\small I} fractional abundance of  $N$(H\,{\small I})/2$N$(H$_2$) = 0.033.

The best-fit  $N$(H$_2$) of the model agrees well with the observational
value $N$(H$_2$)=  $3^{+3}_{-1.5}$\,$10^{20}$cm$^{-2}$ as estimated from the $W(^{12}$CO) map of \citet{toth} 
using the factor $X = N$(H$_2$)/$W(^{12}$CO) = $2\,10^{20}$cm$^{-2}$(K\,km\,s$^{-1}$ )$^{-1}$ \citep{pineda08}.
The derived total volume density of  $n$(H) = 1.4\,$10^{3}$\,cm$^{-3}$ agrees well with its initially adopted value 
of  1\,$10^{3}$\,cm$^{-3}$ which was based on the line-of-sight extinction.

The strength of the incident UV radiation, $I_{\rm UV}$, is a central parameter determining the intensity of  
the H$_2$ fluorescence emission $I$(H$_2$). Observationally, $I$(H$_2$) is seen to decrease towards the cloud 
centre in accord 
with the FUV continuum intensity (see Fig.~\ref{H2_comb}). We show in the right-hand panel of Fig.~\ref{H2_fit} 
the calculated  H$_2$ spectra for
the values  $I_{\rm UV}$ = 0.5, 1.0, 1.5, 1.7, 2.0, 2.5 and 3.0. All the other parameters were the same as in the best-fit model
with $I_{\rm UV}$ = 1.7 described above.% and shown in  Fig.~\ref{H2_fit}\,$a$. 
The fluorescence emission is seen to be roughly proportional to $I_{\rm UV}$. Qualitatively, the observed decrease
of  $I$(H$_2$) from edge towards the centre of LDN~1780 can be understood as a consequence of decreasing  $I_{\rm UV}$.
A quantitative analysis of the attenuation of the radiation field would, however, require a full 3-dimensional modelling 
of the radiative transfer including dust extinction and H$_2$ self-shielding.

\subsubsection{H$_2$ fluorescence emission near LDN~1642 and Draco nebula}

 { The Taurus Molecular Cloud (TMC)} was the first dark nebula where H$_2$ fluorescence emission was detected \citep{martin,hurwitz}. 
\citet{lee06} has analysed a dense core in Taurus as function of optical depth through the core. Their 
result was similar with our case with LDN~1780: a strong H$_2$ fluorescence emission signal was detected in the halo
of the cloud with $A_V \la 1\fm5$\,mag, but only a weak one or none in the core. Also the continuum FUV intensity showed 
a decreasing level as function of $A_V $, again similarly with our results for LDN~1780 (see Fig.~\ref{GALEXvsAV}).

FUV continuum and H$_2$ fluorescence emission maps with SPEAR/FIMS of the Draco nebula area have been presented by \citet{park09}. 
They show a small ($\sim 0\fdg5$) H$_2$ fluorescence emission spot, centred at $(l, b)$ = (89\fdg5, 38\fdg3), displaced
by $\sim$0\fdg2  from Draco's ``Head'' towards the Galactic plane and partly overlapping with it. Its contribution is, 
however, not seen in the $GALEX$ total FUV signal (Fig.~\ref{Draco_GALEX}), nor is there an enhancement in the  
SPEAR/FIMS FUV continuum map.

The Orion-Eridanus superbubble region, including LDN~1642, has been studied by \citet{ryu} and \citet{jo11} 
using SPEAR/FIMS, both with regard to the dust-scattered continuum and the gaseous spectral features. 
The  H$_2$ fluorescent map of \citet{jo11} shows an enhanced intensity region $\sim$5 deg in size, centred at 
R.A.$\approx$ 70\,deg, Dec.$\approx$-13\,deg (J2000), extending from LDN~1642 
towards the Galactic plane.  Interestingly, there is a large ($\ga 5\degr$) plume-like tail associated with LDN~1642 
which is seen in the $IRAS$ and $Planck$ dust maps \citep{oliveira,malinen} as well as in the H{\small I} 21-cm emission 
 \citep{liljestrom}; it coincides relatively well with the H$_2$ emission area both in size and location. 
The situation is similar as in the Draco nebula, especially if one considers that the angular size difference is
partly caused by the different distances, $\sim$800pc for Draco vs. 124 pc for LDN~1642. 
The morphology of  H$_2$ emission in LDN~1642 and Draco differs, however, from that in LDN~1780: separate, possibly 
physically associated batches in LDN~1642 and Draco vs. a surface layer in  LDN~1780.

\section{Conclusions}

We have presented optical surface photometry of selected positions in three translucent high-latitude dark nebulae, 
LDN~1780, LBN~406 (Draco nebula) and LDN~1642. Intermediate-band photometry of LDN~1780 and LBN~406 with
the 2.2-m telescope at Calar Alto and of LDN~1642 with the 1-m telescope at La Silla covered the wavelength 
range 3500 - 5500/5800\,\AA\, and for LDN~1780 up to 8200\,\AA\,. Wide-field imaging of LDN~1780 in the
$B, V, R, i$ bands and of LDN~1642 in the $i$ band were carried out using University 
of Bochum 15-cm VYSOS6 refractor near Cerro Armazones, Chile. Archival data of the spaceborne instruments $GALEX$ and 
SPEAR/FIMS were used to study the FUV (1350 -- 1750\,\AA\,) and NUV (1750 -- 2850\,\AA\,) continuum surface 
brightness of LDN~1780 and LBN~406 and the H$_2$ fluorescence emission in LDN~1780 at $\lambda 1450 - 1650$\,\AA\,. 

For the purpose of modelling the different surface brightness components we have also determined the  visual extinction
distributions in the three nebulae. Besides the direct method utilising the NIR colour-excesses of background stars 
we have made use of the  $ISO$ and $Herschel$ far-IR emission data which have been scaled to visual extinctions. 
The scattered light component of the nebulae was modelled using Monte Carlo radiative transfer simulation, combined with
empirical estimates of impinging radiation field from the Milky Way. For the  H$_2$ fluorescence emission
modelling we used the CLOUD code developed for plane parallel PDR regions \citep{vandishoeck,black}.
The main conclusions of our study are as follows:

\begin{enumerate}
\item 
The dust albedo in the optical rises from $\sim$0.58 at 3500\,\AA\, to  $\sim$0.72 at 7500\,\AA\,. Our
observations support a strongly forward directed scattering function with an asymmetry parameter
$g = \langle{\rm cos} \theta\rangle \approx 0.8$. In the ultraviolet a smaller value of the asymmetry parameter is favoured,
 $g \approx 0.5$, and the resulting FUV and NUV albedo values are in the range $a \approx 0.4 - 0.5$.  

\item
Our albedo values are in good agreement with the predictions \citep{draine03} of the dust model of 
\citet{weingartner} as well as with the THEMIS CMM model \citep{jones,ysard} consisting of evolved core-mantle grains.
The unevolved THEMIS CM grains, not unexpectedly, do not agree with our albedo values.  In the ultraviolet, the 
$g$-values of the grain models WD and THEMIS CMM agree with our preferred value of  $g\approx 0.5$. 
However, all these models predict optical $g$-values that are substantially smaller than the value, 
$g \approx 0.8$, found in this paper and in other studies in the past.

\item
{ We do not agree with the previous suggestion  that attempted to explain the broad maximum at 
$\lambda \sim 6000 - 7000$\,\AA\, in the SED of LDN~1780 in terms of ERE \citep{chlewicki}. We find instead that 
background diffuse galactic light and dust optical depth effects offer a satisfactory explanation.}  

\item
SPEAR/FIMS spectra of LDN~1780 at 1400 -- 1700 \AA\, reveal the presence of H$_2$ fluorescence emission, 
peaking in the southern half of the nebula. This emission is revealed also as an excess component 
in the  FUV total intensity in $GALEX$ and SPEAR/FIMS data; the excess can not be explained in terms of scattered light. 
In an early study of H{\small I} 21-cm emission in LDN~1780 \citep{mattila+sandell} a morphologically similar excess emission
in the  H{\small I} distribution was found in the southern half of LDN~1780. A good correlation is found between the  
21-cm emission excess and the FUV intensity.  We interpret this as evidence for the H{\small I} -- H$_2$ transition zone which 
has been created thanks to the strong extra radiation from the Sco\,OB2 association and the O star $\zeta$\,Oph.

\item
Modelling of the H$_2$ fluorescence emission in LDN~1780 has been made assuming standard gas temperature (100 K) and 
H$_2$ formation coefficient values, and adopting for the impinging 912 -- 1100\,\AA\, radiation $1.7 \times$ 
its local mean ISRF value.
With these assumptions the fitting to the observed fluorescence spectrum resulted for the H$_2$ column density in
a range of log\,N(H$_2$) = 19.9 -- 22.1  and a total gas density of $n({\rm H}) = 1.4\,10^3$ cm$^{-3}$; 
these values are consistent with the estimates derived from CO observations and estimates of the cloud extent
and optical extinction along the line of sight.

\end{enumerate}

\begin{acknowledgements}

We gratefully acknowledge the contribution by the late Gerhard von Appen-Schnur (1940 -- 2013), friend and colleague, 
to the photometry at La Silla as referred to in Section 2.2. We are grateful to Jim Thommes 
for providing us and giving the permission to use his CCD image of LBN~406. 
 We acknowledge the constructive criticism by the referee A.N. Witt.
 Part of this work was supported by the Finnish  {Research Council 
for Natural Sciences and Technology} and the German {Deut\-sche For\-schungs\-ge\-mein\-schaft, DFG}.
PV acknowledges support from the National Research Foundation of South Africa.

\end{acknowledgements}
          
\bibliographystyle{aa} % style aa.bst
\bibliography{L1780_AA.bbl}

%\newpage

\begin{appendix}

\section{ Scattered light model fits in optical}
In this Appendix we present the model fits of the observed surface brightnesses (scattered light)
 for LDN~1780 in the optical wavelength bands $u, B$ and $R$.

\begin{figure*}
%   \resizebox{\hsize}{!}
\vspace{-1.5cm}
\hspace{-0.5cm} 
            {\includegraphics[width=10.5cm,angle=-0]{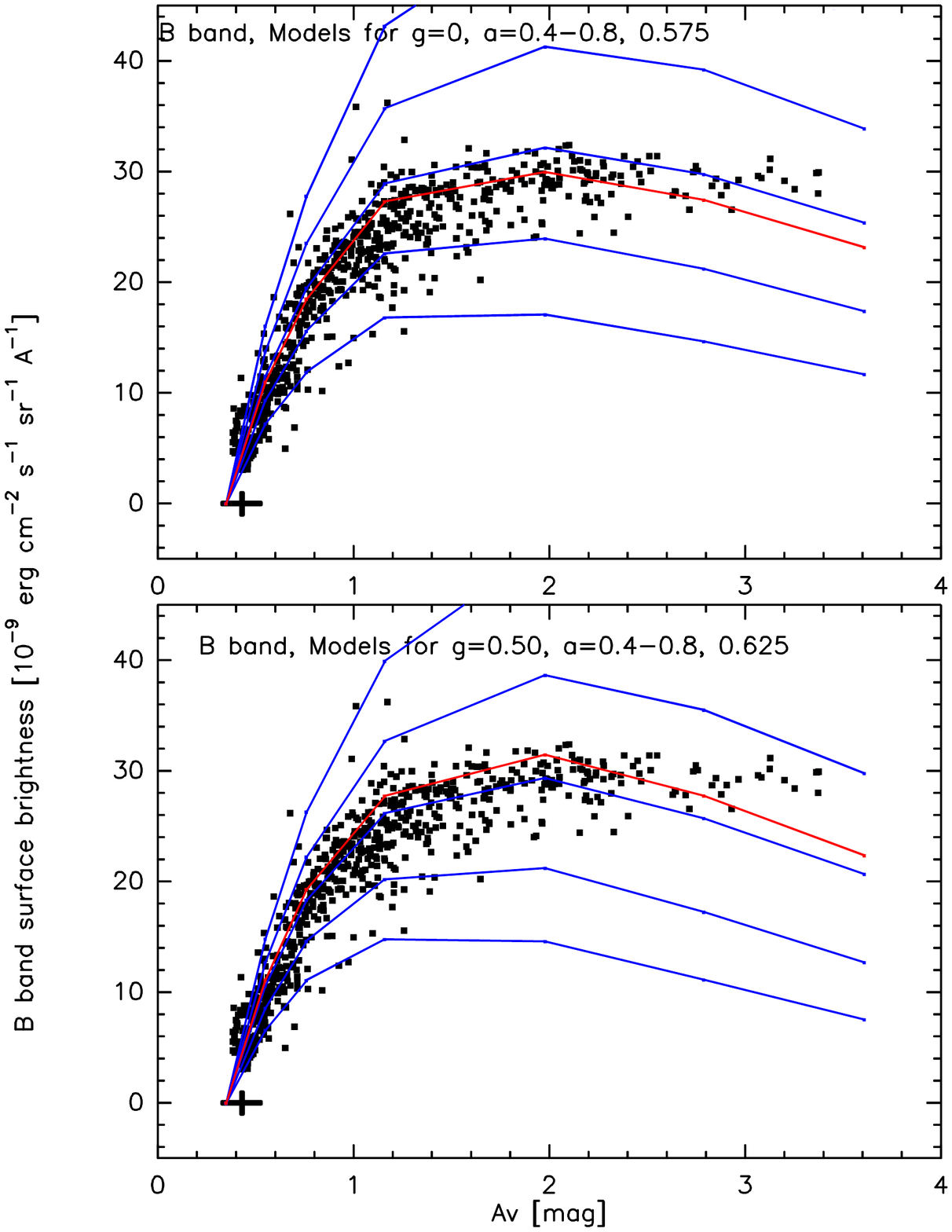}}
\vspace{-1.5cm}
\hspace{-1.5cm} 
%\hspace{0.0cm}
%\vspace{-1.0cm}
            {\includegraphics[width=10.5cm,angle=-0]{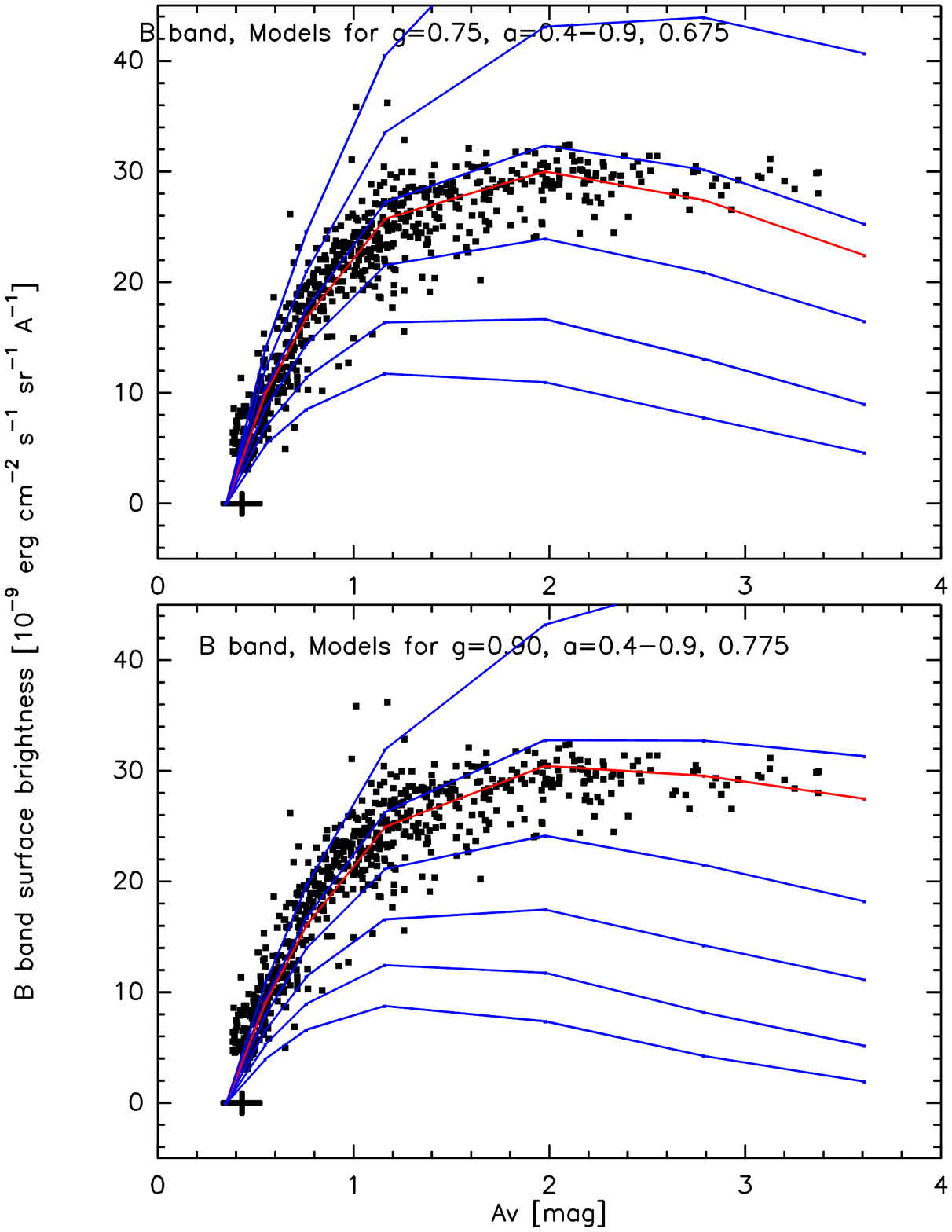}}
\hspace{-1.5cm} 
\vspace{-0.0cm}
%\vspace{-0.5cm}
            {\includegraphics[width=10.5cm,angle=-0]{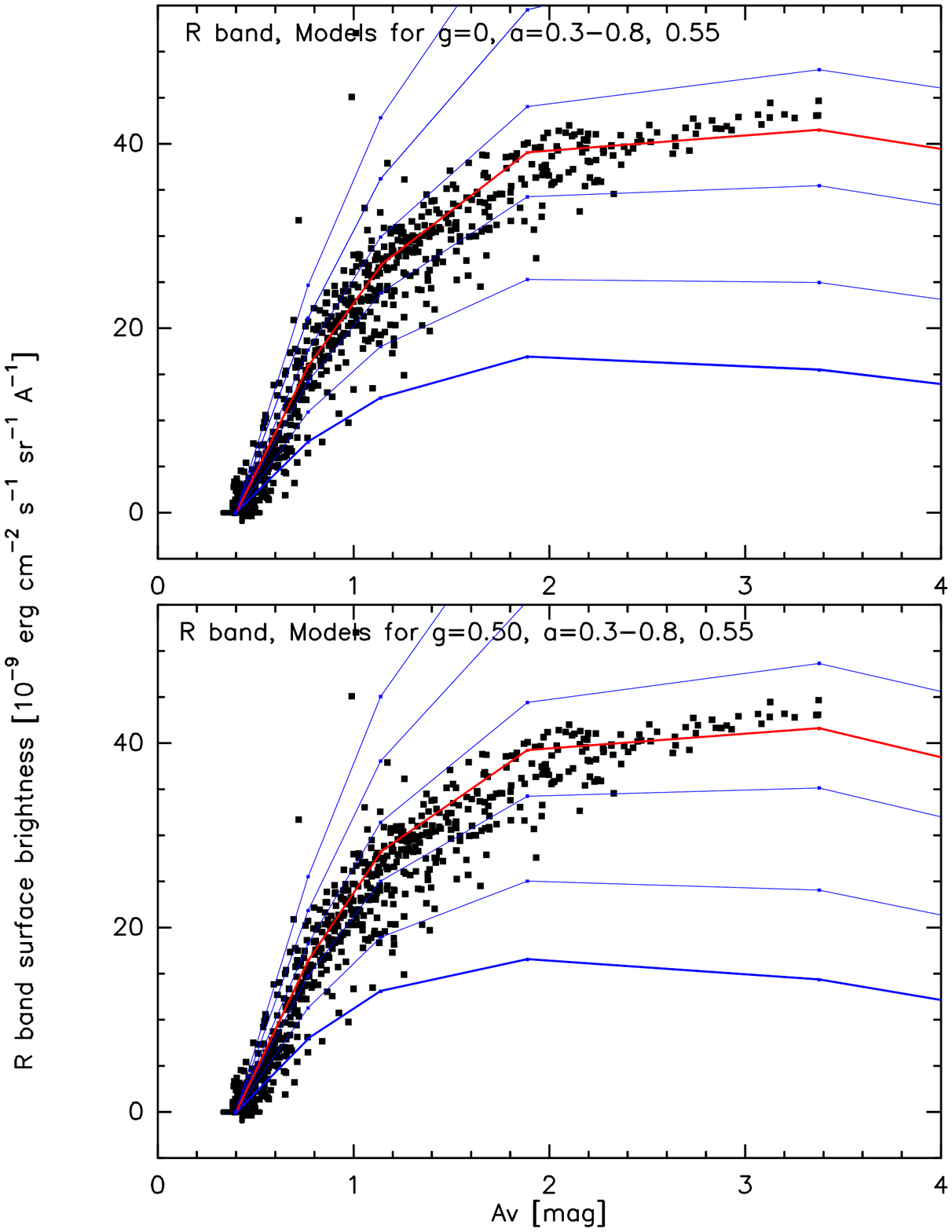}}
\vspace{-.0cm}
\hspace{-1.5cm}
            {\includegraphics[width=10.5cm,angle=-0]{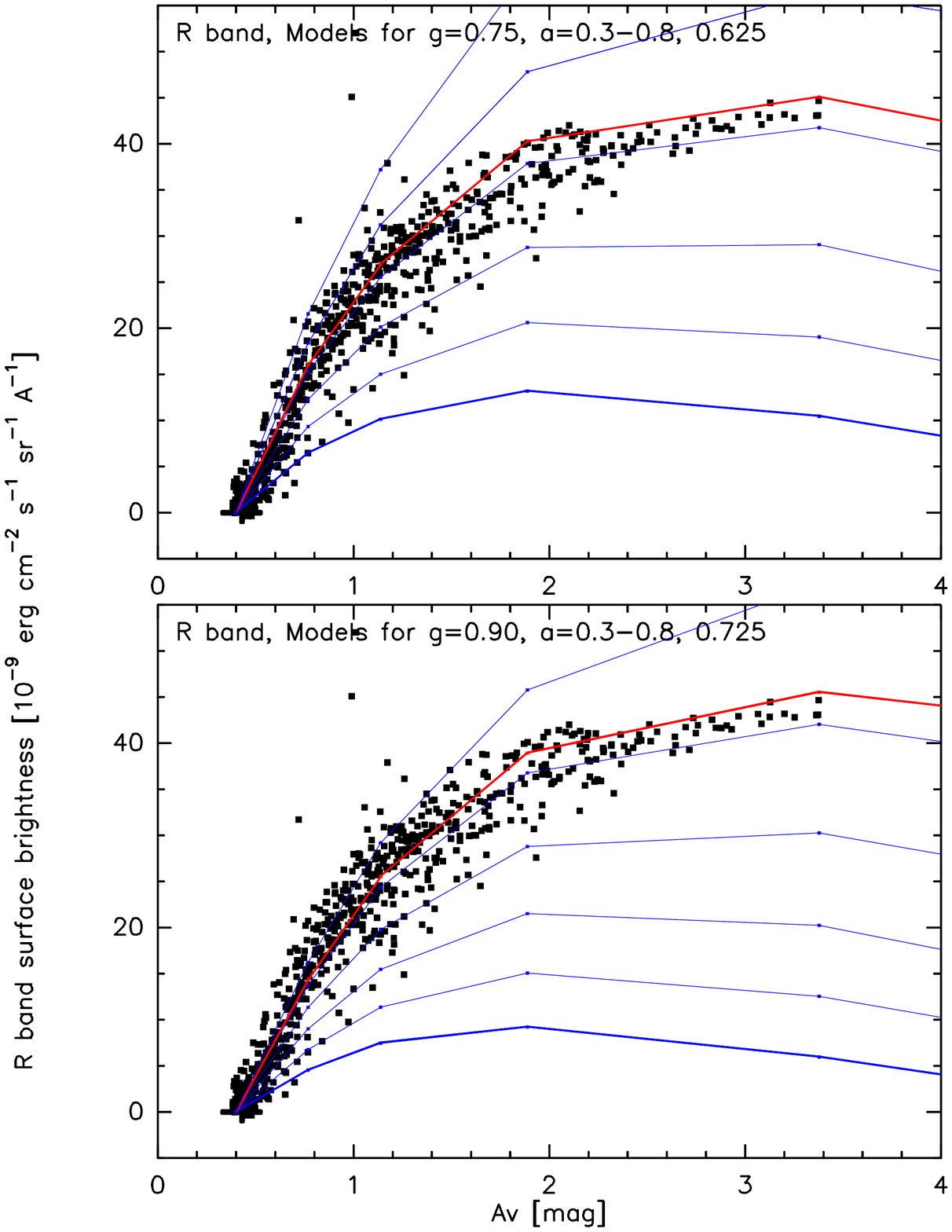}}

\vspace{-0.5cm}

\caption{Model fits of scattered light to the VYSOS $B$ and $R$ band data in LDN~1780. 
Models are shown for four different forward scattering parameters, $g$ = 0, 0.50, 0.75 and 0.90, 
and a range of albedos $a$ as indicated in the figures. The blue lines are for albedos in regular steps of 0.1
and the red lines for an intermediate $a$, given as the last one in the list of  $a$ values.
   The surface brightness is in units of \cgs\,.}
             \hspace{-0cm}
        \label{Isca_models}
   \end{figure*}

\begin{figure*}
\hspace{-1.0cm}
            {\includegraphics[width=10.5cm,angle=-0]{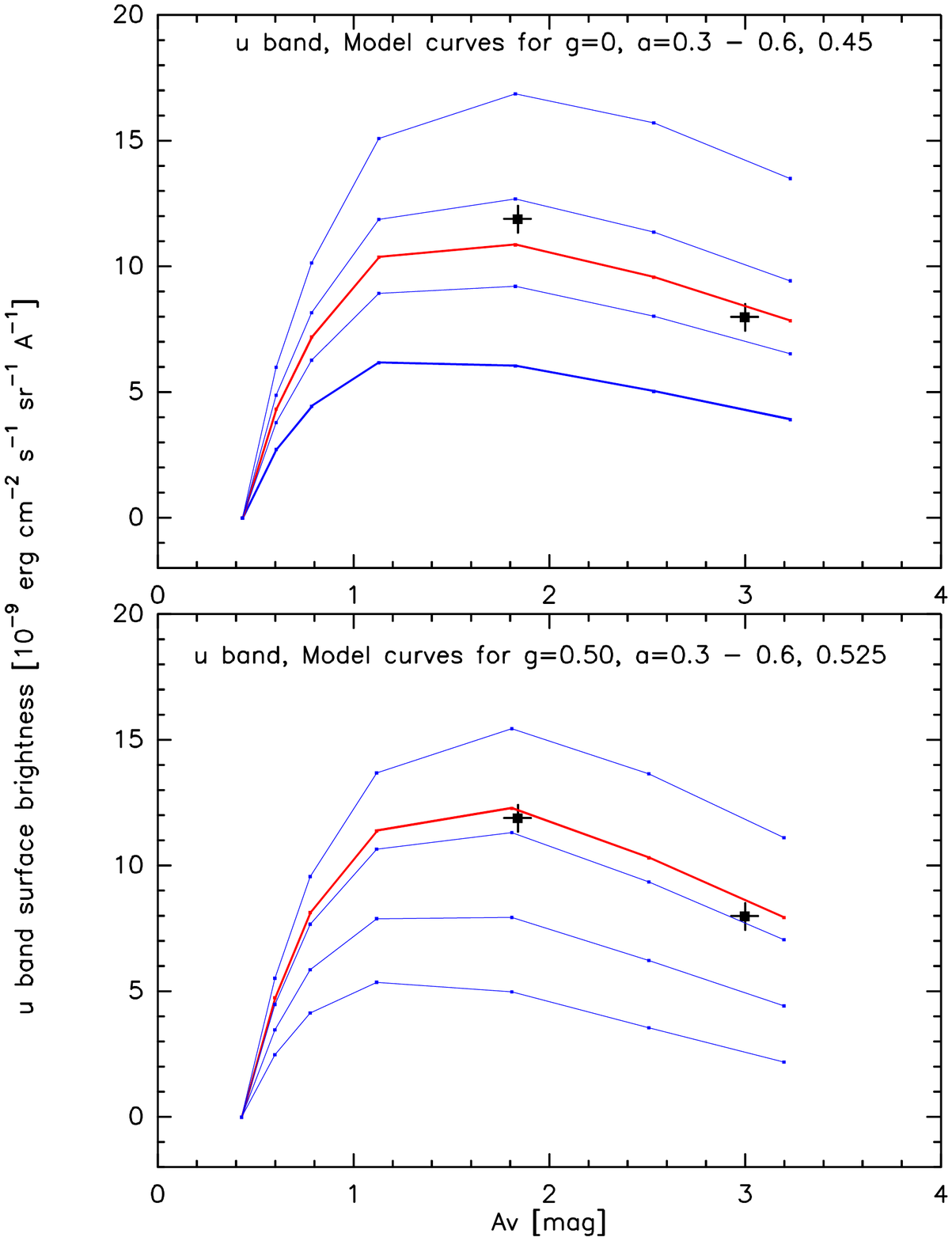}}
\vspace{-0.5cm}
\hspace{-1.5cm}
            {\includegraphics[width=10.5cm,angle=-0]{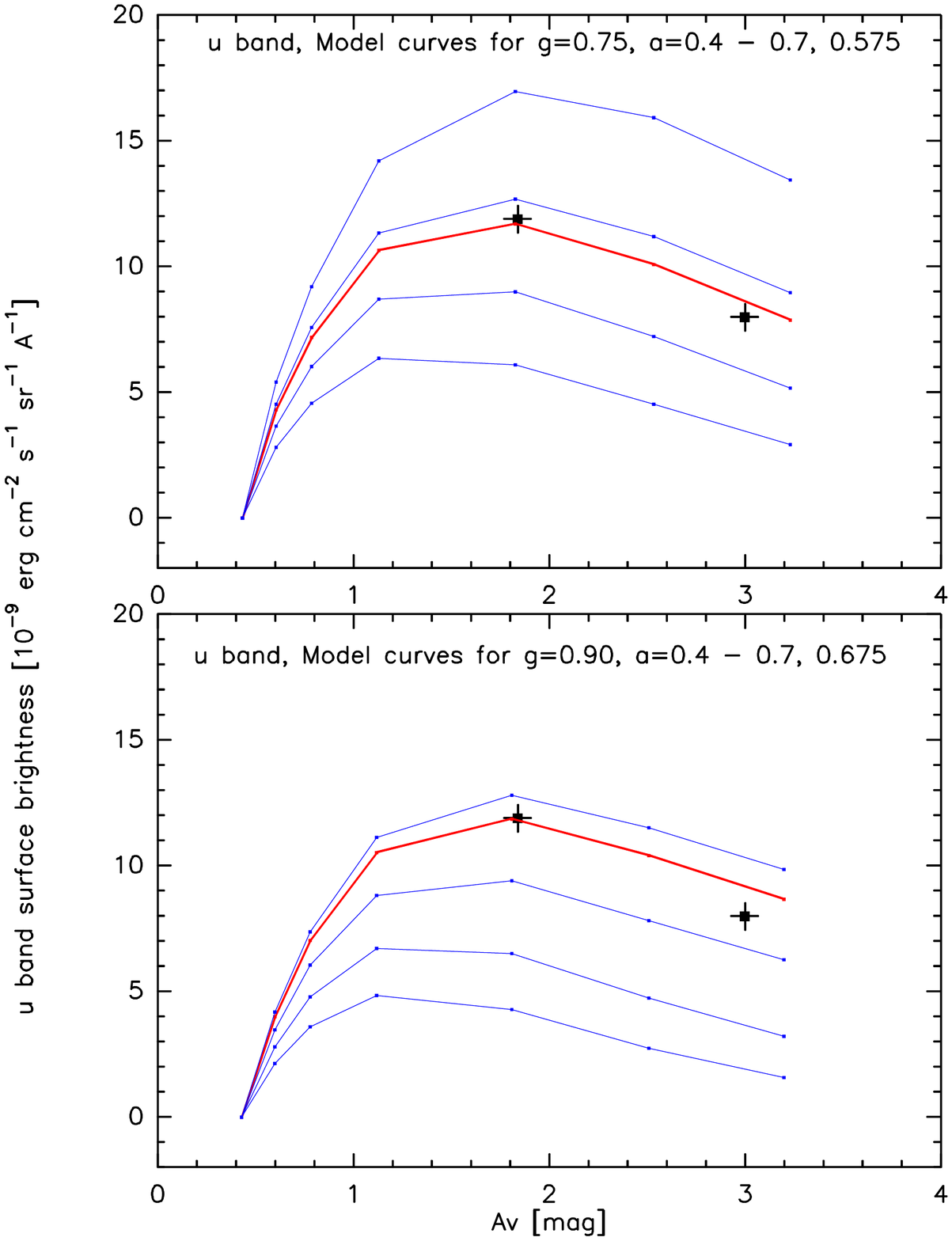}}
\vspace{-0.5cm}
\caption{Model fits of scattered light to the two $u$ band Calar Alto photometric
data points in LDN~1780, indicated with crosses (see Table~3).
Models are shown for four different forward scattering parameters, g =0, 0.5, 0.75 and 0.90, 
and a range of albedos as indicated in the figures. The blue lines are for albedos in regular steps of 0.1
and the red lines for an intermediate $a$, given as the last one in the list of  $a$ values.
The surface brightness is in 
units of \cgs\,.}
%\vspace{0.0cm}
%             \hspace{-0cm}
        \label{Isca_modelU}
\end{figure*}

\section{FUV and NUV scattered light model fits}
In this Appendix we present the model fits of the observed surface brightnesses (scattered light) for 
LDN~1780 and the Draco nebula for the $GALEX$ near and far UV wavelength bands, $\lambda = 1771 - 2831$\,\AA\, 
and  $\lambda = 1344 - 1786$\,\AA\,. 
\begin{figure*}
%   \resizebox{\hsize}{!}
\vspace{-1.5cm}
\hspace{-0.5cm} 
            {\includegraphics[width=10.5cm,angle=-0]{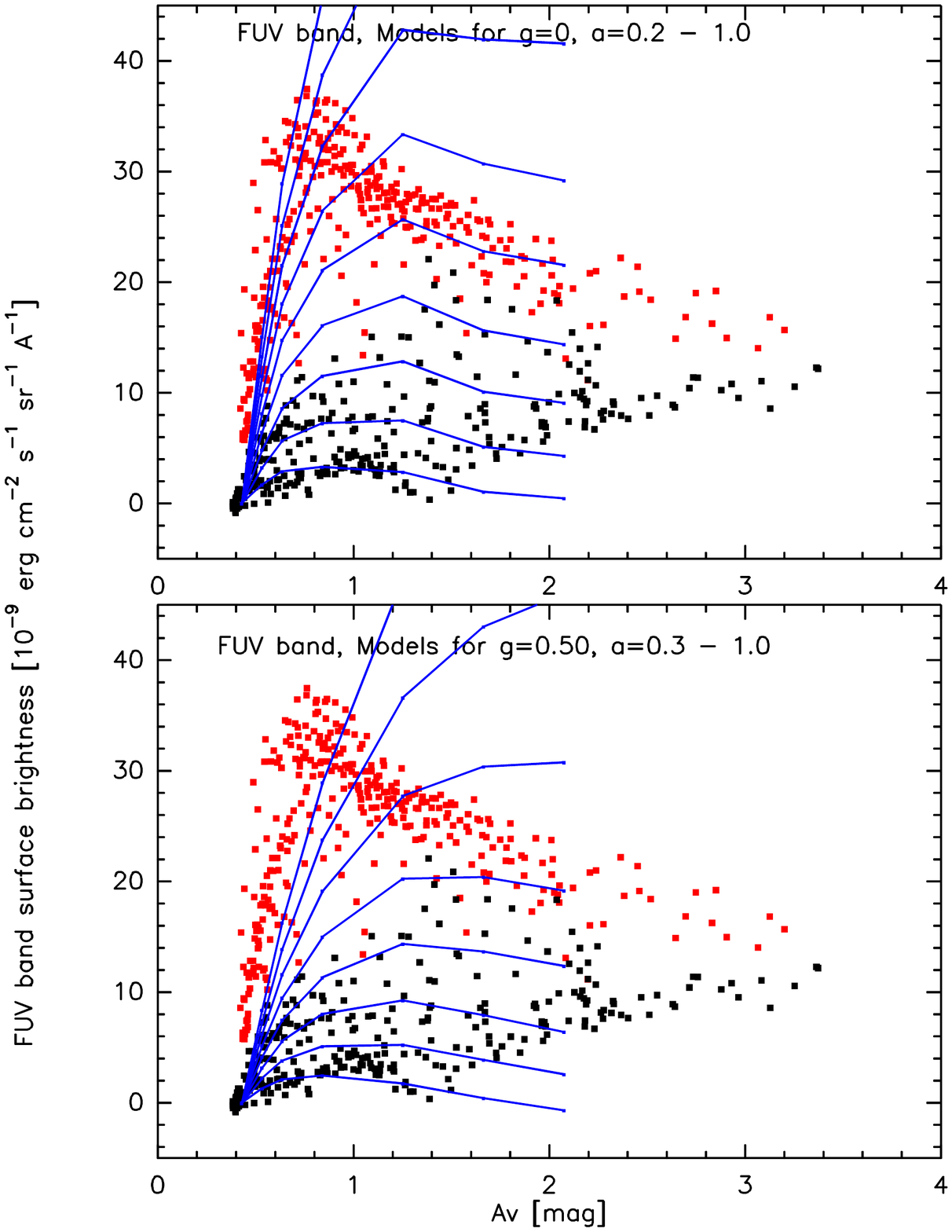}}
\vspace{-1.5cm}
\hspace{-1.5cm} 
%\hspace{0.0cm}
%\vspace{-1.0cm}
            {\includegraphics[width=10.5cm,angle=-0]{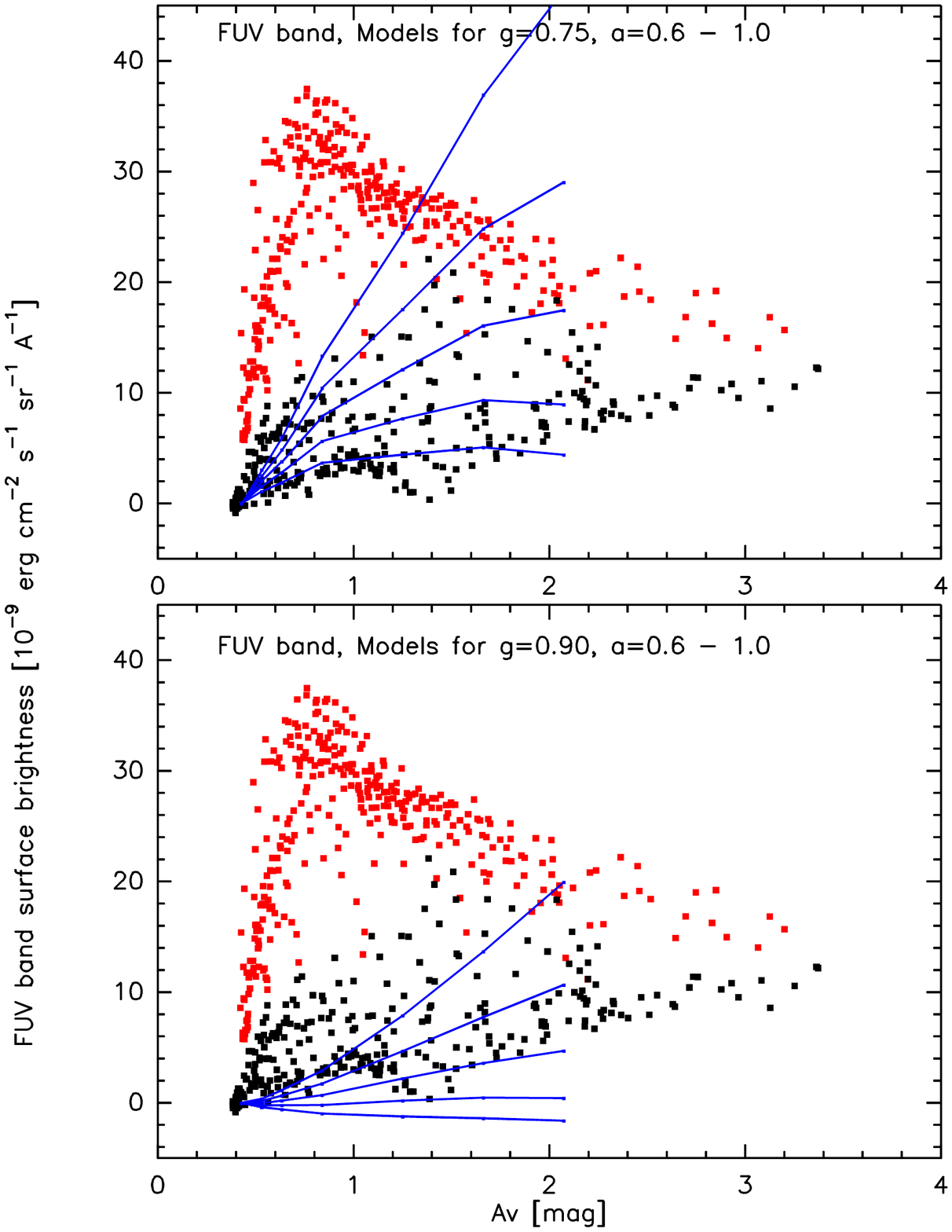}}
\hspace{-1.5cm} 
\vspace{-0.0cm}
%\vspace{-0.5cm}
            {\includegraphics[width=10.5cm,angle=-0]{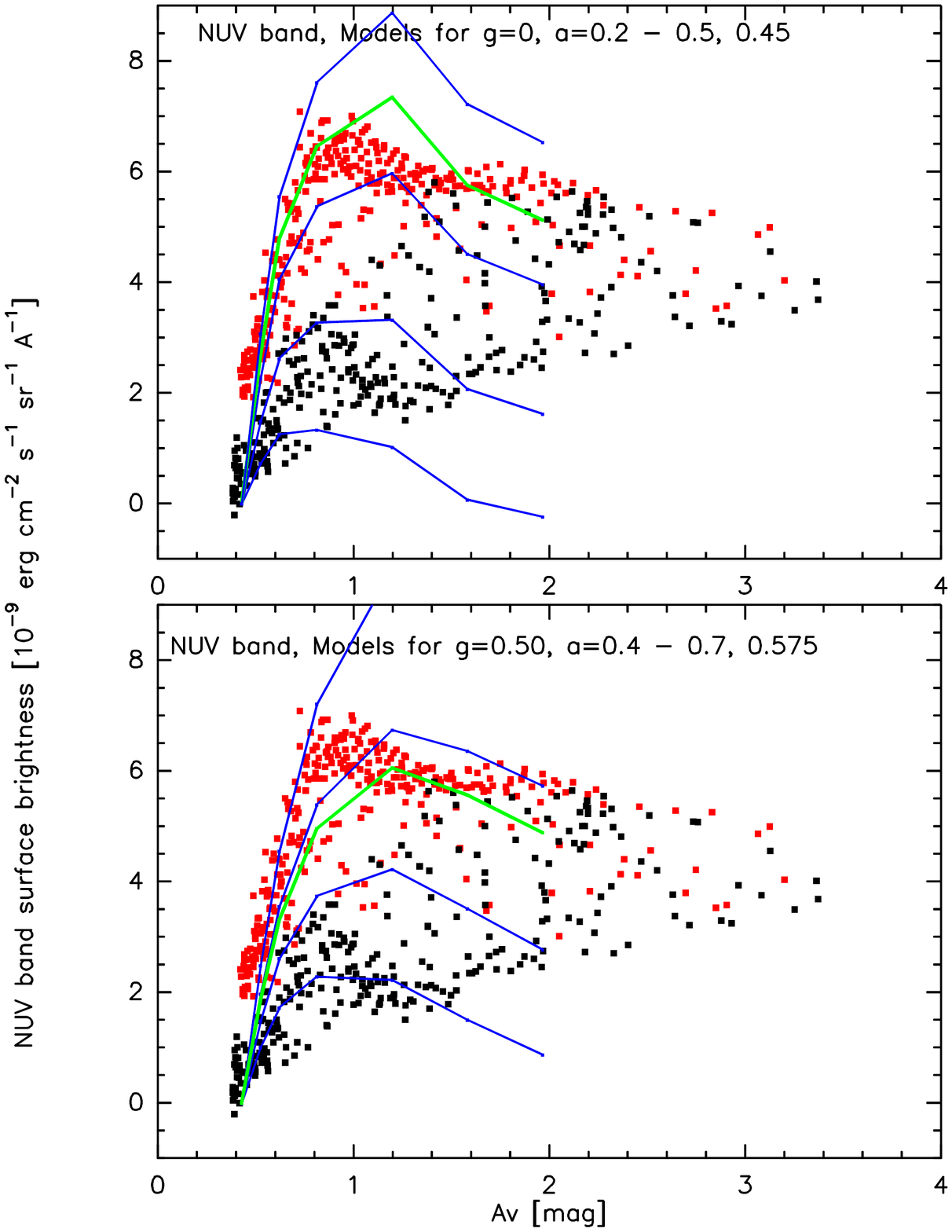}}
\vspace{-.0cm}
\hspace{-1.5cm}
            {\includegraphics[width=10.5cm,angle=-0]{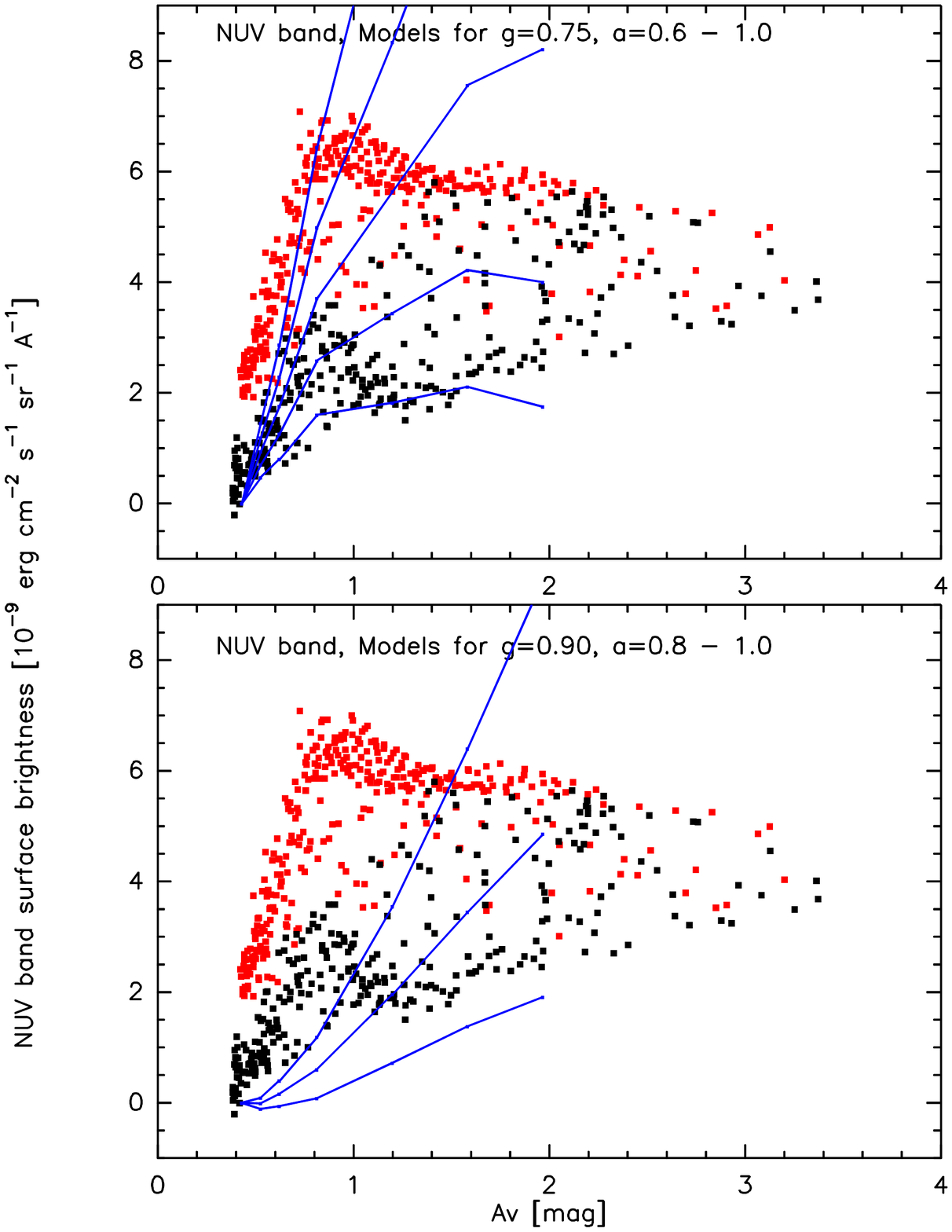}}

\vspace{-0.5cm}

\caption{Model fits of scattered light to the Galex NUV and FUV band data in LDN~1780. 
Models are shown for four different forward scattering parameters, g = 0, 0.50, 0.75 and 0.90, 
 and a range of albedos in regular steps of 0.1 (blue lines) and for an intermediate $a$ (green line), 
given as the last one in the list of  $a$ values.The surface brightness is in units of \cgs\,.}
             \hspace{-0cm}
        \label{L1780_Galex_models}
   \end{figure*}

\begin{figure*}
%   \resizebox{\hsize}{!}
\vspace{-1.5cm}
\hspace{-0.5cm} 
            {\includegraphics[width=10.5cm,angle=-0]{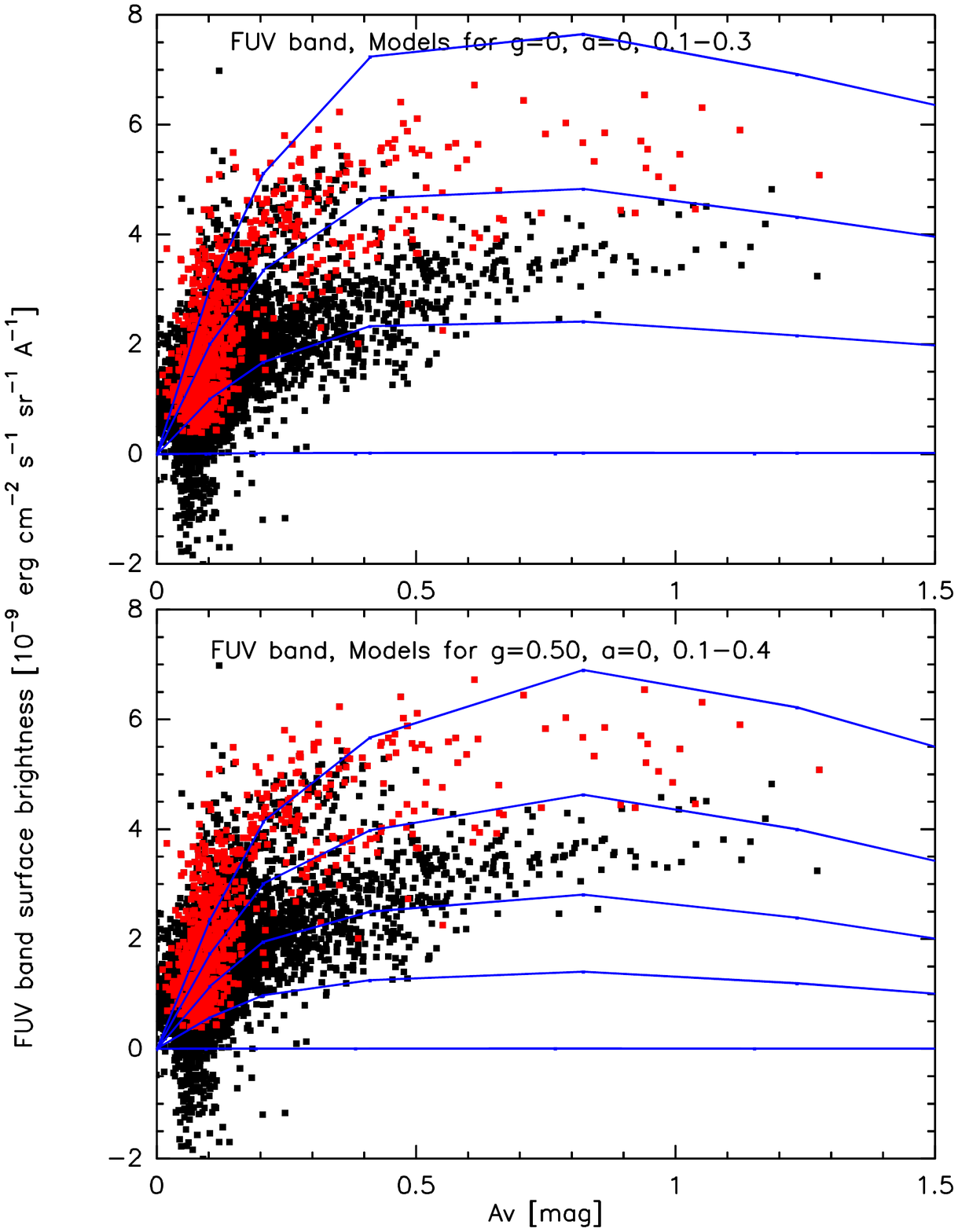}}
\vspace{-1.5cm}
\hspace{-1.5cm} 
%\hspace{0.0cm}
%\vspace{-1.0cm}
            {\includegraphics[width=10.5cm,angle=-0]{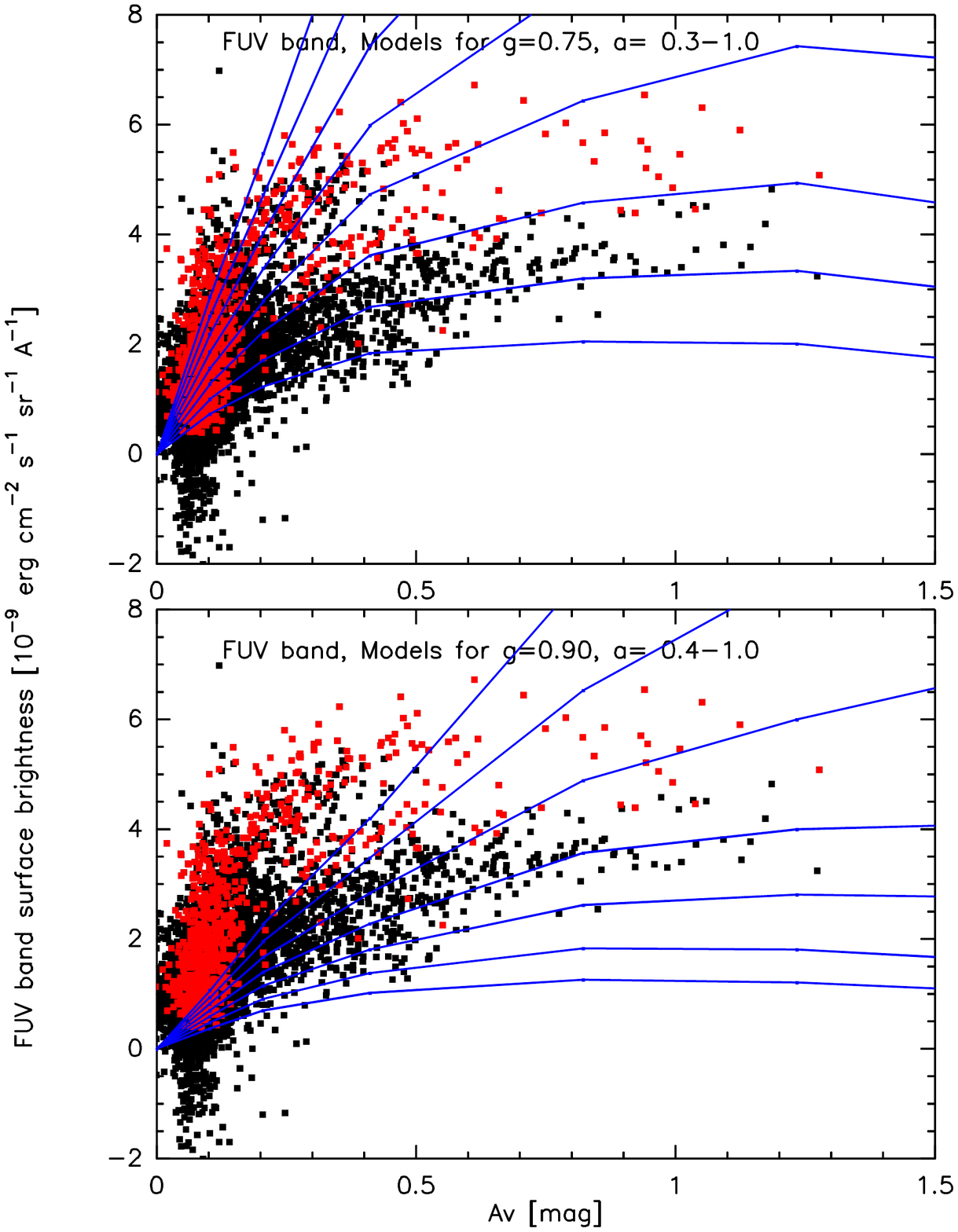}}
\hspace{-1.5cm} 
\vspace{-0.0cm}
%\vspace{-0.5cm}
            {\includegraphics[width=10.5cm,angle=-0]{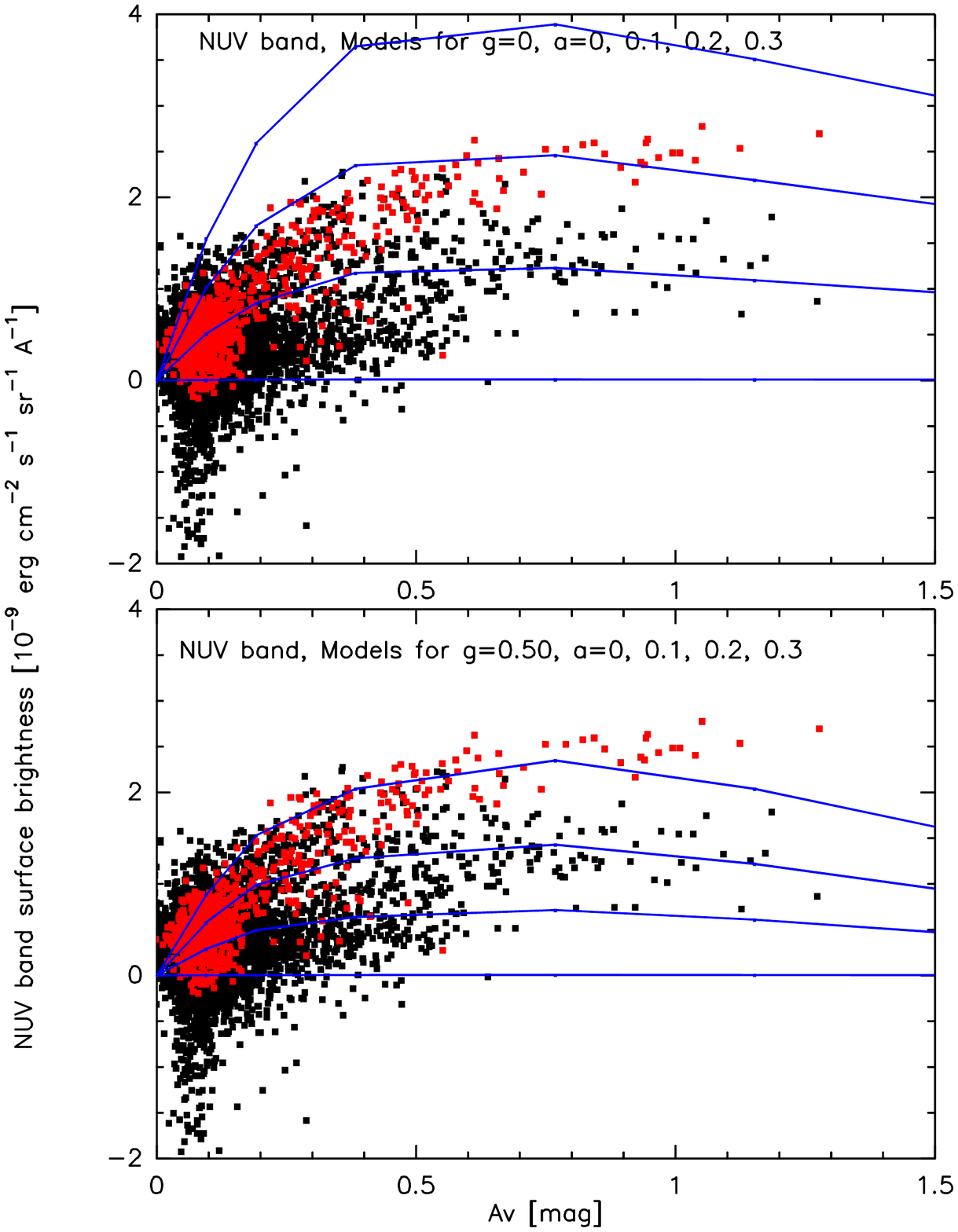}}
\vspace{-.0cm}
\hspace{-1.5cm}
            {\includegraphics[width=10.5cm,angle=-0]{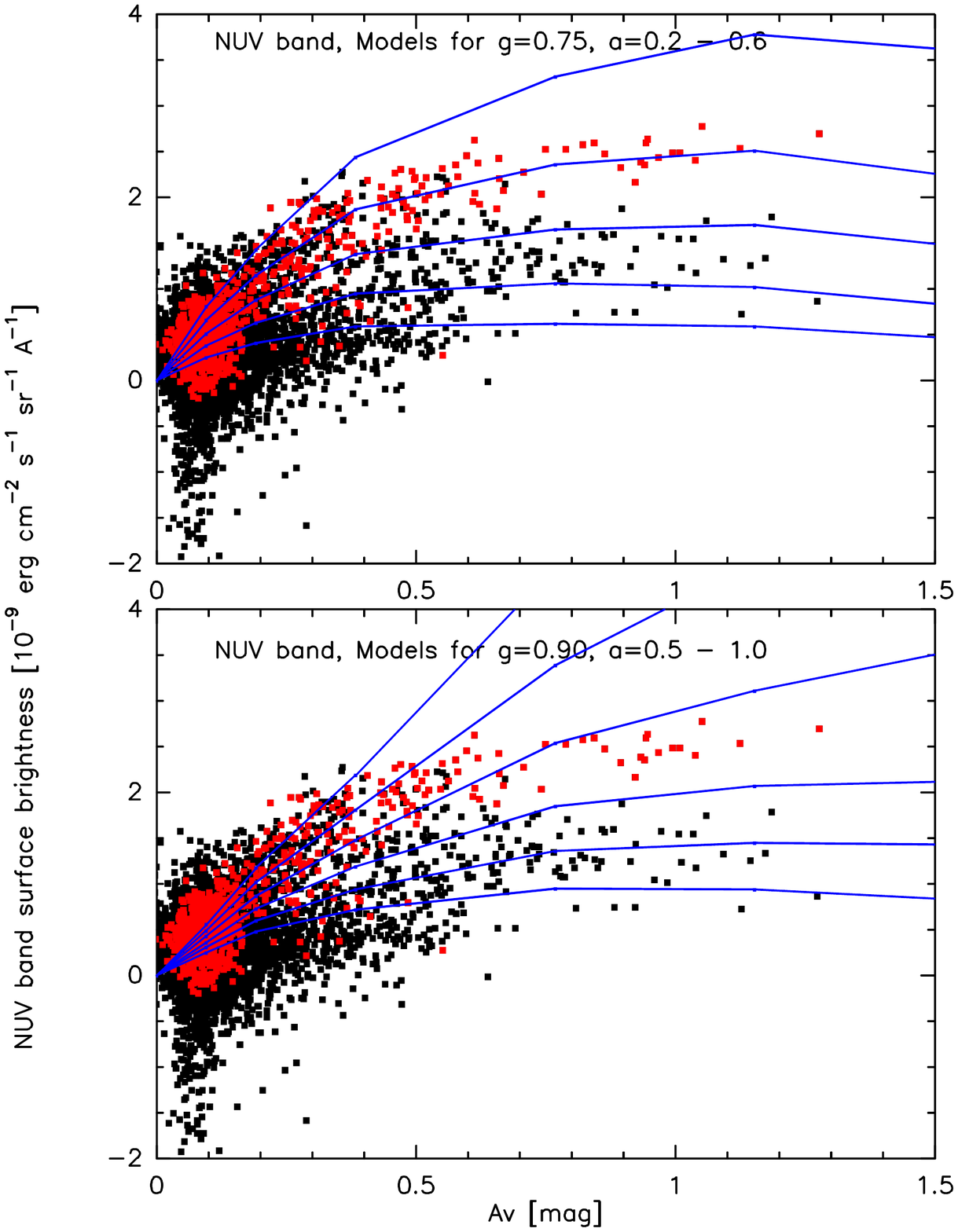}}

\vspace{-0.5cm}

\caption{Model fits of scattered light to the Galex NUV and FUV band data in the Draco nebula. 
Models are shown for four different forward scattering parameters, g = 0, 0.50, 0.75 and 0.90, 
 and a range of albedos in steps of 0.1, as indicated in the figures. 
The blue lines are for albedos in regular steps of 0.1. The surface brightness is in units of \cgs\,.}
             \hspace{-0cm}
        \label{Draco_Galex_models}
   \end{figure*}

\end{appendix}

\end{document}